\documentclass[aps, nofootinbib,superscriptaddress]{revtex4}
\usepackage{graphicx}
\usepackage{amsfonts}
\usepackage{psfrag}
\usepackage{amsmath}
\usepackage{float}
\usepackage{amssymb} 
\usepackage{dsfont}
\newcommand{\va}{\scriptscriptstyle}

\newcommand{\nn}{\sqrt{j(j+1)}}
\def\be{\begin{equation}}
\def\ee{\end{equation}}
\def\ba{\begin{eqnarray}}
\def\ea{\end{eqnarray}}

\def\Tr{\text{Tr}}

\def\ut#1{\rlap{\lower1ex\hbox{$\sim$}}#1{}}

\newcommand{\C}{\mathbb{C}}
\newcommand{\R}{\mathbb{R}}
\DeclareFontFamily{U}{rsfs}{}         % Formal Script            %
\DeclareFontShape{U}{rsfs}{m}{n}{<5> rsfs5 <6><7> rsfs7          %
  <8><9><10><10.95><12><14.4><17.28><20.74><24.88> rsfs10}{}     %
\DeclareMathAlphabet{\mathfs}{U}{rsfs}{m}{n}                     %
\newcommand{\mfs}[1]{\mathfs {#1}}                               %
\newcommand{\n}{{\nonumber}}

\newcommand{\sH}{{\mfs H}}
\newcommand{\sL}{{\mfs L}}

\newcommand{\sN}{{\mfs N}}

\newcommand{\sZ}{{\mfs Z}}

\newcommand{\sI}{{\mfs I}}
\newcommand{\sO}{{\mfs O}}
\newcommand{\h}{{\sH}}

\begin{document}

\title{Black Holes in Loop Quantum Gravity}

\author{Alejandro Perez}
\affiliation{
Centre de Physique Th\'eorique,   Aix Marseille Université, Université de Toulon, CNRS, UMR 7332, 13288 Marseille, France.
}
\begin{abstract}
This is a review of the results on black hole physics in the framework of
loop quantum gravity. The key feature underlying the results is the discreteness of 
geometric quantities at the Planck scale predicted by this approach to quantum gravity.
Quantum discreteness follows directly from the canonical quantization prescription when applied to the 
action of general relativity that is suitable for the coupling of gravity with gauge 
fields and specially with fermions. Planckian discreteness and causal considerations provide the basic structure 
for the understanding of the thermal properties of black holes close to equilibrium. 
Discreteness also provides a fresh new look at more (at the moment) speculative issues such as those concerning the fate 
of information in black hole evaporation. The hypothesis of discreteness leads also to interesting phenomenology with possible
observational consequences. The theory of loop quantum gravity is a developing  program. This review reports its achievements and open questions in a pedagogical manner with an emphasis on quantum aspects of black hole physics.   
\end{abstract}

\maketitle

%\section{Conventions and definitions}
%
%We will often trade use the isomorphism $so(3)\to \R^3$ defined by 
%\be \alpha^{ij}\to \alpha ^i=-\frac{1}{2}
%\epsilon^{i}_{\ jk} \alpha^{jk}.\ee
%This which implies that \be F^i=dA^i+\frac{1}{2}
%\epsilon^i_{\ jk} A^j\wedge A^k\ee and
%\be d_A\alpha^i=d\alpha^i+\epsilon^i_{\ jk} A^j\wedge \alpha^k.\ee  
%Given $\alpha^i$ and $\beta^i$ $p$-form and $q-$form respectively valued in $so(3)$, the exterior covariant derivative $d_A$ satisfies
%\be
%d_A(\alpha_i\wedge \beta^i)=(d_A\alpha_i)\wedge \beta^i+(-1)^{p} \alpha_i\wedge(d_A \beta^i).
%\ee
%The Poisson bracket among basic variables reads
%\be
%\{E^i_{ab},A^j_c\}=\kappa \gamma \epsilon_{abc}\delta^{ij} \delta^{(3)}(x,y)
%\ee
%which is more nicely written in terms of smeared local fields
%\be \left\{\int a_i\wedge E^i, \int b_j\wedge A^j\right\}
%=\kappa \gamma\int a_i \wedge b^i\ee

\section{Introduction} 

Loop quantum gravity (LQG) is an approach to a background independent quantization of the gravitational interaction
based on the non-perturbative canonical quantization of general relativity.   In this framework space-time geometry itself is a dynamical variable that has to be suitably quantized and described in the absence of any background reference geometry. The proposal is still in progress and many important questions remain open. All the same, there are results providing a solid picture of what the quantum nature of space-time at the fundamental scale could be like. One of these key results is that space-time geometric operators acquire discrete spectra: states of the gravitational degrees of freedom can be spanned in terms of spin-network states  each of which admits the interpretation of an eigenstate of geometry which is discrete and atomistic at the fundamental level \cite{Rovelli:1994ge, Ashtekar:1996eg, Ashtekar:1997fb}. Quantum space is made of polymer-like excitations of quantum geometry where one dimensional fluxes of quantised area connect at nodes carrying quantum numbers of volume. The dynamical rules of evolution for these states are also discrete \cite{Perez:2012wv}. Locality and topology are replaced by the relational notions of connectivity of the underlying network of space quanta.  In this framework, the continuum spacetime formulation of general relativity and quantum field theory is seen as the low energy limit of a fundamentally discrete and combinatorial entity.
 
A large body of results in the concrete physical situation defined by quantum aspects of black hole physics have been produced in recent years.  This article aims at presenting these recent developments in an organic and pedagogical way.
The theory of black holes that follows from the LQG underlying model is still incomplete, partly due to the technical difficulties in defining the notion of black holes in the quantum realm, partly because of the intrinsic difficulties associated with the definition of the dynamics and the low energy limit of the fundamental theory.  Nevertheless, the theory indicates a solid conceptual perspective that produces promising insights into the nature of quantum gravity in general. This is an account of a research program in progress.  

The approach that we will describe can seem quite peculiar from (what sometimes can appear as) the main stream of thought in the holographic tumult of the high energy community. However, we will see, the perspective that arises from our analysis is actually quite conservative and presents many analogies with the behaviour of standard physical systems. 
The tension with other more popular approaches resides in the complete lack of compliance with any fundamental  notion of holographic principle \cite{Bousso:2002ju}.  Despite this, it can be shown that the theory of black holes stemming from LQG is indeed consistent with the black hole phenomenology derived from semiclassical analysis (we could call this an emergent weak holography). Thus, the  picture of LQG is very different from the bulk-boundary-duality type of quantum gravity scenario  proposed by  the ADS-CFT correspondence \cite{Maldacena:1997re}.  We will see that the alternative offered by the LQG treatment may present important advantages in avoiding certain inconsistencies in the description of gravitational collapse and subsequent black hole evaporation. 

The article will also review the theoretical basis leading to the prediction of discreteness of quantum geometry by LQG.
In Section \ref{two} we will briefly review the construction of the phase space of general relativity starting form an action and variables that satisfy a criterion of naturality once some general principles are stressed. We will see that the roots of discreteness of quantum geometry are found in Heisenberg's uncertainty  relations for geometric quantities.  The inclusion of black holes in terms of boundaries satisfying suitable boundary conditions will be described in Section \ref{qg}. The quantisation of the volume and area operators will be sketched in Section \ref{quantum-geometry}. In Section \ref{BHE}  we will apply the formalism to the problem of computing black hole entropy. In Section \ref{loss} we will discuss the problem of the fate of information in black hole evaporation, and some phenomenological ideas with possible observational consequences that are motivated by the discussion of information loss. 

Throughout this paper there might be sections that seem too technical for a general reader not necessarily  interested in all the mathematical details.  Equations are written to guide the argumentation and, for general readers, are important only in this sense. 
Once equations are written they call for technical precision (important for those that might be interested in detailed derivations); however, in spite of their apparent complexity due to the presence of indices and other tensorial operations that are often necessary in the presentation of field theoretical notions in the context of general relativity, their message should be transparent when ignoring these details. The reader more interested in the conceptual line  should read these equations without paying too much of attention to the details of the index structure and concentrate rather on their algebraic form. This is specially so for the construction of the phase space of general relativity; Section \ref{qg} (very important for us as it implies the Poisson non-commutativity of geometry behind quantum discreteness). Classical mechanics is briefly described in its symplectic formulation at the beginning so that all the equations that follow, and are important for gravity, can be interpreted by analogy with these initial equations.  Geometric units ($G_N=c=1$) are used in discussions so that energy, mass, and time are all measured in the same units as legth.
 
\subsection{Black hole thermodynamics: an invitation to quantum gravity} \label{twoty}

Black holes are remarkable solutions of general relativity describing the classical  aspects of the late stages
of gravitational collapse. Their existence in our nearby
universe is by now supported by a great amount of observational evidence \cite{Narayan:2013gca}. When isolated, these systems
become very simple as seen by late and distant observers. Once the initial very dynamical phase of collapse has passed (according to physical expectation and the validity of the `no-hair theorem'\footnote{The no-hair theorem is   a collection of results by Hawking, Israel, Carter and others implying that a stationary (axisymmetric) black hole solution of Einstein's equations coupled with Maxwell fields must be Kerr-Newman \cite{Israel:1967wq, Israel:1967za, Carter:1971zc}. Some aspects of this result remain without complete proof and some authors refer to is at the no-hair conjecture (for more details see \cite{Dafermos:2008en} and references therein). The physical relevance of Einstein-Maxwell resides in the fact that gravity and electromagnetism are the only long range interactions. Other forces might be relevant for the description of the matter dynamics during collapse but play no role in describing the final result where matter has already crossed the BH horizon. }) the system settles
down to a stationary situation completely described by a member of the Kerr-Newman family. These are solutions of Einstein's equations coupled with electromagnetism representing a stationary and axisymetric black hole characterised by three parameters only:  its mass $M$, its the angular momentum $J$, and its electromagnetic charge $Q$.

The fact that the final state of gravitational collapse is described by only a few macroscopic parameters, independently of the details of the initial conditions leading to the collapse, is perhaps the first reminiscence of their thermodynamical nature
of black holes.   As we will review here, there is a vast degeneracy of configurations (microstates) that can lead to a same final stationary macroscopic state, and the nature of these microstates becomes manifest only when quantum gravity effects are considered. 
Another classical indication of the thermodynamical nature of black holes (BHs) emerged from the limitations on amount of energy that could be gained from interactions with BHs in thought experiments such as the Penrose mechanism \cite{Penrose:1971uk} and the phenomenon of BH superradiance \cite{Starobinsky:1973aij}; its field theoretical analog. Later it became clear that such limitations where special instances of the very general Hawking's area theorem \cite{Hawking:1971tu} stating that for natural energy conditions (satisfied by classical matter fields) the area $a$ of a black hole horizon can only increase in any physical process. This is the so-called second law of black hole mechanics which reads:
\be\label{2nd}
\delta  a\ge 0.
\ee
This brings in the irreversibility proper of thermodynamical systems to the context of black hole physics and motivated Bekenstein \cite{Bekenstein:1973ur, Bekenstein:1972tm}
to associate to BHs a notion of entropy proportional so their area.
Classically, black holes also satisfy the so-called first law of BH mechanics \cite{Bardeen:1973gs} which is an energy balance equation
relating different nearby stationary BH spacetimes according to
\begin{eqnarray}\label{1st}
\delta M=\underbrace{\frac{\kappa}{8\pi} \delta a }_{heat?}+\Omega \delta J+\Phi  \delta Q,
\end{eqnarray}
where $\Omega$ is the angular velocity of the horizon, $\Phi$ is the horizon electric potential, and $\kappa$ is the surface gravity which plays the role 
of a temperature in the analogy with thermodynamics. The surface gravity, defined only in equilibrium,  can be related to an intrinsic local geometric quantity associated with the BH horizon; it takes a constant value on the horizon  depending only on the macroscopic parameters $M, Q$ and $J$
(for the simplest non-rotating and uncharged BH $\kappa=1/(4M)$). The homogeneity of $\kappa$ on the horizon is
called the zeroth law of BH mechanics. The other intensive parameters $\Omega$ and $\Phi$ are also functions of $M, Q$ and $J$ only (their explicit expression can be found for instance in \cite{wald}).

With the exception of the horizon area $a$, all the quantities appearing in the first law have an unambiguous  physical meaning  for asymptotic inertial observers at rest at infinity: $M$ is the total mass defined in terms of the Hamiltonian generating time translation for these observers, $J$ is the generator of rotations around the BH symmetry axis, etc. The quantity $\Phi$ is the electrostatic potential difference between the horizon and infinity, $\Omega$ is the angular velocity of
the horizon as seen from infinity, and $\kappa$ (if
extrapolated from the non-rotating case) is the acceleration of the stationary
observers as they approach the horizon as seen from infinity \cite{wald}. It is possible however to translate the first law in terms of physical quantities measures by quasi-local observers close to the BH horizon \cite{Frodden:2011eb}.  This  clarifies the role of the horizon and its near spacetime geometry as the genuine thermodynamical system. 

\begin{figure}[t]
%\begin{center}
\centerline{\hspace{0.5cm} \(
\begin{array}{c}
\includegraphics[height=12cm]{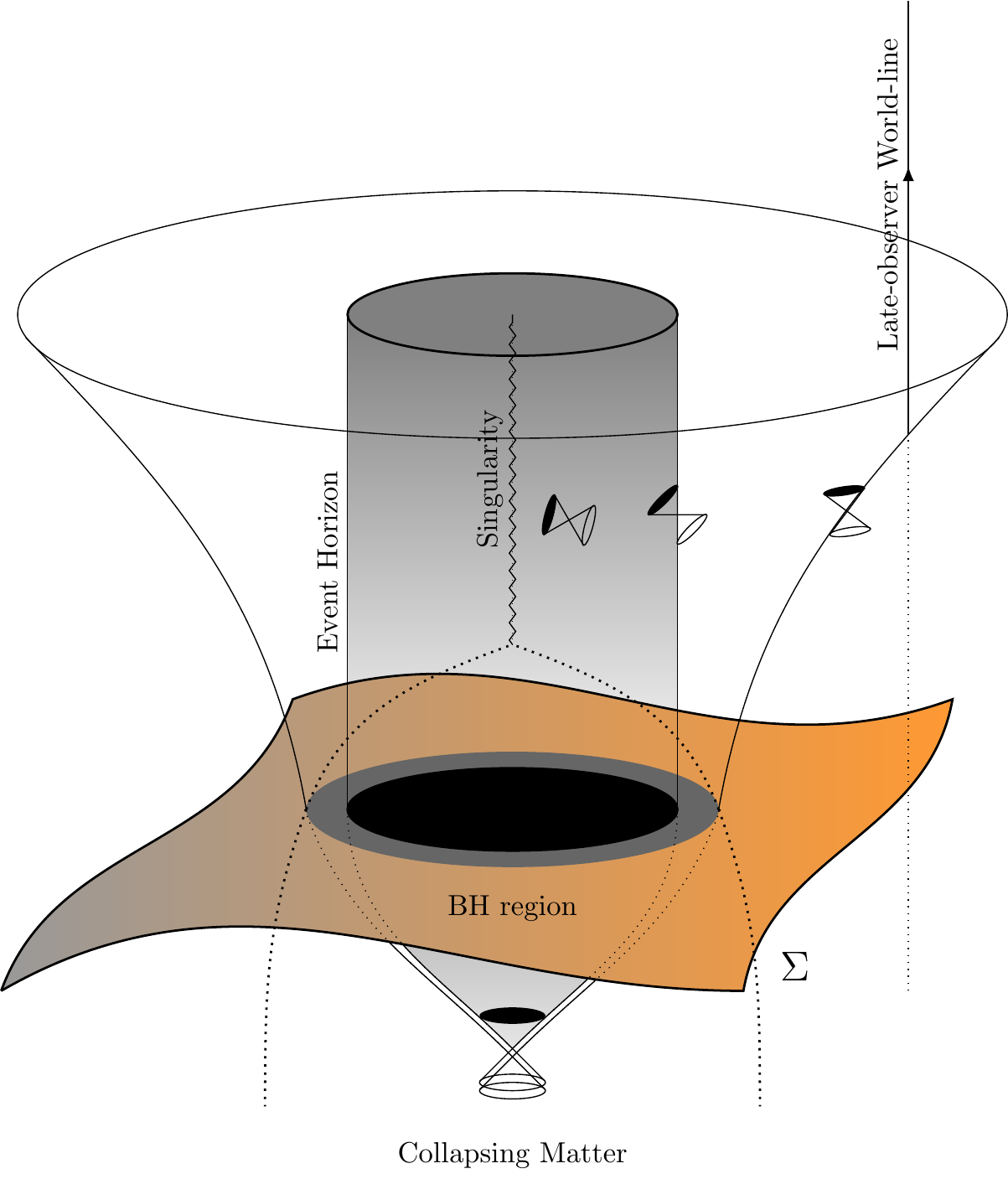} 
\end{array}\) } \caption{Spacetime representing gravitational collapse (time direction upwards). The matter of a compact objet (dotted lines) collapses and forms a singularity inside a black hole event horizon: a region where classical general relativity breaks down. The system settles down to a stationary black hole spacetime for late observers ($\delta t \gg M$). An outgoing light wave-front is shown separating the early (very dynamical) phase from the late (equilibrium) phase. Late stationary observers can ``see'' only a shell-like region of the matter and spacetime outside the black hole with a volume on a Cauchy surface $\Sigma$ (representing an instant around the collapsing moment) that is exponentially squeezed in the outward direction. The system of interest (grey shell-like region) for these observers is effectively 2-dimensional!  Light cones are shown to make manifest the main features of the causal structure.}
\label{figu}
%\end{center}
\end{figure}

The realization that black holes can indeed be considered (in the semiclassical regime) as thermodynamical systems came with the
discovery of black hole radiation. In the mid 70's Hawking considered the scattering of a quantum test field on a space time background geometry representing gravitational collapse of a compact source \cite{Hawking:1974sw}. Assuming that very early observers far away from the source prepare the field in the vacuum state, he showed that---after the very dynamical phase of gravitational collapse has ended and the space time settles down to a geometry well described by that of a stationary black hole---late and faraway observers in the future (see Figure \ref{figu}) measure an afterglow of particles of the test field coming from the horizon with a temperature
\be
T =\frac{\kappa \hbar}{2\pi}.
\ee  
For the case Schwarzschild  BH ($Q=J=0$), the radiation temperature is $T=\hbar/(8\pi M)$.
As black holes radiate the immediate conclusion is that they must evaporate through the (quantum phenomenon of) emission of Hawking radiation. This expectation is confirmed by the study of the the expectation value of the energy-momentum tensor in the corresponding quantum state that shows that there is a net flux of energy out of the BH horizon (see for instance \cite{Birrell:1982ix, Parker:2009uva} and references therein). The quantum energy-momentum tensor violates the energy conditions assumed in Hawkings area theorem and allows the violations of \eqref{2nd}: the horizon can shrink. The calculation of Hawking assumes the field to be a test field and thus neglects by construction the back reaction of such radiation.
However, it provides a good approximation for the description of black holes that are sufficiently large in order for  the radiated power to be arbitrarily small. 

This result together with the validity of the first and second laws imply that semiclassical black holes should be associated an 
entropy (here referred to as Bekenstein-Hawking entropy) given by 
\be
S_{H}=\frac{a}{4\ell_p^2}+S_0
\ee
where $\ell_p=\sqrt{\hbar G_N/c^3}$ is the Planck scale,   $S_0$ is an integration constant that cannot be fixed by the sole use of the first law. 
In fact, as in any thermodynamical system, entropy cannot be determined by sole thermodynamical considerations. 
Entropy can either be measured in an experimental setup (this was the initial way in which the concept was introduced) or calculated from basic degrees of freedom using statistical mechanical methods once a model for these fundamental building 
blocks of the system is available. Remarkably, the functional dependence of the entropy of a BH on the area was argued 
by Bekenstein first on statistical mechanical terms \cite{Bekenstein:1973ur}. 

%%%%%
In this way thermodynamics shows once more its profound insights into the physics of  
more fundamental degrees of freedom behind macroscopic variables. In this case by shedding light on the nature of the quantum gravitational building blocks of spacetime geometry. As in standard systems, the first law for BHs implies that---when considering energy changes due to the action of macroscopic variables
(e.g. work done by changing the volume)---there is a part of the energy that goes into the microscopic
molecular chaos (i.e. heat). Thus the first law describing the physics of steam machines and internal combustion engines of the nineteenth century
reveals the existence of the microscopic physics of molecules and atoms. Moreover, it is by trying to construct a consistent description of the 
thermodynamics of photons that Planck made the founding postulate of quantum mechanics \cite{Planck:1901tja} (more explicit in Einstein \cite{Einstein:1905cc}) that radiation too
is made of fundamental building blocks called photons. {\bf Similarly, in the present context, equation (\ref{1st}) is a clear physical indication that the smooth spacetime geometry    
description of the gravitational field must be replaced by some more fundamental atomistic picture. }
In this way, black holes offer a privileged window for learning about quantum gravity.

%%%%%
Mathematically, even though the thermodynamical nature of semiclassical black holes is a
robust prediction of the combination of general relativity and quantum field theory as a first approximation of quantum gravity, 
the precise expression for the entropy of black holes is a question that can only be answered within the
framework of quantum gravity. This is a central question for any 
proposal of a quantum gravity theory.

%The discovery that black holes are thermodynamical systems is therefore, the most solid guiding 
%principle revealing the existence of an underlying fundamentally `atomistic' structure of spacetime geometry at the Planck scale. 
%The calculation of black hole entropy  in the semiclassical regime is only the tip of a great mountain of puzzles that have 
%followed Hawking's discovery.  

\subsection{Weak Holography}\label{intro-holo}

A surprising property of the Bekenstein-Hawking entropy of a black hole is that it is proportional to the area $a$ of the event horizon instead of scaling linearly with some three-dimensional volumetric measure of the systems size. The fact that black hole entropy scales as in a lower dimensional system together with the discovery of bounds on the entropy of compact objects (conjectured via the analysis of thought experiments involving black holes and conventional objects; see \cite{Bekenstein:1980jp, Bekenstein:1984vm, Bekenstein:1983iq, Unruh:1983ir, Marolf:2003wu}) has led an important part of the quantum gravity community to believe in the so-called {\em holographic principle} \cite{Bousso:2002ju}. In its crudest form the principle states that the classical  physical world should admit a fundamental description in terms of a hologram on a lower dimensional screen. This is a view that the  ADS-CFT formulation of string theory incarnates \cite{Maldacena:1997re}.

In LQG we do not see any convincing evidence for the need for such a radical principle, and subscribe to some weaker 
notion that has been described as {\em weak holography} \cite{Smolin:2000ag}. 
The reason for this view is that all the apparently puzzling properties of black holes and their interactions with external agents appear to be completely consistent once  the following two ingredient are combined:
discreteness at Planckian scales, and compatibility with the causal structure predicted by general relativity in the continuum limit. Both ingredients are expected features in LQG.  The holographic principle plays no role in the construction of the theory.

Causality is one of the keys for understanding the system at hand. This can be clearly illustrated in an intuitive manner with the help of the spacetime representation of gravitational collapse shown in Figure \ref{figu}. Concretely,  consider a BH of mass $M$ and an external observer that becomes a stationary observer\footnote{A stationary observer is an observer at constant $r,\theta$, and $\phi$ in a Kerr-Newman spacetime in Boyer-Lindquist coordinates \cite{wald}. More generally, is an observer whose $4$-velocity is parallel to a timelike Killing field for a stationary spacetime. } in the asymptotic future when its proper time becomes large, $\tau\gg M$, when measured starting at some arbitrary instant around the moment of collapse. In Figure \ref{figu} this is defined by a Cauchy surface $\Sigma$ placed around the region where the horizon settles down to a stationary one. Due to the presence of the event horizon (itself an outgoing light-like surface trapped with zero radial expansion in the stationary region), an outgoing light wave-front leaving the collapsing matter, outside but close to the event horizon, remains close to the event horizon for a long ``time" (more precisely for long values of an affine parameter along the outward-pointing null geodesics) until it finally escapes the strong gravitational region towards infinity (see Figure \ref{figu}). This implies that, in order for the wave front to reach the observer at late times $\tau\gg M$, it must have left the collapsing body from a proper distance $\ell$ to the horizon at instant $\Sigma$ that scales as $\ell\approx M \exp(-\tau/(8M))$ (as a simple calculation in the spherically symmetric case would show). In other words, the portion of the collapsing body ``seen'' by a late observer---for whom the spacetime looks stationary and hence the laws of BH mechanics apply---corresponds to an exponentially-thin hyper annulus given by the region contained between the surface at constant proper distance $\ell$ from the horizon and the horizon itself.  

The previous exponential relationship implies that $\ell$ becomes quickly smaller than $\ell_p$ for late proper time $\tau$ of the external observer. However,  we cannot trust the classical expression for $\ell$ all the way down to transplanckian scales. If we think of these fluctuations as affecting the position of the outgoing wave-front from the boundary of the hyper annulus, then uncertainty in its position sets a natural lower bound for $\ell$ of the order of Planck's length $\ell\approx \ell_p$. Thus the volume outside the BH that  the very late outside observer can actually see is given by
\be
v=a\ell_p.
\ee  
We know the system radiates and it is in a close-to-thermal equilibrium state (at least for large BH masses $M$ in Planck units). Statistical mechanical arguments based on equipartition of probability for volumetric fundamental bits imply that the systems entropy should scale linearly with $V$ in Planck units from which we get that 
\be S\approx \frac{v}{\ell_p^3}=\frac{a}{\ell_p^2}, \ee
which is in agreement the Bekenstein-Hawking area entropy law and based on a completely standard statistical mechanical 
rationale with no need to invoking an hypothetical holographic principle.  

One could objet to the above argument that standard statistical mechanical reasoning also suggests that the entropy should grow linearly with the energy of the system. Remarkably, it turns out that energy and area are proportional to each other when one considers the system described above. Useful notions of energy are scarce in general relativity due to its necessary link to a time translational symmetry that is not always available in arbitrary gravitational configurations \cite{Szabados:2009eka}. When the black hole spacetime is asymptotically flat then there are standard definitions of its total energy content such as the ADM mass \cite{Arnowitt:1959ah}, or the Bondi mass \cite{Bondi:1962px}, when the spacetime is stationary there is also the Komar mass \cite{Komar:1958wp}. For the Kerr-Newman BHs (the most general stationary BH solutions classically expected to represent the end result of gravitational collapse) the previous three notions coincide and correspond to the quantity called $M$ in \eqref{1st}. However, non of these energy notions are appropriate for describing the system at hand as they are global notions referring to the energy content of the entire spacetime.  For stationary black holes one can  show \cite{Frodden:2011eb}, using perturbation theory and Einstein's equations, that exchanges of energy (as defined by local stationary observers) with the system, defined as the 
annulus around the horizon mentioned above, are directly related to changes of its area according to the simple law
\be
\delta E=\frac{\delta a}{8\pi\ell},
\ee
where $\delta E$ is the standard notion of energy content of the matter falling into BH for local stationary observers, i.e., the one that a calorimeter held stationary close to the horizon would register if captured by the devise. This implies that the natural measure of the internal energy $E$ of the system of interest behaves linearly with the area $E\propto a$. Once the appropiate local notion of energy is invoked the apparent tension between the area scaling of the entropy and standard thermodynamics 
disappears. 

In addition to the area-scaling of BH entropy, the holographic hypothesis is said to be supported by entropy bounds
for weakly gravitating systems. These bounds where originally proposed by Bekenstein \cite{Bekenstein:1980jp} who studied suitable thought experiments designed to test the validity of the so-called {\em generalized second law} \cite{Bekenstein:1974ax} of thermodynamics in situations where black holes would be fed with regular matter.  Covariant versions of these bounds were constructed by Bousso \cite{Bousso:1999xy}. However, recent results \cite{Casini:2008cr, Bousso:2014sda} strongly suggest that these bounds, when defined 
in a precise manner, turn out to be valid in the context of standard quantum field theory semiclassically coupled to gravity. Thus their validity in the
setting of a theory that is by no means holographic (in the sense of the holographic principle \cite{Bousso:2002ju}) confirms that these bounds cannot be used as physical evidence for the alluded fundamental principle of quantum gravity.
Holographic-like behaviour is simply there in standard physics when the situation is befitting.  

Finally, the generalized second law (GSL) states that the total entropy defined by the Bekenstein-Hawking BH entropy plus the entropy of the external matter can only increase in any physical process. As the BH entropy is expected to arise from  standard statistical mechanical considerations, which are not different at the fundamental level from those leading to the definition of the entropy of the rest of matter fields, it is widely accepted that the GSL must hold. As in standard statistical mechanics, the second law is hard to prove rigorously (mainly due to the difficulty in defining entropy of matter fields precisely). Nevertheless, versions of the GSL constructed in terms of geometric notions of matter entropy (e.g. entanglement entropy, mutual information, etc.) exist \cite{Wall:2011hj} and capture a physical meaning that is closely related to the GSL formulated in terms of the standard coarse graining definition of entropy. As in the case of entropy bounds, these proofs rely only on the validity of general relativity, quantum field theory, and the semiclassical formulation where the gravitational field couples to the expectation value of the stress-energy-momentum tensor. Once more no holographic principle needs to be invoked, the GSL (used to motivate holography) is just valid for standard 3+1 dimensional theories carrying genuine bulk degrees of freedom.

In conclusion, the black hole system is effectively a 2+1 dimensional system when analyzed by external stationary (and therefore late) observers. The dimension transversal to the horizon is exponentially squeezed by the redshift effect near the horizon and the system becomes effectively 2-dimensional.   Consequently, according to a view that enjoys some consensus in the LQG community, there is no need for {\em fundamental} screens and {\em fundamental} holographic ideas when considering the statistical mechanical origin of the Bekenstein-Hawking area entropy law or any of the black hole phenomenology associated with thought experiments involving interactions with matter and fields in the semiclassical regime. Black holes are special and their thermal properties are encoded in a lower dimensional system: their horizon. Holography, in this weaker sense, is not a {\em fundamental} property of quantum gravity but simply a property of BHs (and suitable null surfaces); simply a special behaviour of a very special situation.

\section{The classical basis of loop quantum gravity} \label{two}

In this section we briefly review the main features of the classical theory and the parametrisation of its phase space that defines the starting point for the quantisation program of LQG. The main message of this section is that the action of general relativity when formulated in terms of first order variables (which are suitable for the implementation of the non perturbative quantisation program of LQG) imposes non trivial canonical commutation relations for  geometric quantities. The consequence of this is that suitable geometric observables have discrete spectra in the quantum theory. 

\subsection{Where to start? The choice of the basic fields and action principle} \label{caca}

The starting point is the choice of the fundamental field variables in terms of which one describes
the dynamics of gravity. In the original formulation one uses the metric tensor $g_{ab}$ (encoding the spacetime geometry) and its dynamics is described by the Einstein-Hilbert action \cite{Einstein:1916vd}
\be
S[g_{ab}]=\frac{1}{2\kappa} \int \sqrt{|g|} R(g_{ab}) dx^4,
\ee
where $\kappa=8\pi G c^{-4}$, $g$ is the determinant of the metric ($dv=\sqrt{|g|} dx^4$ is simply the spacetime volume element), and $R(g_{ab})$ is the Ricci scalar: the `trace' $R=g^{\mu \rho}g^{\nu \sigma}R_{\mu \nu \rho \sigma}$ of the Riemann curvature  tensor $R_{abcd}$ where $g^{ab}$ is the inverse metric ($g^{\mu\nu}g_{\mu\sigma}=\delta^\nu_\sigma$). The vacuum Einstein's equations are \be\label{ee} R_{ac}=0,\ee
where $R_{ac}=g^{\mu\nu}R_{a\mu c\nu}$ is the Ricci tensor (we are using here abstract index notation, latin indices denote abstract spacetime indices, greek letters coordinate indices; see \cite{wald}).  Even though the Ricci scalar (or scalar curvature) has a very simple geometric meaning, its dependence on the dynamical field $g_{ab}$ is quite complicated, namely
\be
R(g_{ab})=g^{\mu\rho} \left[\partial_{\nu} \Gamma^{\nu}_{\mu\rho}-\partial_{\mu} \Gamma^{\nu}_{\nu\rho}+\Gamma^{\alpha}_{\mu\rho} \Gamma^{\nu}_{\nu\alpha}-\Gamma^{\alpha}_{\nu\rho} \Gamma^{\nu}_{\mu\alpha} \right],
\ee 
where \be \label{GG}\Gamma^{\rho}_{\mu\nu}=\frac{1}{2}g^{\rho\sigma}\left[\partial_{\mu} g_{\nu\sigma} +\partial_{\nu} g_{\mu \sigma} -\partial_{\sigma} g_{\mu\nu} \right]\ee are the Christoffel symbols. Thus, despite the simple geometric meaning of the Einstein-Hilbert action, the Lagrangian of general relativity is quite complicated in terms of metric variables. The algebraic structure of the action can be simplified (in the so-called Palatini formulation) by 
declaring the Christoffel symbols as independent variables. Such modification goes in a good direction; however, there is another, more important, disadvantage of the present choice of variables in the Eintein-Hilbert action or the Palatini modification: one cannot couple fermion fields to gravity described in this form (we come back to this point below).

Another disadvantage of the choice of the metric $g_{ab}$ as a basic variable is the huge (naively infinite) dimensionality of the space of actions that are
related to the Einstein-Hilbert action via the renormalization group flow. According to the Wilsonian perspective \cite{Wilson:1973jj} there is an intrinsic uncertainty in the 
selection of an action principle due to the flow in the space of action principles induced by the integration of quantum fluctuations at scales that are not relevant for the physics of interest. In this sense there is an ambiguity in naming the action principle of a theory: the set of suitable action principles is only limited by the field and symmetry content of the theory. In the case of general relativity in metric variables this corresponds to all possible general covariant functionals of $g_{ab}$. This set is characterized by infinitely many coupling constants, concretely
\be\label{action1}
S[g_{ab}]=\frac{1}{2 \kappa} \int \sqrt{|g|} \left(R+\Lambda+\alpha_1 R^2+\alpha_2 R^3+\cdots+\beta_1 R_{\mu\nu\alpha\sigma} R^{\mu\nu\alpha\sigma}\cdots\right) dx^4,
\ee 
where  only some representative terms have been written with couplings  $\alpha_1, \alpha_2,\cdots$, $\beta_1, \beta_2, \cdots$, etc.  If all the infinite dimensional set of couplings defining the above family of metric variable actions would be relevant then it would be impossible to decide what the correct starting point for canonical quantization would be and quantum gravity predictability would be compromised. It is possible, however, that the renormalization group flow selects a final dimensional 
space in this infinite dimensional world of metric gravity actions \cite{Weinberg:1980gg}. Such possibility, known as the {\em asymptotic safety} scenario, 
is under present active exploration \cite{Niedermaier:2006wt}.     

The necessity of having an action principle that is suitable for the coupling with fermions leads to the type of variables that define the starting point for quantization in LQG. As we will see below the new variables allow for the introduction of natural extended observables which transform covariantly under diffeomorphism, and lead to algebraically simpler action principles (simpler field equations) in a space of actions whose dimensionality is drastically reduced: for pure gravity the space of actions is finite dimensional.

\subsubsection{The first order formalism}

In order to couple fermions to general relativity one needs variables where a local action of the rotation group (and more generally Lorentz transformations) is defined. This is naturally achieved by describing the spacetime geometry in terms of an orthonormal frame instead of a metric. Local Lorentz transformations are realized as the set of transformations relating different orthonormal frames.   This subsection might seem a bit technical for those that are not familiar with the formalism.
Those readers should go through the equations without paying too much of attention to the index structure. The intended message of this part is the algebraic simplicity of the new formulation in comparison the previous one.

Concretely one can introduce an orthonormal frame field defined by four co-vectors $e_a^I$ (with the index $I=0,\cdots, 3$; $a$ and other latin indices denote spacetime indices) and write  the spacetime metric  as a composite object
\ba\label{jjj}
g_{ab}&=&-e^0_{a}e^0_{b}+e^1_{a}e^1_{b}+e^2_{a}e^2_{b}+e^3_{a}e^3_{b}\n \\
&=&e^I_{a}e^J_{b}\eta_{IJ},
\ea
where in the second line the internal Minkowski metric $\eta_{IJ}={\rm diag}(-1,1,1,1)$ is explicitly written.
In the familiar three dimensional space there are infinitely many frame-fields related by local rotations; in the present four dimensional Lorentzian setting the choice of an orthonormal frame is also ambiguous. Indeed  the previous (defining) equation is invariant under  Lorentz transformations: both $e$ and $\tilde e$ are solutions with $e_a^I\to \tilde e^I_a=\Lambda^I_J e^J_a$
which we will write in matrix notation as
\be\label{gt} e_a \to \tilde e_a=\Lambda e_a,\ee where $\Lambda^I_J$  satisfies $\eta_{KM}=\eta_{IJ}\Lambda^I_K\Lambda^J_M$. The physics cannot fix such freedom in the choice of a tetrad; this new symmetry is an  additional gauge symmetry of general relativity when formulated in these variables. As in any gauge theory, derivatives of covariant fields require the introduction of the notion of a connection $\omega^{IJ}=-\omega^{JI}$ (a one-form called the Lorentz connection in this case) defining the covariant derivative. More precisely, if $\lambda^I$ is an object with internal index transforming covariantly $\lambda\to  \tilde \lambda=\Lambda \lambda$ under a Lorentz transformation $\Lambda^I_J$ then its covariant (exterior) derivative, defined by 
\be
d_{\omega}\lambda^I=d\lambda^I+\omega^{IJ}\wedge \lambda_J,
\ee
also transforms covariantly because $\omega_a^{AB}$ transforms inhomogeneusly under internal Lorentz transformations \eqref{gt}, namely  
\be \omega \to \tilde \omega=\Lambda \omega  \Lambda^{-1} + \Lambda d\Lambda^{-1}.\label{ccg}\ee
Thus, the Lorentz connection $\omega^{IJ}$ is an additional field that is necessary in the tetrad formulation to define derivatives in a context where frames can be locally changed by a local Lorentz transformation.
In a suitable sense the Lorentz connection plays a role that is similar to that of the Christoffel symbol of the metric formulation. When the gravity field equations are satisfied, this connection is fixed in terms of derivatives of the tetrad field by equations that resemble equation \eqref{GG}. 

In terms of  $e^I$ and $\omega^{IJ}$ the action principle of gravity drastically simplifies becoming
\be\label{action2}
S[e_a^A,\omega_a^{AB}]=\frac{1}{2 \kappa}\int \epsilon_{IJKL} e^I\wedge e^J\wedge F^{KL}(\omega), 
\ee
where  $F^{AB}_{ab}$ the curvature of the connection $\omega_a^{AB}$; a two-form valued in the Lie algebra of the Lorentz group with a simple dependence on the connection given by  
\be \label{FF} F^{AB}=d\omega^{AB}+\omega^{AM}\wedge\omega_{M}^{\ B}.\ee 
The curvature transforms covariantly under a local Lorentz transformation $F\to \Lambda F \Lambda^{-1}$. The internal Levi-Civita symbol $\epsilon_{ABCD}$---a totally antisymmetric internal tensor such that $\epsilon_{0123}=1$---is invariant under the simultanneus action of the Lorentz group on its four entries. The action is in this way invariant under the Lorentz gauge transformations \eqref{gt} and \eqref{ccg}. Equations \eqref{gt} and \eqref{ccg} define the (internal) Lorentz gauge transformations of the basic fields entering the action. Nevertheless, the  gauge transformations (\ref{gt}) and (\ref{ccg}) need not be listed in addition to \eqref{action2}; the very field equations stemming from the action know about these symmetries. This is specially explicit in the Hamiltonian formulation where gauge symmetries are in direct correspondence with constraints (restrictions among the phase space fields) which in turn are the canonical generators of gauge transformations. These constraints (generators of gauge transformations) are part of the field equations \cite{Dirac:1964:LQM} (see also \cite{Henneaux:1992ig}). We will write them explicitly in Section \ref{haha}.

In addition to internal Lorentz transformations the action \eqref{action2} is invariant under diffeomorphisms  (general covariance). At the technical level this comes from the fact that the action
\eqref{action2} is the integral of a 4-form (a completely antisymmetric tensor with 4 contravariant indices): under coordinate transformation $x^{\mu}\to y^{\mu}(x)$ fields transform as tensors
\ba\label{diffy}
e^{J}_{\mu} dx^\mu &=& e^J_{\mu} \frac{\partial x^\mu }{\partial y^\alpha} dy^{\alpha} \n \\
\omega^{JK}_{\mu} dx^\mu &=& \omega^{JK}_{\mu} \frac{\partial x^\mu }{\partial y^\alpha} dy^{\alpha},
\ea
 while the integral remains unchanged as the 4-form  transforms precisely by multiplication by the Jacobian $\left| \frac{\partial x^\mu }{\partial y^\alpha} \right|$. 
 
 Once more such symmetry will be dictated to us by the equations of motion coming from the action if not explicitly taken into account.
 This is in fact how Einstein himself was confronted with general covariance: his equations would seem to violate determinism as certain field components 
 would not be entirely determined by the evolution equations. After some struggling with (what became to be know as) the hole argument he 
 realized that the action (\ref{action1}) implied that coordinates have no physical meaning and that only coordinate independent  statements (diffeomorphism invariant in modern jargon) contain physical information (see \cite{Rovelli:2004tv} for a modern account). In the present case, these are functions of the basic fields $e$ and $\omega$ invariant under the transformations \eqref{diffy} in addition to  
 \eqref{gt} and \eqref{ccg}.
 
 The equations of motion coming from \eqref{action2} follow from $\delta_e S=0$ and $\delta_\omega S=0$ respectively
 \ba\label{eom}
&&  \epsilon_{IJKL} e^J \wedge F(\omega)^{KL}=0 \\ 
 && d_{\omega}(e^I\wedge e^J)=0. \label{eom2}
 \ea
Notice their algebraic simplicity. If the tetrad field is invertible (which basically means that a non degenerate metric can be constructed from it according to  \eqref{jjj}) then the previous equations are equivalent to Einstein's equation \eqref{ee}.  However, the field equations, as well as the action (\ref{action2}) continue to make sense for degenerate tetrads. For example the {\em no-geometry state}  $e=0$---diffeomorphism invariant vacuum---solves the equations and makes perfect sense in terms of the new variables. 
 
In this way, guided by the necessity of coupling gravity with fermions, the first order variables and the action \eqref{action2} introduce a paradigm shift that will be crucial in the quantum theory:  the space of solutions (elements of the phase space of the theory (\ref{action2})) contain degenerate configurations. These configurations are {\em pregeometric} in the sense of Wheeler \cite{Misner:1974qy} and will play a central role in the state space of LQG. Even when these are not important for the description of classical gravitational phenomena they are expected to dominate the physics at the deep Planckian regime. We will see in what follows that these pre-geometric configurations (in the form of quantum excitations) are responsible for the quantum gravitational phenomena associated to black holes (BHs); ranging from their thermal behavior, the relationship of their entropy with their area, to a possible natural explanation the information loss paradox.    

Another striking property of the tetrad formulation is the radical reduction of the space of actions (formally\footnote{The renormalization group flow in first order variables cannot be defined in terms of the usual background field perturbation techniques. The problem is that no well-defined gauge fixing for diffeomorphisms is know around the natural degenerate  background $e=0$. If instead a non degenerate background is used then arbitrary terms can be generated by the symmetry breaking that it introduces (see \cite{Rovelli:2005qb} for an example in Yang-Mills context, and \cite{Daum:2010qt} for a discussion in the gravitational case).} expected to be probed 
by the renormalization group flow. Concretely, if one restricts to the pure gravitational sector the most general action that is compatible with the field content of (\ref{action2}) and its symmetries has only 6 different terms. Indeed all possible gauge invariant 4-forms that can be constructed out of the tetrad $e^I$ and the Lorentz connection $\omega^{IJ}$ are  
\ba\label{actiong}
S[e_a^A,\omega_a^{AB}]&=&\frac{1}{2 \kappa} \int \overbrace{ \epsilon_{IJKL} e^I\wedge e^J\wedge F^{KL}(\omega)}^{\rm Einstein}+\overbrace{{\Lambda} \, \epsilon_{IJKL} e^I\wedge e^J\wedge e^K\wedge e^L }^{\rm Cosmological\, Constant}+\overbrace{\alpha_1 \ e_I\wedge e_J\wedge F^{IJ}(\omega)}^{\rm Holst} \\\n  &+&\underbrace{\alpha_2 \ (d_{\omega} e^I \wedge d_{\omega} e_I \,- \ e_I\wedge e_J\wedge F^{IJ}(\omega))}_{\rm Nieh-Yan}  + \underbrace{\alpha_3 \ F(\omega)_{IJ} \wedge F^{IJ}(\omega)}_{\rm Pontrjagin}  +\underbrace{\alpha_4 \ \epsilon_{IJKL} F(\omega)^ {IJ} \wedge F^{KL}(\omega)}_{\rm Euler}, 
\ea 
where $d_\omega e^I$ is the covariant exterior derivative of $e^I$ and $\alpha_1\cdots \alpha_4$ are coupling constants.  For non-degenerate
tetrads Einstein's field equations follow from the previous action independently of the values of the $\alpha$'s:
the additional terms are called topological invariants describing global properties of the field configurations in spacetime. The $\alpha_1$-term is called the Holst term \cite{Holst:1995pc}, the $\alpha_2$-term is the Nieh-Yan invariant, the $\alpha_3$-term is the Pontryagin invariant, and the $\alpha_4$-term is the Euler invariant.  Inspite of not changing the equation of motion these terms can actually be interpreted as producing  canonical transformations in the phase space of gravity \footnote{In the presence of Fermions $\gamma$ controls the strength of an emergent four-fermion interaction \cite{Perez:2005pm, Freidel:2005sn, Mercuri:2006um}.}. In such a context the so-called Immirzi parameter \cite{Immirzi:1996di} corresponds to the combination \cite{Rezende:2009sv}
\be
\gamma\equiv \frac{1}{(\alpha_1+2\alpha_2)}.
\ee
The parameter $\gamma$  will be particularly important in what follows.

\subsubsection{Extended variables}\label{ev}

General covariance is the distinctive feature of general relativity and we have recalled how this is explicitly encoded in the action principles for gravity. The central difficulty of quantum gravity is how generalize what we have learnt about quantum field theory (in the description of other interactions) in order to understand the generally covariant physics of gravity. In general relativity, measurable quantities cannot be defined with the help of coordinates or any non dynamical background  as both concepts stop carrying any physical meaning. Localisation of spacetime events is possible only in a relational manner where some degrees of freedom are related to others to produce a generally covariant observable: one that is well defined independently of the coordinates we choose to label events. 

In the classical theory these observables are always non-local. Localisation in general relativity is always done in a relational fashion using the notion of test observers. Test observers are key in the spacetime interpretation of general relativity; the observables that follow from them are always non local in spacetime. An illustrating example is the case of two free test observers with world lines---geodesics in the spacetime---that meet at some event $A$, then separate and meet again at an event $B$. The proper time $\tau_{AB}$ measured by one of the observers between these two events  is a genuine coordinate independent quantity but is non-local. Another example is the definition of a black hole event horizon which separates those observers that can in principle escape out to infinity from those that cannot: test photons are used to define the horizon in a coordinate independent fashion. All observables are non-local in general relativity.

These thoughts led to the idea that extended variables might be best suited for the definition of a 
quantum theory of gravity. Even when the motivations are sometimes different non local objects are also central in other approaches such as strings, branes \cite{Polchinski:1998rq}, twistor theory \cite{Penrose:1972ia}, or causal sets \cite{Bombelli:1987aa}.
 
\begin{figure}[t]
%\begin{center}
\centerline{\hspace{0.5cm} \(
\begin{array}{c}
\includegraphics[height=2cm]{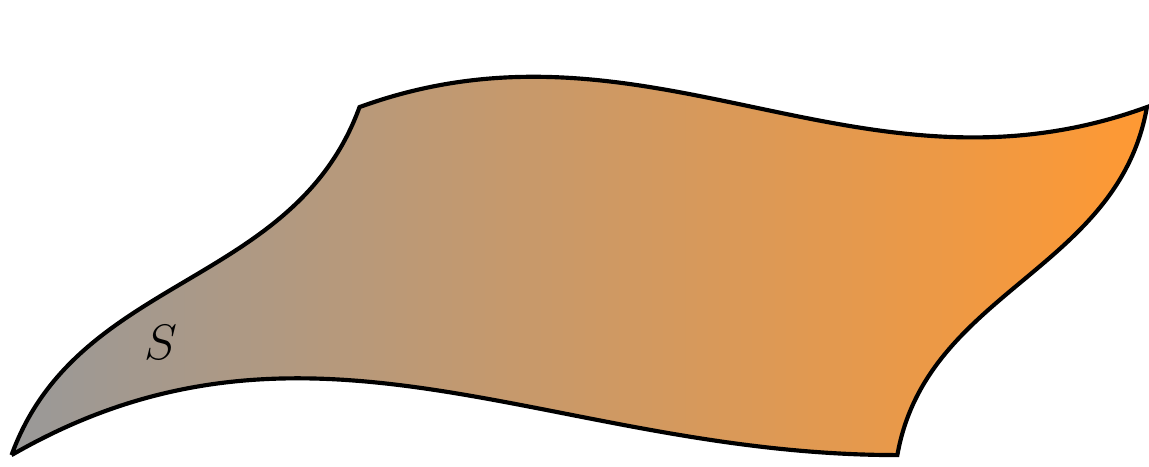}\ \ \ \ \ \ \ \ \ \ \ \ \ \ \ \ \ \ \ \ \ \ \ \ \ \ \ 
\includegraphics[height=2cm]{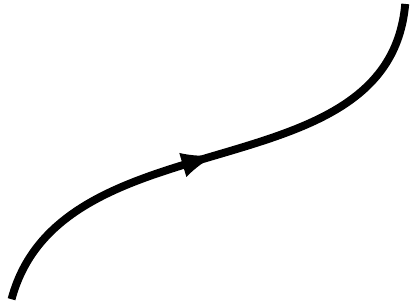} 
\end{array}\) } \caption{First order variables can be naturally associated with extended variables behaving covariantly under diffeomorphisms. The exterior product $e\wedge e$ of frame fields can be naturally smeared on two dimensional surfaces; equation \eqref{23}. The connection can be integrated along a one dimensional path to produce a group element defining parallel transport; this object is called the holonomy and in given in (\ref{24}). The first 2d extended object plays the role of `momentum' conjugate to the holonomy in the path integral regularisation of gravity provided by the spin foam representation. }
\label{fig:penrose-2}
%\end{center}
\end{figure}

An advantage of the new variables in (\ref{action2}) over the metric variables in (\ref{action1}) is that they allow for the introduction of natural quantities associated to extended
subsets (submanifolds) of the spacetime. These quantities are the fluxes of $e\wedge e$ and the holonomies
of the Lorentz connection $\omega$. More precisely the fluxes are 
\ba E(\alpha,S)\equiv \int_{S} \alpha_{IJ} e^I \wedge e^J,
\label{23} \ea
where $\alpha_{IJ}$ is a smearing field and  $S$ is a two-dimensional surface.
The holonomy assigns an element $\Lambda(\ell, \omega)$ of the Lorentz group to any one dimensional path in spacetime, by the rule
\be\label{24}
\Lambda(\ell, \omega)\equiv {\rm P} \exp-\int_{\ell} \omega,
\ee
where ${\rm P}\exp$ denotes the path ordered exponential.
None of these extended variables are diffeomorphism invariant; however, they transform in a very simple way under coordinate transformations: the action of a diffeomorphism on them amounts to the deformation of the surface $S$ and the path $\ell$ by the action of the diffeomorphism on spacetime points. This behaviour makes these extended variables suitable for the construction of covariant non local operators for the quantum theory.  These extended variables are represented in Figure \ref{fig:penrose-2}.

The above non-local variables are the basic building blocks in the attempts of giving a meaning to the path integral definition of quantum gravity based on action (\ref{action2}). Such research direction is known as the spin foam approach \cite{Perez:2003vx, Perez:2012wv, Perez:2012db}. Even though some applications of spin foams to black holes are available; most of the developments have been achieved in the canonical  (or Hamiltonian) formulation. We will see in what follows that the above type of extended variables are also available in the Hamilatonian formulation, but for that we have to briefly describe the phase space structure of general relativity when written in first order variables.

\subsection{First step towards the quantum theory: the Hamiltonian formulation}\label{newvariables}

We need to study the Hamiltonian formulation of gravity formulated in terms of (\ref{actiong}). In particular we are interested in obtaining the Poisson brackets between suitable basic variables in terms of which we shall parametrize the phase space of the theory. These Poisson brackets will become the canonical commutation relations in the quantum theory that are responsible for the discreteness of geometric quantities in LQG. In this way, the origin of the Planckian discreteness of geometry is easily seen from the Hamiltonian analysis. We only need to recall a shortcut for the construction of the canonical variables in mechanics, due to the simplicity of the action of gravity in the first order formalism we will be able to derive, via simple algebraic steps, the from of the Poisson brackets for gravity and foresee the seeds of discreteness.

\subsubsection{The covariant phase space formulation in a nut-shell}

\begin{figure}[h!!!!!!!!!!!!!!!]\!\!\!\!\!\!
\centerline{\hspace{0.5cm} \(
\begin{array}{c}
\includegraphics[height=4cm]{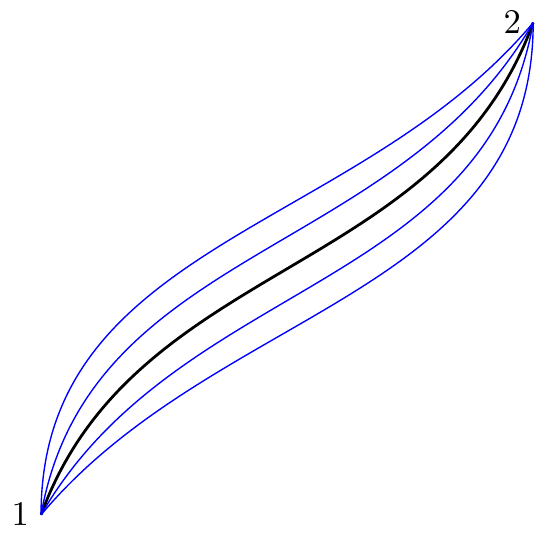}\ \ \ \ \ \ \ \ \ 
\includegraphics[height=4.25cm]{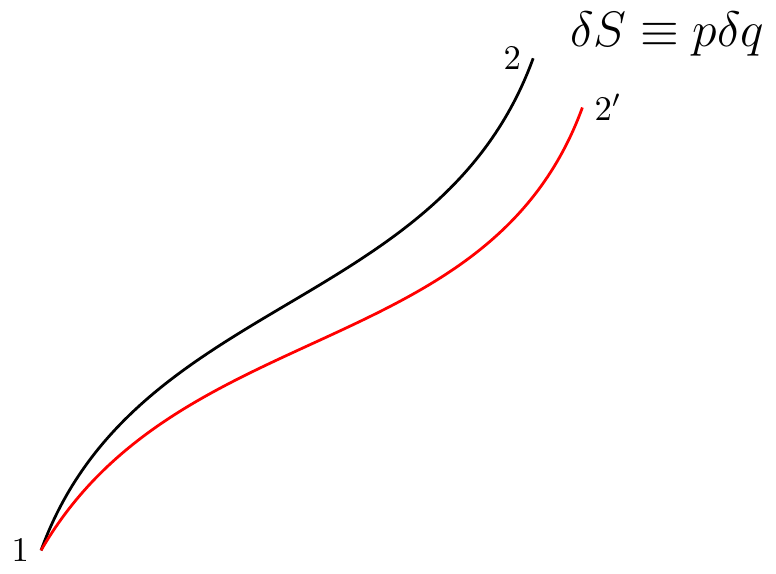} 
\end{array}\) } \caption{The action contains information about both the equations of motion and the phase space structure. Stationarity of the action under variations that vanish at the initial and final point give the equations of motion (left panel). Changes of the action under on-shell variations (solutions of the e.o.m.) encode the phase space structure (right panel).
These two features of the action are stated in equation (\ref{covph-eq}) that shows the form of a general variation. }
\label{covph}
\end{figure}

 There is a direct way for obtaining the phase space structure of a field theory from the action principle.
The method is easily illustrated by a simple mechanical system with a single degree of freedom and 
Lagrangian $L(q,\dot q)$. Under general variations the action changes according to
\be\label{covph-eq}
\delta S=\int\limits_1^2 \underbrace{\left[\frac{\partial L}{\partial q}-\frac{d}{dt}\left(\frac{\partial L}{\partial \dot q}\right)\right] }_{\rm e.o.m.} \delta q  dt +  \underbrace{ \left.\frac{\partial L}{\partial \dot q} \delta q \right|_1^2}_{p \delta q},  
\ee  
where the boundary term comes from the integration by parts that is necessary to arrive at the equations of motion in the first term. The previous equation contains important information encoded in  the type of variations $\delta q(t)$ and its boundary conditions (Figure \ref{covph}). If $\delta q(t)$ is arbitrary for intermediate times but it vanishes at the boundary instants $1$ and $2$, then $\delta S=0$ for those variations gives the equations of motion. If instead $\delta q(t)$ are variations defined by infinitesimal differences between solutions of the equations of motion---not necessarily vanishing at the boundary times---, then the first term in (\ref{covph-eq}) vanishes and $\delta S= p\delta q|_1^2$. These boundary contributions to the on-shell variation of the action tell us 
what the phase space structure of the system is, i.e., what the momentum $p$ conjugate to $q$ is. In this simple example such method for obtaining the momentum conjugate to $q$ might seem excessive as in this case we already know the recipe $p=\partial L/\partial \dot q$; however, it often shows to be the simplest and most direct method when dealing with generally covariant field theories such as the one defined by our action (\ref{action2}). We will use this method to directly access the Poisson commutation relations of geometric variables in gravity. 

On a slightly more technical level, the boundary term $\Theta(\delta) \equiv p\delta q$ is called the symplectic potential and is a function of $\delta$ in the sense that it depends on the specific form of the on shell variation at the boundary---where $\delta$ denotes the infinitesimal difference between two solutions, it can be seen as a vector with components $\delta\equiv(\delta q, \delta p)$. From the symplectic potential $\Theta(\delta)$ one can obtain the symplectic form $\Omega(\delta, \delta')$ by an additional independent variation $\delta^\prime$ according to 
\be\label{ome}
\Omega(\delta, \delta')\equiv \delta \Theta(\delta')-\delta' \Theta(\delta)=\delta p\ \delta' \! q-\delta'\!p\ \delta q,
\ee   
i.e., the on-shell antisymetrized variation (exterior field derivative) of the symplectic potential gives the symplectic form. 
In one simple step, the on-shell antisymetrized variation of the action leads (from (\ref{covph-eq})) to the conservation of the symplectic form
\be\label{liv}
0=(\delta\delta'-\delta'\delta)S=\Omega(\delta, \delta')|_2-\Omega(\delta, \delta')|_1,
\ee
and its corollary: Liouville's theorem on the conservation of phase space volume \footnote{In the case of $N$ degrees of freedom the volume form in phase space is ${\rm vol}\equiv -\frac{1}{2^{N}} \wedge^N \Omega$ .}.
All these, standard properties of the phase space of a dynamical system with finitely many degrees of freedom 
carry over to the field theories with several mathematical subtleties that are not important here. This is the great power of the covariant phase space formalism (see \cite{Ashtekar:1990gc, Crnkovic:1986ex, Lee:1990nz} for further reading).

The symplectic form carries the information about the phase space structure of the system: it defines the dynamically invariant phase space volume measure (Liouville's theorem) and the Poisson brackets of observables (the starting point for quantization). The previous relation between the symplectic form and the symplectic potential also says that $\Theta(\delta)\to \Theta(\delta)+\delta \mu$ for some function $\mu$ does not change the symplectic structure as $\delta\delta'\mu-\delta'\!\delta\ \mu=0$. The possibility of changing the symplectic potential by the addition of the variation of a function $\mu$ can be shown to encode the notion of canonical transformations. 

\subsubsection{Implementation in gravity}
 
\begin{figure}[h!!!!!!!!!!!!!!!]\!\!\!\!\!\!
\centerline{\hspace{0.5cm} \(
\begin{array}{c} \includegraphics[height=6cm]{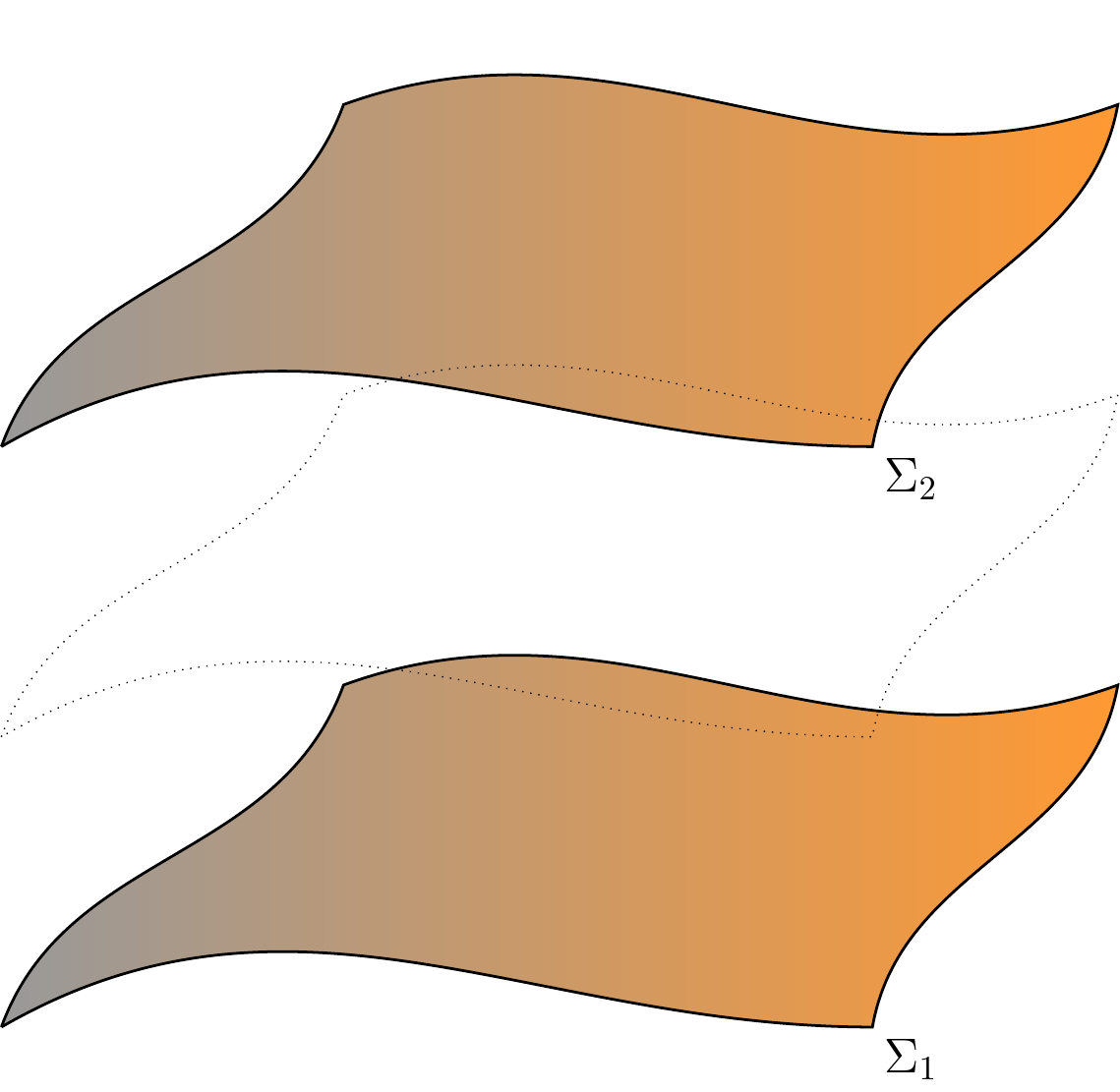}\ \ \ \ \ \ \ \ \ \ \ \ \ \ \ \ \ \ 
\includegraphics[height=6cm]{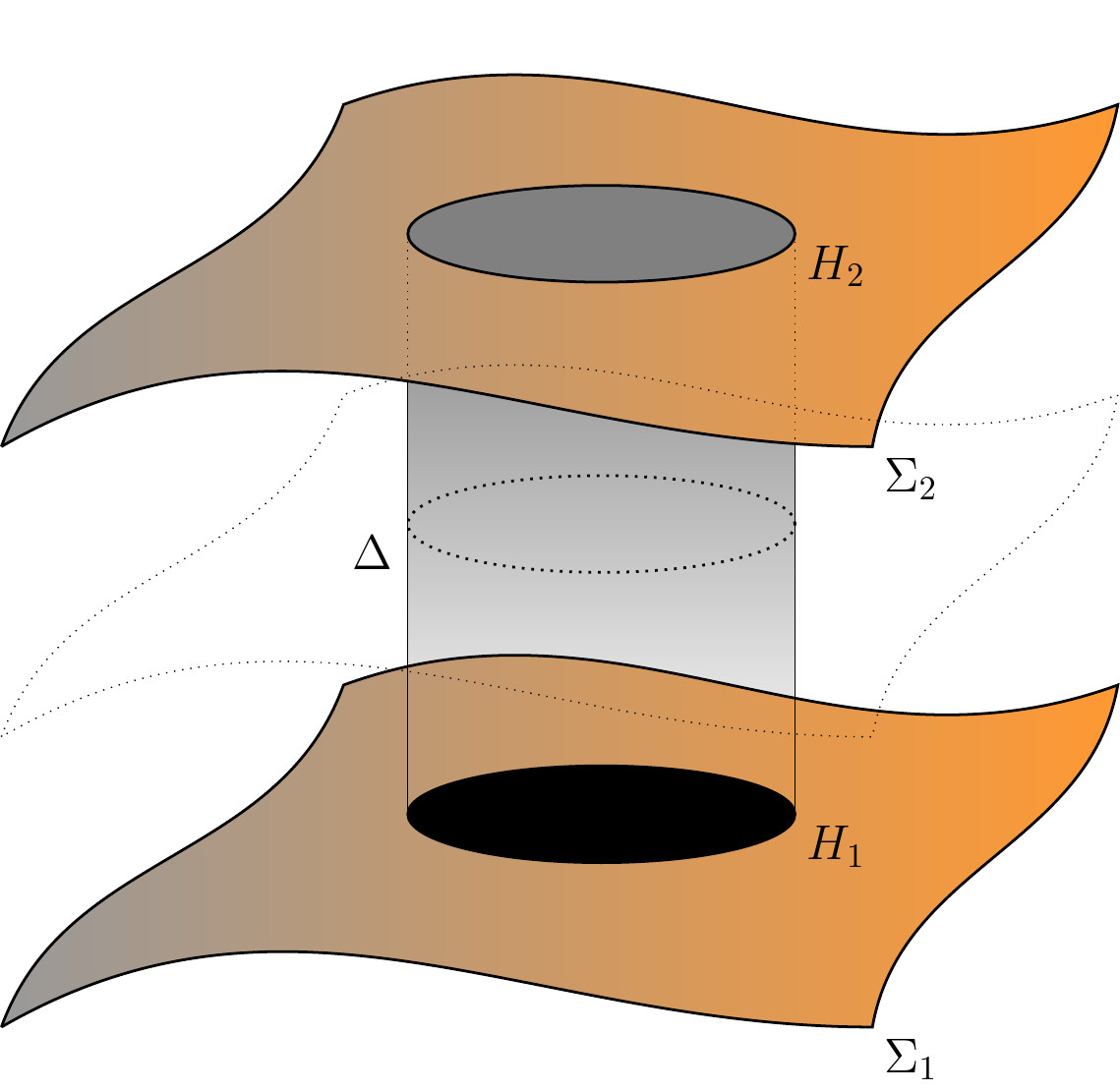}
\end{array}\) } \caption{{\bf Left panel:} Foliation of a spacetime region $M$ without internal boundaries. The space-like hyper surfaces $\Sigma_1$ and $\Sigma_2$ (Cauchy hypersurfaces) are the analog of instants $1$ and $2$ in our mechanical analog depicted in Figure \ref{covph}. {\bf Right panel:} Such space-like surfaces (where the gravitational field at a given instant is represented) can have a boundary $H=\partial \Sigma$. Black holes in loop quantum gravity are treated as a boundary where fields satisfy suitable boundary conditions; the so-called {\em isolated horizon} boundary condition. }
\label{covph-4d}
\end{figure}

Now we are ready to apply the previous techniques to the case of interest.
In order to simplify the following analysis we set $\Lambda$, $\alpha_2, \alpha_3$ and $\alpha_4$ to zero in (\ref{actiong}) and get the simpler (Holst) action 
\be\label{actionH}
S=\frac{1}{2 \kappa}\int (\epsilon_{IJKL}+\frac{1}{\gamma}\eta_{IK}\eta_{JL})  \left( e^I\wedge e^J\wedge F^{KL}(\omega)\right),
\ee
which defines our starting point.  The result is not affected if we drop this assumption but the proof becomes more technical \cite{Date:2008rb, Rezende:2009sv}. Following our recipe, in analogy with \eqref{covph-eq}, we simply need to consider the most general variation of (\ref{actionH}) in order to obtain the phase space structure of general relativity in first order variables. As discussed before it will be important to express
\be
\alpha_1=\frac{1}{\gamma}
\ee
as $\gamma$---the Barbero-Immirzi parameter---will play a central role in what follows.
Replacing in \eqref{covph-eq} and varying we obtain
\ba
\delta S=\frac{1}{2 \kappa}\int (\epsilon_{IJKL}+\frac{1}{\gamma}\eta_{IK}\eta_{JL})  \left(2\delta e^I\wedge e^J\wedge F^{KL}(\omega)+ e^I\wedge e^J\wedge \delta F^{KL}(\omega)\right).
\ea
The first term does not involve variations of derivatives of the fundamental fields, while the second term does. In fact a well know property of the field strength of a gauge theory is that $\delta F^{KL}(\omega)=d_{\omega} (\delta\omega^{IJ})$ which directly follows from \eqref{FF}.  Using this and defining 
\be
p_{IJKL}\equiv (\epsilon_{IJKL}+\frac{1}{\gamma}\eta_{IK}\eta_{JL}), 
\ee
we get to the result  by  integrations by parts as explicitly shown in the following three lines:
\ba
&& \delta S=\frac{1}{2 \kappa}\int_M 2 p_{IJKL} \delta e^I\wedge e^J\wedge F^{KL}(\omega)+  p_{IJKL}  e^I\wedge e^J\wedge d_{\omega}(\delta \omega^{KL})\n \\ && =\frac{1}{2 \kappa}\int_M 2p_{IJKL} \delta e^I\wedge e^J\wedge F^{KL}(\omega)-  p_{IJKL}  d_{\omega}(e^I\wedge e^J)\wedge \delta \omega^{KL}+d([p_{IJKL}  e^I\wedge e^J]\wedge \delta \omega^{KL})\n \\
&& =\frac{1}{2 \kappa}\int_M \underbrace{2p_{IJKL} \delta e^I\wedge e^J\wedge F^{KL}(\omega)-  p_{IJKL}  d_{\omega}(e^I\wedge e^J)\wedge \delta \omega^{KL}}_{\rm e.o.m.}+\int_{\partial M} \underbrace{\frac{1}{2 \kappa}[p_{IJKL}  e^I\wedge e^J]\wedge \delta \omega^{KL}}_{p\delta q},
\ea
where in the first line we substituted $\delta F^{KL}(\omega)=d_{\omega} (\delta\omega^{IJ})$ in the second term and then integrated by parts. In the first term (the bulk integral) of the last line we recognise the field equations \eqref{eom} while the second term (the boundary integral)  tells us that $P_{KL}\equiv -2 \kappa^{-1}p_{IJKL} e^I\wedge e^J$ is the momentum density conjugate to the Lorentz connection $\omega^{KL}$. In the Language of the symplectic potential we have
\be\label{sympo}
\Theta(\delta)=\int_{\Sigma} \frac{1}{2 \kappa}[p_{IJKL}  e^I\wedge e^J]\wedge \delta \omega^{KL},
\ee
where $\Sigma$ is a spacelike hypersurface (one of the two components of the boundary $\partial M$ in Figure \ref{covph-4d}) representing the analog of an {\em instant}.

However, there is a problem: there are $18$ independent components in the $\omega_a^{IJ}$ ($6$ independent internal configurations of the antisymetric ${IJ}$-indices times $3$ values of the $a$-index for the three space coordinates of the spacial boundary $\Sigma$) while, naively, the same amount of component are present in the $P_{ab}^{IJ}$ only $12$ are independent as they are all function of $e_a^I$! This can be stated by saying that the $P_{ab}^{IJ}$'s must satisfy constraints. These constraints (which in the literature a known as the {\em simplicity} constraints) complicate the identification of the genuine phase space variables and must be taken care of. There are two prescriptions for doing this, one is to solve them in some way before going on, the other is the Dirac modification of the Poisson brackets \cite{Dirac:1964:LQM}. In the present case it will be the easiest to simply solve these constraints by introducing a gauge fixing of the gauge freedom  (\ref{gt}).

\begin{figure}[h!!!!!!!!!!!!!!!]\!\!\!\!\!\!
\centerline{\hspace{0.5cm} \(
\begin{array}{c}
\includegraphics[height=3cm]{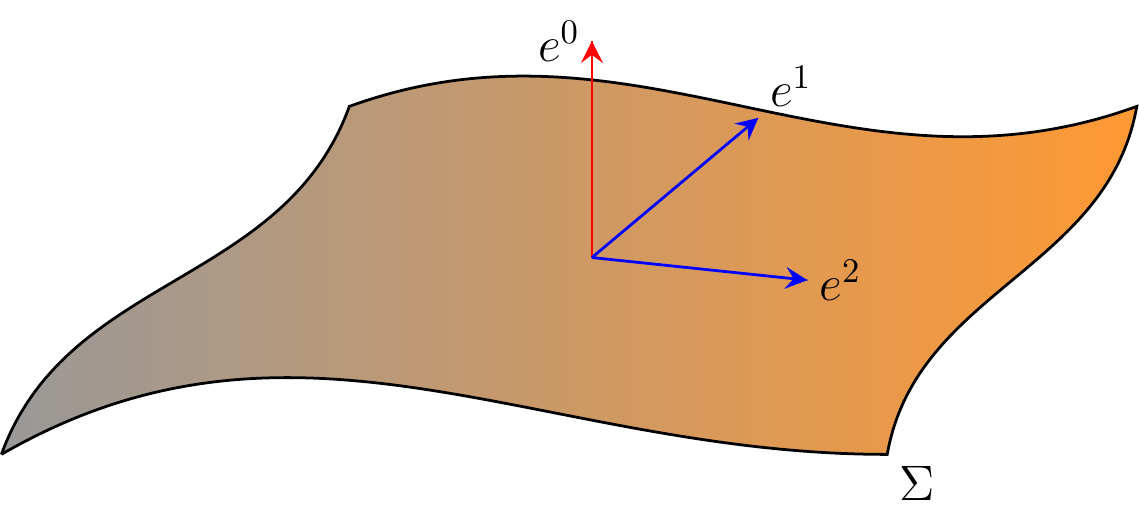}
\end{array}\) } \caption{{\bf Time-gauge and simplicity constraints:} The mismatch between the number of independent components of $P_{KL}\equiv (2\kappa)^{-1}\epsilon_{IJKL} e^I\wedge e^J$ and the configuration variable $\omega^{IJ}$ implies constraints among the $P^{IJ}$'s; the so-called {\em simplicity constraints}. These constraints can be solved by restricting the Lorentz gauge symmetry (\ref{gt}) to the $SO(3)$ subgroup defined by the condition that $e^0$ is normal to the time slices $\Sigma$. This gauge choice implies that, as an induced covectors on $\Sigma$, $e^0=0$ and so the $P^{ij}$ with $i,j=1,2,3$ are all vanishing. The only non trivial components entering the symplectic potential are then the nine $P^{i0}$---which are functionals of the nine $e^i$ and hence independent---and the conjugate nine $\omega^{i0}$. As the mismatch in the number of components has been resolved, no additional constraints on the phase space variables remain in the time gauge. }
\label{timegauge}
\end{figure}

The idea is to reduce the Lorentz symmetry in (\ref{gt}) by demanding  the co-vector $e^0$ (which defines the time axis of the frame field; the only timeline member of the tetrad) to  
be perpendicular to the time slices $\Sigma$, or equivalently to be aligned with the unit normal $n$ to $\Sigma$, namely 
\be\label{tgfix}
e^0_a=n_a.
\ee
This reduces the Lorentz gauge freedom to the rotation sub-group of the Lorentz group that leave invariant the normal to $\Sigma$; we denote this $SU(2)\subset SL(2,\C)$ \footnote{At this stage the rationale would imply that the original gauge group is $SO^+(3,1)$, the proper orthochronous Lorentz group with  $SO(3)$ the subgroup obtained via the time gauge.
However, for applications including fermions and other features that become clear in the quantum theory it is more convenient to work with the universal coverings $SL(2,\C)$ and $SU(2)$. }. This partial gauge fixing is known as the {\em time-gauge}, see Figure \ref{timegauge}. Such choice is very natural in the Hamiltonian formulation of gravity where the slicing of spacetime in terms of space-like hypersurfaces is already available. The time-gauge amounts to adjusting the time axis in our frame field to the one that is singled out by the foliation. 

The previous gauge fixing solves the problem of the mismatch of the number of independent components in the momenta as defined in \eqref{sympo}. If we explicitly separate the $0$ from the $i=1,2,3$ internal indices then the symplectic potential \eqref{sympo} becomes
\ba\label{sympotg}
\Theta(\delta)&=&\frac{1}{\kappa} \int_{\Sigma}\left( \epsilon_{0jkl}  e^0\wedge e^j\wedge \delta \omega^{kl}+ \frac{1}{\gamma} e^0\wedge e^i  \wedge \delta \omega_{0i}\right)-\frac{1}{\kappa} \int_{\Sigma}\left( \epsilon_{0jkl}  e^j\wedge e^k \wedge \delta \omega^{l0} +\frac{1}{\gamma} e^i\wedge e^j  \wedge \delta \omega_{ij}\right) \n \\ &=& 
-\frac{1}{\gamma\kappa} \int_{\Sigma} [\epsilon_{jkl}  e^j\wedge e^k]\wedge \delta\!\!\! \!\!\! \!\underbrace{(\gamma \omega^{l0}+\epsilon^{lmn}\omega_{mn})}_{\textrm{Ashtekar-Barbero connection}} \n \\ &=& 
-\frac{1}{\gamma\kappa} \int_{\Sigma} [\epsilon_{jkl}  e^j\wedge e^k]\wedge \delta A^l, 
\ea
where the first term in the first line vanishes because $e^0$ is normal to the space slice $\Sigma$ (it has no space components due to \eqref{tgfix} or more precisely its pull back to $\Sigma$ vanishes). In the second line we used that 
$\epsilon_{0ijk}=\epsilon_{ijk}$, simple algebraic properties of $\epsilon_{ijk}$, and we have factored out $\gamma^{-1}$. In the third line we have defined a new configuration variable
\be \label{coco}A^i\equiv \gamma \omega^{i0}+\underbrace{\epsilon^{ijk} \omega_{jk}}_\textrm{Holst},\ee which transforms as a gauge connection under the $SU(2)$ gauge symmetry that remains after the imposition of the time-gauge and is called the Ashtekar-Barbero connection. Now we have 9 $A^i_a$ configuration variables for the 9 conjugate momenta $\epsilon_{jkl}  e^j\wedge e^k$ depending of the 9 components of $e_a^i$. The strategy of the gauge fixing has worked as there are no additional constraints on momentum variables.   Recall our previous discussion on how important it was for the framework to have a connection formulation. For that the factor in front of the second term in the definition of $A^i$ must be precisely $1$; this is why one obtains a factor $\gamma^{-1}$ in front of the symplectic potential. 
 
From now on we adopt the more compact notation
\be\label{sigma} E^i=\epsilon^{i}_{jk} e^j\wedge e^k, \ee
and write \eqref{sympotg} as
\ba\label{sympotgo}
\Theta(\delta)&=& 
-\frac{1}{\gamma\kappa} \int_{\Sigma} E_i \wedge \delta A^i.
\ea

Notice that the term that makes the connection $A^i$ transform as a connection is the second term in \eqref{coco} (the first transforms as a vector under an $SU(2)$ rotation) which actually comes directly form the contribution of the Holst terms in (\ref{actiong}) to the symplectic potential (as mentioned above there is also a contribution to this term coming from the Nieh-Yan invariant in \eqref{actiong}). Further analysis shows that $\omega^{ik}$ is not free; indeed part of the field equations---equation \eqref{eom2}---imply Cartan's first structure equation
\be\label{ce}
d e^i+\omega^{ik} \wedge e_k=0, 
\ee
whose solution is a unique function of the triad $e^i$ and we denote $\omega^{ij}=\omega(e)^{ij}$. 
%The field equations also imply that $\omega^{0i}=$
%\be
%K^i_a=\frac{1}{2}e_i^b \sL_{n} h_{ab} 
%\ee

The symplectic structure that follows from the recipe (\ref{ome}) and the symplectic potential \eqref{sympotg} 
\ba \label{symp-sans}
\Omega(\delta, \delta')
&=&\frac{1}{2\kappa\gamma }\int_{\Sigma}  \delta A^{i}\wedge \delta'
E^{\va}_{i}-\delta' A^{i}\wedge \delta
E^{\va}_{i},
\ea
The associated Poisson brackets relations are
\ba
&& \{E^i(x),E^j(y)\}=0 \label{uno}\n \\
&& \{A^i(x),A^j(y)\}=0 \label{dos} \n \\
 && \{E^i(x),A^j(y),\}=\kappa \gamma\, \boldsymbol{\epsilon}^{\va (3)}\delta^{ij} \delta^{\va (3)}(x,y), \label{tres}
\ea
where $\boldsymbol{\epsilon}^{\va (3)} \delta^{\va (3)}(x,y)$ is the Dirac delta distribution with the usual properties that are familiar in the non gravitational context
when integrated against text functions with the additional feature of being defined on arbitrary coordinates \footnote{More precicely, $\boldsymbol{\epsilon}^{\va (3)}_{abc}$ is a three form
Levi-Civita density. Its tensor structure matches the one of the left hand side where we have the (one-form) connection $A^i_a$ times the (two form) $E_{ab}^j$. We have dropped tensor indices to improve readability.}. The phase space structure of gravity in connection variables is exactly that of a non-Abelian $SU(2)$ Yang-Mills theory. This combined with background independence will lead to the discreteness of geometric observables in the quantum theory as we will soon show \footnote{If one drops the Einstein term in \eqref{actiong} the theory becomes topological (with no local propagating degrees of freedom), but still admits the $(E,A)$ phase space parametrization \cite{Liu:2009em}. }.

\subsubsection{An alternative derivation: the importance of being in $3+1$ dimensions}

But before let us make a little detour that emphasises the importance of the dimensionality of spacetime in the present treatment. There is a peculiar feature of three dimensional space playing a central role in the existence of the $E$ and $A$ canonical variables.
If we had started from the simplest action \eqref{action2}, namely 
\be
S=\frac{1}{2 \kappa}\int \epsilon_{IJKL}  \left( e^I\wedge e^J\wedge F^{KL}(\omega)\right)
\ee
then the symplectic potential would have resulted in
\be\label{sympo-bis}
\Theta(\delta)=\int_{\Sigma} \frac{1}{2 \kappa}[p_{IJKL}  e^I\wedge e^J]\wedge \delta \omega^{KL},
\ee
which after the time-gauge fixing would have become
\ba\label{sympotg-bis}
\Theta(\delta)&=&
-\frac{1}{\kappa} \int_{\Sigma} \epsilon_{jkl}  e^j\wedge e^k\wedge \delta{ \omega^{i0}}
%\\ &=&
%-\frac{1}{\kappa} \int_{\Sigma} E_i\wedge \delta{ \omega^{i0}}.
\ea
which tells us that $E^i=\epsilon^{i}_{\ jk} e^j\wedge e^k$ and $\omega^{i0}$ are canonical pairs. However, non of these variables transforms as a connection under the remaining $SU(2)$ gauge symmetry. 

The $SU(2)$ connection formulation found in the previous subsection can be recovered via a canonical transformation  thanks to a remarkable property of
the geometry of frame fields in three dimensions. Given a frame field (which in 3d corresponds to our field $e_a^i$) there is a unique solution of the Cartan first structure equation \eqref{ce} (recall that this equation comes from the field equation \eqref{eom2}) that we call the spin connection $\omega^{ij}(e)$.  This is true in any dimension $d$ with the range of index $i,j=1,\cdots, d$. The antisymmetry of the connection in $ij$ implies that there are $d_c=d(d-1)/2$ independent internal components.  The case $d=3$ is special because only in this case one has $d_c=d$: the connection has exactly the right amount of components to be added to an object that transforms as a vector under the action of the frame rotation group ($SU(2)$ in this case). Indeed for $d=3$ we can express this algebraic property by encoding the components of the connection $\omega^{ij}$ in terms of an object with only one internal index (like a vector) using the Levi-Civita internal tensor, namely
\be
\omega^i(e)\equiv \epsilon^{ijk}\omega_{jk}(e);
\ee
which can be inverted to give $\omega^{ij}(e)\equiv \epsilon^{ijk}\omega_{k}(e)$. Now in three dimensions only, and in  terms of this definition, we can write the Cartan equation (\ref{ce}) as
\be\label{ce}
d e^i+\epsilon^{ijk}\omega_{j}(e)\wedge e_k=0.
\ee
and, most importantly for what follows, taking the variation of Cartan equation and then computing the wedge product with $e_i$ on gets
\be
d (\delta e^i)\wedge e_i+\epsilon^{ijk} \delta \omega_{j}(e)\wedge e_k\wedge e_i+\epsilon^{ijk}\omega_{j}(e)\wedge \delta e_k\wedge e_i=0.
\ee
using (\ref{ce}) to rewrite the third term, and renaming dummy indices, one gets a key result for the foundations of LQG, namely that 
\be \label{niuns}
\epsilon^{ijk}  e_i\wedge e_j \wedge \delta \omega_{k}(e)=d(\delta e^i\wedge e_i).
\ee
We can use the previous identity to now manipulate \eqref{sympotg-bis} and get
\ba\label{lance}
\Theta(\delta)&=&\n
-\frac{1}{\kappa} \int_{\Sigma} \epsilon_{jkl}  e^j\wedge e^k\wedge \delta{ \omega^{i0}}
%\\ &=& \n -\frac{1}{\gamma \kappa} \int_{\Sigma} \epsilon_{jkl}  e^j\wedge e^k\wedge \gamma \delta{ \omega^{i0}}
\\ &=&  -\frac{1}{\gamma \kappa} \int_{\Sigma} \epsilon_{jkl}  e^j\wedge e^k\wedge \delta{ \left(\gamma \omega^{i0}+\omega^i(e)\right)}+\frac{1}{\gamma \kappa}\underbrace{ \int_{\Sigma} \epsilon_{jkl}  e^j\wedge e^k\wedge \delta{\omega^i(e)}}_{\int_{\partial \Sigma} \delta e^i\wedge e_i},
\ea
where we have introduce the Immirzi parameter by adding and substracting a term proportional to the left hand side of  \eqref{niuns}. Assuming that $\Sigma$ is compact the last term in the previous expression vanishes due to \eqref{niuns} and we get 
\ba
\Theta(\delta)&=&
-\frac{1}{\kappa} \int_{\Sigma} E_i\wedge \delta{A^i}\ea 
in agreement with the previous derivation \eqref{sympotgo}.  If $\partial \Sigma\not=0$ then the last term contributes to the symplectic structure with a boundary term; this will be important in the presence of a BH in Section \ref{pqg}. A canonical transformation available in $3+1$ dimensions is the way  by which we find the Yang-Mills like parametrisation of the phase space of gravity (the Immirzi parameter labels a one parameter family of these).

Some general comments: as it becomes clear from the previous discussion the construction of the phase space of connection variables presented here works naturally only in $3+1$ dimensions. There is another possible canonical transformation which leads to the analog of the $\theta$ parameter in QCD; its effects on the phase space structure and black holes is studied in \cite{Rezende:2007mt}.
Connection variables are also natural in $2+1$ dimension where the absence of the simplicity constraints implies that one does not need to introduce the time gauge and can keep manifest Lorentz invariance. It is possible to avoid the time gauge and keep Lorentz invariance in the $3+1$ dimensional setting at the price of having non commutative connections due to the contributions of the simplicity constraints to the Dirac brackets \cite{Alexandrov:2001wt, Alexandrov:2000jw}. Because of this, the quantisation program has not been rigorously realised in this case (see \cite{Alexandrov:2002br} for an heuristic approach). The connection parametrisation of higher dimensional gravity is possible but more complicated (due to the presence of simplicity constraints) as has been shown in \cite{Bodendorfer:2011nw, Bodendorfer:2011nv, Bodendorfer:2011ny}. 
For a discussion of its quantisation see \cite{Bodendorfer:2011nx, Bodendorfer:2011pa}. The formalism has been generalised in order to include supergravity in \cite{Bodendorfer:2011xe, Bodendorfer:2011hs, Bodendorfer:2011pb, Bodendorfer:2011pc}. The calculation of BH entropy in higher dimensions has been studied in \cite{Wang:2014cga}.

\subsection{Constraints: the Hamiltonian form of Einstein's equations}\label{haha}

We have seen how the covariant phase space formulation offers a direct road to obtaining 
the phase space structure of general relativity. The Poisson brackets we have obtained in \eqref{uno}
are key in understanding the prediction of Planckian discreteness of geometry (we postpone this discussion to Section \ref{qg}). However, for simplicity we have not discussed in any details the dynamical equation of gravity in the Hamiltonian framework. It is possible to show (for more details see for instance \cite{Perez:2004hj}) that Einsteins equations split into 
the following three constraints on the initial field configuration $(E, A)$ given on a slice  $\Sigma$, and the Hamiltonian evolution equation for these data. The constraints are
\be\label{tresl}
 G^i(E,A)=d_AE^i= 0,
\end{equation} 
\be\label{unol}
 V_d (E,A)=\epsilon^{abc}E_{ab}\cdot F_{cd}= 0
\end{equation}
\be\label{dosl}
S(E,A)= \frac{(E_{ab}\times E_{de})}{\sqrt{{\rm
    det}(E)}} \cdot F_{cf} \ \epsilon^{abc} \epsilon^{def}+\cdots= 0,
\end{equation}
which are called the Gauss constraint, the diffeomorphism constraint, and the scalar constraint (there is an additional term in the last expression that we have omitted for simplicity, the full expression can be found in \cite{Ashtekar:2004eh}). The $\cdot$ and $\times$ denotes the scalar and the exterior product in the internal space.  For any quantity $O(A,E)$ (this includes in particular the phase space variables $A$ and $E$) its evolution is given by the canonical equation $\dot O=\{O, H[ \alpha, \vec N,M ]\}$ with the Hamiltonian
\be\label{dina}
H[ \alpha, \vec N,M ]\equiv\int_\Sigma \alpha_i G^i+N^a V_a+M S,
\ee
where the fields $\alpha$, $\vec N$, and $M$ are completely arbitrary in their space and time dependence; the freedom is associated with coordinate invariance of gravity (encoded in the four free fields $\vec N$, and $M$) and the additional $SU(2)$ internal gauge symmetry of the first order formulation in the time gauge. Time evolution is therefore not uniquely defined but all different spacetimes and fields reproduced from one particular initial data satisfying the constraints via \eqref{dina} can be shown to solve Einsteins equations and be related to each other via a diffeomorphims (coordinate transformation) and  gauge transformations. A moment of reflection shows that an initial data $(A,E)$ (i.e., solving the constraints) and $(A+\delta A, E+\delta E)$ such that $\delta A=\{A, H[ \alpha, \vec N,M ]\}$ and $\delta E=\{E, H[ \alpha, \vec N,M ]\}$ lead to solutions which are related by a diffeomorphism and gauge transformations. The interpretation of this fact is that   $(A,E)$ and $(A+\delta A, E+\delta E)$ are the very same data in different gauges. Thus, the Hamiltonian generates both time evolution and gauge transformations in a  generally covariant theory \cite{Dirac:1964:LQM}.  Notice also that the Hamiltonian vanishes identically on solutions.

The Gauss constraint (\ref{tresl}) is specially important. In our gravity context it follows from the covariant field equation \eqref{eom2} but on a more general basis it has a completely geometric origin: it arises from the presence of the underlying $SU(2)$ gauge symmetry (what remains of the original Lorentz symmetry after the time gauge-fixing \eqref{tgfix}).  Equation \eqref{tresl} is the strict analog of the Gauss law of electromagnetism and Yang-Mills theory.  It can be shown  that the smeared version
\be\label{gauss}
G[\alpha]\equiv\int_{\Sigma} \alpha_i d_A E^i, 
\ee
generates (via Hamilton's equations) $SU(2)$ gauge transformations of the basic variables and hence of any phase space quantity.
Explicitly:
\ba\label{transfi}
&& \delta_{\alpha} A=\{A,G[\alpha]\}=-d_A\alpha \n \\
&& \delta_{\alpha} E=\{E,G[\alpha]\}=[\alpha,E].
\ea 
These transformations are the generalization of the gauge transformations of electromagnetism to the non-Abelian case.
The Gauss law and the transformations it generates  are strictly the same as those of an $SU(2)$ Yang-Mills theory. The phase space parametrization in terms of a non-Abelian electric field $E$ (with a geometric interpretation in this case) and its conjugate $SU(2)$ connection also mimics the natural phase space parametrization of an $SU(2)$ Yang-Mills phase space. This are key features of the variables that follow naturally form \eqref{actiong}; they have central importance in the construction of the non-perturbative techniques used to define LQG.

In the presence of boundaries some subtleties arise when considering the gauge 
transformations generated by \eqref{gauss}. We will return to this important point in Section \ref{cattaneo}.

\section{Quantum geometry: the horizon and the outside}\label{qg}

In this Section we analyse further the commutation relations we found in \eqref{tres}. On the one hand we will see what kind of commutation relations they imply for geometric observables on the boundary (that can represent a BH horizon), on the other hand we will also derive commutation relations for geometric observables inside (in the bulk). We sketch the quantisation of the observables and present the basics of the theory quantum geometry on which LQG is based. 
We also review the basic facts about the isolated horizon boundary condition \cite{Ashtekar:2000sz} representing black holes in the formalism.

\subsection{Modeling a black hole horizon in equilibrium: the Phase space of an Isolated Horizon}\label{mbhs}

As discussed in Section \ref{intro-holo}, the physics of BHs in equilibrium as seen by external, late stationary observers
is the physics of an infinitesimal hyper annulus around the horizon with a width that tends exponentially to zero with the proper 
time of the external observers and becomes quickly shorter than the Planck length. This suggests that the entity 
encoding the relevant degrees of freedom for the description of the statistical mechanical nature of BH entropy is the 
$2+1$ dimensional null hyper-surface defining the BH horizon. With the prospect of quantizing the later degrees of freedom in the canonical framework where LQG techniques have been developed, it is necessary to rethink the Hamiltonian formulation of general relativity in the presence of a null boundary 
with suitable boundary conditions incorporating the physical notion of a BH horizon in equilibrium. 
This has led to the development of  the so-called {\em isolated horizon} (IH) formalism which we briefly describe below (see \cite{Ashtekar:2004cn} for a specific review).
 
 The definition of the 
phase space of gravity (as for any field theory) needs special care when boundaries are present. For illustration consider the two situations depicted in figure \ref{covph}: (assuming that the spacetimes of interest are asymptotically flat) on the left panel there is only a boundary at infinity, while on the left panel  the presence of the black hole is modelled by an additional  internal null boundary $\Delta$. In the presence of such null or time-like boundaries the local conservation of the symplectic structure (encoding the basic Poisson brackets structure of the field theory) does not suffice to grant the conservation of the symplectic structure from one initial Cauchy surface $\Sigma_1$ to a later one $\Sigma_2$. 
This is due to the fact that non trivial degrees of freedom excited when specified on $\Sigma_1$ can in turn excite those on the timelike or null boundaries of spacetime without being registered on $\Sigma_2$. Physically, energy can be carried in or out of the system via the boundaries. This implies that generically there is non trivial symplectic flux leaking across the timelike of null components of the boundary of the spacetime region of interest and hence a lack of conservation of the symplectic form defined on spacelike sections like $\Sigma$. The field theory defined on a bounded region is naturally that of an open system.

With suitable restrictions of the behaviour of fields on the boundaries one can recover a closed-system field theoretical model (i.e. with a conserved symplectic structure) which  can represent physically interesting situations.
When these boundaries are at infinity the conservation of the symplectic structure follows from the vanishing of the sympletic flux across the boundary at infinity due to suitable fall off conditions  on solutions of the field equations (defining asymptotically flat spacetimes). In the presence of the internal boundary $\Delta$ (see figure \ref{covph})  boundary conditions must be specified in order to: on the one hand capture the geometry and other degrees of freedom on the boundary in the desired regime (stationary black holes in our case), and, on the other hand, grant the vanishing of the symplectic flux across $\Delta$. If these two conditions are satisfied then one can view the system as a close hamiltonian system with a conserved symplectic structure, and thus contemplate its possible quantization.

In order to achieve this goal the definition of isolated horizons declares the boundary $\Delta$ to be a null surface with the same topology of that corresponding to a stationary BH horizon, namely $\Delta=S^2\times R$, and then (appart from some technical subtleties) fixes the part of the geometry and matter fields that can be freely specified on such a characteristic surface to match those of the corresponding stationary BH horizon, i.e., those of the Kerr-Newman family. Among other  properties, isolated horizons admit a preferred slicing in terms of spheres $S^2$ with an intrinsic geometry that is `time' independent in the sense that it does not depend on the $S^2$ spheres corresponding to the intersection of the spacelike hypersurfaces $\Sigma$ with $\Delta$.  This implies that the area and shape of the horizon are time independent and forbids,  by consistency with Einstein's equation, the flux of matter fields and gravitational radiation across the boundary. The definition of isolated horizons is independent of coordinates preserving the slicing, and admits a formulation which does not break $SU(2)$ gauge transformations \cite{Engle:2009vc, Engle:2010kt, Engle:2009zz}. For that reason fields on the boundary $\Delta$ are only fixed up to these two gauge symmetries, hence the boundary condition allows for field variations which are only pure gauge on $\Delta$ (a combination of tangent diffeomorphisms and $SU(2)$ gauge transformations). The definition of isolated horizons admits null horizons with local distortion \cite{Ashtekar:2004nd}. The number of degrees of freedom specifying this distortion is infinite and can be encoded in multipole moments \cite{Ashtekar:2004gp}. These cases are thought to represent BH horizons in equilibrium with exterior matter distributions causing the distortion due to tidal effects.

The local nature of the boundary condition implies a certain ambiguity in the characterisation of time evolution right at the boundary. For stationary black holes the normal is uniquely fixed by requiring it to correspond to the Killing fields whose normalization are fixed by demanding that they generate the symmetries of inertial observers at infinity. Such relationship between symmetries at infinity (which are generated from the Hamiltonian perspective by mass $M$ and angular momentum $J$) and the symmetry of the horizon along the null normal $\ell$ is the central ingredient in the validity of the first law (\ref{1st}). In the case of isolated horizons such link is lost due to the very local nature of the definition. 
A direct technical consequence of that is that one can no longer fix the null normal to the horizon $\ell_{\va IH}$ by simple global symmetry requirements and so  the  null generator $\ell_{\va IH}$ is defined only up to multiplication by a constant. 
This ambiguity has important consequences for the first law of BH mechanics which we discussed below.

\subsection{The laws of isolated horizons}\label{lih}

The restrictions on the boundary condition that are mentioned above capture the essential features of BH horizons in geometric terms. Despite the limitations described at the end of the previous section it has been shown that isolated horizons satisfy similar mechanical laws as BH horizons \cite{Ashtekar:1998sp, Ashtekar:1999yj, Ashtekar:2001is, Ashtekar:2001jb}. The first step is to define the notion of surface gravity $\kappa_{\va \rm IH}$ which is achieved by the conditions that define a  {\em weakly isolated horizon}.  A slight strengthening of these conditions leads to the notion of {\em isolated horizons} for which the only freedom that remains resides in the value of a constant rescaling of the null normal $\ell_{\va \rm IH}$ mentioned before. Surface gravity is defined via the intrinsic equation $\ell^a_{\va \rm IH}\nabla_a \ell_{\va\rm  IH}^b=\kappa_{\va\rm  IH} \ell_{\va \rm IH}^b$. Thus, the scaling ambiguity in  $\ell^a_{\va \rm{IH}}$ implies that the surface gravity $\kappa_{\va\rm IH}$ is up defined up multiplication by a constant as well. Nevertheless, one can show (as for stationary BHs) that $\kappa_{\va \rm IH}$ is indeed a constant on the horizon even when these are not necessarily spherically symmetric. This is known as the {\em zeroth law of black hole mechanics} in analogy with the zeroth law of thermodynamics stating that temperature is uniform in a body at thermal equilibrium. 

Despite the fact that the definition of isolated horizons does not allow for the flow of matter across the null surface representing the BH horizon, it is possible to prove the validity of the {\em first law} for isolated horizon (i.e. a balance law analogous to \eqref{1st}).  The way around is the integrability conditions that follow from requiring the existence of a consistent time evolution.  Time evolution in general relativity is defined by a timelike vector field (a time flow). When considering time evolution in the canonical framework the flow is generated via Hamilton's equations by a suitable Hamiltonian. In general relativity this Hamiltonian has non trivial contributions (i.e. not vanishing when Einsteins equations are satisfied) only coming from the boundaries \footnote{The vanishing of the Hamiltonian density in the bulk---a direct consequence of Einstein's equations---is rooted in general covariance. Due to the absence of any preferred time notion evolution is pure gauge in a generally covariant field theory. Boundary conditions break general covariance by introducing additional structure such as the inertial time flow at infinity or the particular choice of the null normal at the isolated horizon. A consequence of this is the appearance of boundary contributions to the Hamiltonian (see for instance \cite{wald}; appendix E).}. Consistency of the time evolution demands the boundary contribution to the Hamiltonian coming from the BH horizon to depend on other boundary charges---the area of the horizon $a$, angular momentum $J$ (which exists when the isolated horizon is axisymmetric \cite{Ashtekar:2001is}), and possibly other matter charges such as the electromagnetic charge $Q$---in a way that is encoded in a differential relation that amounts for the first law. Explicitly,
\be
dE_{\va \rm IH}=\frac{\kappa_{\rm \va IH} }{8\pi} da+\Omega_{\va \rm IH} dJ+ \Phi_{\va \rm IH} dQ,
\ee
where $\Omega_{\va \rm IH}$ and $\Phi_{\va\rm IH}$ are the angular velocity and electromagnetic potential of the isolated horizon respectively. All intensive quantities are defined up to a constant rescaling inheriting the freedom of the choice of the null normal $\ell_{\rm \va IH}$. This freedom precludes the integration of the previous relation to get a unique `state function' $E_{\va \rm IH}(a,J,Q)$. As a consequence there is an infinite family of energy notions for an isolated horizons; one need extra structure in order to extract physics form the previous form of the first law.  We will get back to this important point in Section \ref{nhg}. For an extensive review on properties of isolated and dynamical horizons see \cite{Ashtekar:2004cn, Ashtekar:2002ag}. Isolated horizons have been defined in $2+1$ \cite{Ashtekar:2002qc} and higher dimensions \cite{Liko:2007th, Bodendorfer:2013jba, Korzynski:2004gr}.

\subsection{Pre-quantum geometry I: Poisson brackets of geometric quantities on the Horizon} \label{pqg}

It turns out that under the restrictions imposed by the isolated horizon boundary conditionthe symplectic flux across the boundary defining the black hole horizon $\Delta$ is not zero but is given by the integral of a total differential; hence, it can be written as integrals on the boundary $\partial \Delta$. These boundary fluxes can be absorbed in the definition of  the symplectic structure of a closed system (i.e. conserved). Concretely, the symplectic structure \eqref{symp-sans} acquires a boundary term encoding the presence of the internal boundary with its isolated horizon degrees of freedom 
\ba \label{symp}
\Omega(\delta,\delta')
=\frac{1}{2\kappa\gamma }\int_{\Sigma}  \delta A^{i}\wedge \delta'
E^{\va}_{i}&-&\delta' A^{i}\wedge \delta
E^{\va}_{i},
  \n \\ &+& \frac{1}{\kappa \gamma } \int_{H\subset \partial \Sigma} \delta e_i\wedge\delta' e^i,
\ea
where $H=\Delta\cup\Sigma$ is a cross section of the isolated horizon (see right panel in Figure \ref{covph-4d}). 
This boundary term comes from the last term in \eqref{lance}. Thus, in addition to the Poisson brackets (\ref{tres})
for the bulk basic variables, the boundary term in the symplectic structure implies the following boundary fields commutation relations:
\be\label{cr}
\{e^i_a(x), e^j_b(y)\}= { \kappa \gamma}\, \delta^{ij} \boldsymbol{\epsilon}^{\va (2)}_{ab}\delta^{\va (2)}(x, y)\,s
\ee
where $\epsilon^{\va (2)}_{ab}$ is the 2d Levi-Civita density.
Recall that the $e$-fields encode the metric information, so the previous equation (which we have shown to come directly from the gravity action \eqref{actiong}) anticipates the non-commutativity of the boundary horizon geometry in the quantum theory.  More explicitly, if we consider the two dimensional induced metric tensor $g^{\va (2)}_{ab}\equiv e_a^ie^j_b\delta_{ij}$ on $\partial \Sigma$ the commutation relations that follow from \eqref{cr} are
\be\label{metric}
\{g^{\va (2)}_{ab}(x), g^{\va (2)}_{cd}(y)\}= { \kappa \gamma}\left(g^{\va (2)}_{ac}(x)\epsilon^{\va (2)}_{bd}+g^{\va (2)}_{bc}(x)\epsilon^{\va (2)}_{ad}+g^{\va (2)}_{ad}(x)\epsilon^{\va (2)}_{bc}+g^{\va (2)}_{bc}(x)\epsilon^{\va (2)}_{ad} \right)\delta^{\va (2)}(x,y).
\ee
The previous equation predicts quantum fuzziness of the geometry of the BH horizon (not all the components of the metric can be determined simultaneously due to Heisenberg uncertainty principle). 

For later application it is instructive to consider the bi-vectors 
\be\label{sigmaB} E_{B}^i=\epsilon^{i}_{jk} e^j\wedge e^k, \ee
where $E_{B}^i$ carries a subindex that stands for {\em boundary} to distinguish it from the
analogous looking object (\ref{sigma}) defined in terms of bulk fields instead (recall discussion on extended variables in Section \ref{ev}).  A straightforward calculation shows that (\ref{cr}) implies
\be\label{www}
\{E_{B}^i(x), E^j_{B}(y)\}= { \kappa \gamma}\, \epsilon^{ij}_{\ \ k} E_{B}^k(x) \ \ \delta^{\va (2)}(x,y),
\ee
which is remarkably simple: it corresponds to the algebra of angular momentum generators in standard mechanics.
By introducing the smeared version 
\be\label{fifi}
E_{B}(\alpha)\equiv \int_H\alpha_iE_B^i,
\ee
for an arbitrary smearing field $\alpha_i$ then we can express the previous commutation relations as
\be\label{eee}
\{E_{B}(\alpha),E_{B}(\beta)\}=\kappa \gamma E_{B}([\alpha,\beta]),
\ee
where $[\alpha,\beta]_k\equiv \epsilon^{ij}_{\ \ k} \alpha_i\beta_j$.
In the quantum theory this will lead to the discreteness of the BH horizon area spectrum\footnote{Similar commutation relations have been proposed via an independent argument in \cite{'tHooft:1996tq}.}.  It is possible to generalise the construction in order to describe higher dimensional BHs \cite{Bodendorfer:2013jba}.

\subsubsection{The gauge theory way: Chern-Simons formulation}\label{CSF}

Non rotating BHs in equilibrium can be modelled by the isolated horizon boundary condition. When one assumes the 
horizon to be spherically symmetric one finds that
the curvature of the Ashtekar-Barbero connection in the bulk is related to the `electric' fields on the horizon as
\be\label{ih}
\frac{a_{H}}{2\pi}\  F^i(A)= E_{B}^i,
\ee  
where $a_{H}$ is the area of the horizon (taken as non dynamical parameter characterizing the isolated horizon), and we assume 
$K^i=0$ which corresponds to the time-symmetric slicing where the analysis of Section \ref{nhg} holds \footnote{The quasilocal treatment leading to the area Hamiltonian was not available when the first derivation of the CS formulation was proposed \cite{Engle:2010kt}. A different slicing was then used which lead a factor $a_H/(\pi (1-\gamma^2))$ on the {\em l.h.s} of (\ref{ih}).}. As shown in \cite{Engle:2010kt}, the constraint \eqref{ih} amounts for the imposition of both diffeomorphism and the Gauss constraint at the boundary, i.e. it is in a precise sense the analog of (\ref{unol}) and (\ref{tresl}) together.  Because of the boundary condition, no scalar constraint (\ref{dosl}) needs to be imposed. All the boundary dynamics is coded by \eqref{ih}.  
 
Using the previous equation,  and under the assumption that fields satisfy the IH boundary conditions conditions described in Section \ref{mbhs}, the boundary contribution to symplectic structure \eqref{symp} can be rewritten in terms of the Ashtekar-Barbero connection on the boundary
as
\be 
-\frac{a_{H}}{2\pi \kappa \gamma}\int_{\partial \Sigma} \delta A_i\wedge \delta' A^i.
\ee
This implies that (on the boundary) the Ashtekar-Barbero connection does not commute with itself, namely
\be
\{A^i_a(x), A^j_b(y)\}= \frac{2\pi \kappa \gamma}{a_{H}}\, \delta^{ij} \boldsymbol{\epsilon}^{\va (2)}\delta^{\va (2)}(x, y)\,.
\ee 
The previous Poisson structure corresponds to that of an $SU(2)$ Chern-Simons theory 
with level $k=a_{H}/(2\pi \gamma)$.  It can be shown  that in this framework
equation \eqref{ih} is a constraint that completes the Gauss constraint (\ref{tresl}) in the presence of a boundary: via Hamiltons equation 
it is the generator of boundary gauge transformations \cite{Engle:2010kt} (the relation between $E_B$ and the generators of internal gauge transformations will be clarified further in Section \ref{cattaneo}). 

Thus, the classical degrees of freedom on an isolated horizon can be described dynamically by a Chern-Simons theory. Historically, the Chern-Simons formulation of isolated horizons was first found in its $U(1)$ gauge fixed form in \cite{Ashtekar:1997yu, Ashtekar:2000eq}. However, the $U(1)$ gauge fixed theory in the quantum theory when one tries to impose  the gauge fixed version of (\ref{ih}).  The  obstacle is that the classical gauge fixing becomes incompatible with Heisenberg's uncertainty principle due to the commutation relations \eqref{www}. This difficulty is circumvented in the $SU(2)$ formulation which was put forward later \cite{Engle:2009vc, Engle:2010kt}.  The formulation was extended to static BHs with distorsion in \cite{Perez:2010pq}. There are other parametrizations of the phase space of isolated horizons in the literature establishing a link with BF theories \cite{Celada:2016jdt}; see for instance \cite{Pranzetti:2014tla, Wang:2014oua}.

Rotating black holes do not satisfy the boundary condition \eqref{ih} \cite{Roken:2013mqa}. Technical difficulties related to the action of diffeomorphisms also arise. For a discussion of these issues and a proposed model \cite{Frodden:2012en}.  Isolated horizons which are not spherically symmetric and not rotating can be mapped to new variables so that the analog of \eqref{ih} (in the $U(1)$ gauge) is satisfied \cite{Ashtekar:2004nd, Beetle:2010rd}. For simplicity we will concentrate on spherically symmetric black holes in this article.

\subsection{Pre-quantum geometry II: Poisson brackets of geometric quantities in the bulk} 

Here we show how the Poisson non-commutativity of the geometric variables on a boundary is not a peculiar feature of boundary variables but a generic property of metric observables which remains valid in the bulk. This leads to the non commutativity of the associated quantum operators in LQG and to its main prediction: the fundamental discreteness of the eigenvalues of geometry. This prediction is central for the description of the quantum properties of black holes in this approach to quantum gravity. 

\subsubsection{Fluxes: the building block of quantum geometry}
Given an arbitrary surface $S$ in space $\Sigma$  one can define the following classical object which we call the {\em flux} of (geometry) $E$---in analogy with the equivalent quantity in electromagnetism or Yang-Mills theory---by the following expression
\be\label{ffflux}
E(S,\alpha)\equiv \int_S \alpha_iE^i,
\ee
where the smearing field $\alpha^i$ is assumed to have compact support in $\Sigma$. This quantity is central in the construction
of quantum operators capturing geometric notions in LQG. It is an extended variable (as discussed in Section \ref{ev}) which, through its non locality, allows for the necessary point-splitting regularization of non linear observables in the quantum theory \footnote{When applying the canonical quantization recipe, the basic variables $E$ and $A$ mut be promoted to suitable operators acting in a Hilbert space. Because of the distributional nature of the Poisson brackets (\ref{tres}), these operators make sense as distributions as well. Products of these operators at a same point are mathematically ill-defined and lead to the UV divergencies that plague quantum field theories. The extended variables used in LQG are natural regulating structures that resolve this mathematical problem in the definition of (non-linear) geometric observables.}.  Among the simplest geometric observables one has the {\em area} of a surface $S$, which can be shown to be given by 
\be\label{area}
a(S)=\int_S \sqrt{E_{xy}\cdot E_{xy}} \ dx dy, 
\ee
where $\cdot$ denotes the contraction of internal indices (inner product in the internal space) of the $E$'s and $x,y$ are local coordinates on $S$. The fact that area is given by the previous expression is a simple consequence of the definition (\ref{sigma}) and the relationship of the triad $e$ with the metric. 
Similarly, one can define the {\em volume} of a region $R \in \Sigma$ as 
\be\label{volume}
v(R)=\int_R \sqrt{E_{xy}\cdot(E_{yz}\times E_{zx})} \ dxdydz. 
\ee
Both of which are potentially $\rm UV$-divergent in the quantum theory due to the fact that they involve the multiplication of operator-valued distributions at the same space point. The statement, that we give here without a proof, is that the quantum operators $\widehat a(S)$ and $\widehat v(R)$ for arbitrary surfaces $S$ and arbitrary regions $R$ can be defined on the Hilbert space of LQG as functionals of the fluxes (\ref{ffflux}) for families of regulating surfaces which are removed via a suitable limiting procedure (for details see \cite{Ashtekar:1996eg, Ashtekar:1997fb}). In this way the fluxes (\ref{fifi})---which arise naturally in the context of the boundary geometry---are also very important when defined in the bulk 
in terms of  an arbitrary $2$-surface $S\subset \Sigma$. We will see in what follows that the bulk fluxes also satisfy 
commutation relations of the type (\ref{eee}).     

\subsubsection{ Non-commutativity of fluxes; the heart of Planckian discreteness}\label{cattaneo} 

Here we show that the Poisson brackets among fluxes \eqref{ffflux} reproduce the algebra of angular momentum generators at every single point on the surface.  Here we also show how the appearance of the rotation algebra is related to the $SU(2)$ gauge transformations generated by the Gauss law. Such non commutativity might seem at first paradoxical from the fact that the $E^i$ Poisson commute according to \eqref{tres}.
The apparent tension is resolved when one appropriately takes into account the Gauss law (\ref{tresl}) and studies carefully the mathematical subtlelties associated with computing the Poisson bracket of an observable smeared on a $2$-dimensional surface surface---as \eqref{ffflux}---in the context of the field theory on $3+1$ dimensions. This subtlety has been dealt with in at least two related ways some time ago \cite{Ashtekar:1998ak, Freidel:2011ue}. Here we follow a simpler and more geometric account recently introduced in  \cite{Cattaneo:2016zsq}. 
We present it in what follows for the interested reader.

Without loss of generality we assume $S$ to be a close surface---if the $2$-surface $S$ does not close we can extend it to a new surface $S'$ in some arbitrary way in the region outside the support of $\alpha$ to have it closed so that $E(S,\alpha)=E(S',\alpha)$.  Using Stokes theorem we can write (\ref{ffflux}) as a 3-dimensional integral in the interior of $S$
\ba
\n E(S,\alpha)&=& \int_{{\rm int}[S]} d(\alpha_iE^i)=\int_{{\rm int}[S]} (d_A\alpha_i) E^i+\alpha_i (d_AE^i)\\
&\approx& \int_{{\rm int}[S]} (d_A\alpha_i)\wedge E^i,\label{gggi}
\ea 
where in the second line the symbol $\approx$ reminds us that we have used the Gauss law \eqref{tresl}.
More precisely, this implies that the Poisson bracket of any gauge invariant observable\footnote{A quantity is gauge invariant if $O(E,A)=O(E+\delta E, A+\delta A)$ with $\delta$'s given by \eqref{transfi} which is equivalent to saying that $O(E,A)$ Poisson commutes with the Gauss generator \eqref{gauss}. } and $E(S,\alpha)$,  and Poisson bracket of the same observable and the expression of the right hand side of $\approx$ coincide. In other words, when considering gauge invariant quantities $\approx$ amounts to an $=$ sign. 
 
It is only at this point---after writing the fluxes in terms of a $3$ dimensional smearing of local fields---that we can use the Poisson brackets (\ref{tres}) (whose meaning is a distribution in three dimensions as the Dirac delta functions in (\ref{tres}) explicitly show). 
But now the new expression of the fluxes  \eqref{gggi} explicitly depends on the connection $A^i$ via the covariant derivative $d_A$. This is the reason at the origin of the non trivial Poisson bracket between fluxes. Direct evaluation of the Poisson brackets  using (\ref{tres}) yields
\ba
&& \left\{E(S,\alpha),E(S,\beta)\right\}\approx \int\int dx^3 dy^3 \left\{d\alpha_i\wedge E^i+\epsilon_{ijk} A^j\wedge \alpha^k\wedge E^i, 
d\beta_l\wedge E^l+\epsilon_{lmn} A^m\wedge \beta^n\wedge E^l\right\}\n \\
&&\approx \int\int  dx^3 dy^3 \left\{d\alpha_i\wedge E^i, 
\epsilon_{lmn} A^m\wedge \beta^n\wedge E^l\right\}
+\left\{\epsilon_{ijk} A^j\wedge \alpha^k\wedge E^i, 
d\beta_l\wedge E^l\right\}
+\left\{\epsilon_{ijk} A^j\wedge \alpha^k\wedge E^i, 
\epsilon_{lmn} A^m\wedge \beta^n\wedge E^l\right\}\n \\
&&\approx\kappa \gamma \int  dx^3   
\epsilon_{ijk} d\alpha^i\wedge \beta^j\wedge E^k
+\epsilon_{ijk} \alpha^i \wedge d\beta^j \wedge E^k+\cdots \n \\
&&\approx\kappa \gamma \int  dx^3  d_A([\alpha,\beta])_k\wedge E^k\n\\
&&\approx \kappa \gamma E[[\alpha,\beta], S],  \n
\ea
where $[\alpha,\beta]_k\equiv \epsilon_{kij} \alpha^i\beta^k$, and in the third line we have omitted the explicit computation of the third term of the second line as this one can be guessed from the fact that the result must be gauge invariant. This leads to the sought result: the non commutativity of the fluxes that is at the heart of the discreteness of geometric kinematical observables in LQG. Namely:
\be\label{npc1}
\left\{E(S,\alpha),E(S,\beta)\right\}\approx\kappa \gamma E[[\alpha,\beta], S].
\ee
We recover in this way in the bulk for the smeared fluxes
the same result found in \eqref{eee} for the boundary. The observables $e^i$ and the Poisson brackets \eqref{cr} and \eqref{metric}
are not known to be available in the bulk of space $\Sigma$. However, recent results \cite{Freidel:2016bxd} indicate that there might be a way to extending their to the interior of the space. This could have very important consequences as it would allow for the definition of a new set of observables that could, one the one hand, lead to a natural geometrization of matter degrees of freedom, and, on the other hand, reduce some quantization ambiguities in the definition of the dynamics of LQG. We will comment on these developments in Section \ref{outlook}. 

%
%\section{Conceptual discussion}
%
%\begin{enumerate}
%\item {\em What are black holes in quantum gravity.}
%\item {\em Holography and quantum gravity.}
%\item {\em The entanglement entropy approach.}
%\item {\em Addressing the entropy problem.}
%\end{enumerate}
\subsection{Quantum geometry}\label{quantum-geometry}

 We are now ready to sketch the construction of the quantum theory. 
 LQG was born from the convergence of two main set of ideas: the old ideas
about background independence formulated by Dirac, Wheeler, DeWitt
and Misner in the context of Hamiltonian general relativity, and
the observation by Wilson, Migdal, among others, that Wilson loops
are natural variables in the non perturbative formulation of gauge
theories. The relevance of these two ideas is manifest if one
formulates classical gravity in terms of the variables that we introduced in Section \ref{newvariables} 
that render the some of the equations of general relativity similar to those of
standard electromagnetism or Yang-Mills theory (Section \ref{haha}).

What is the physical meaning of the {\em new variables}? The triplet
of vector potentials $A^i$ have an interpretation that is similar
to that of $\vec A$ in electromagnetism: they encode the `Aharonov-Bohm phase'
acquired by matter when parallel transported along a path $\gamma$ in
space---affecting all forms of matter due to the universality of
gravity. Unlike in electromagnetism, here the `phase' is replaced by
an element of $SU(2)$ associated to the action of a real rotation in
space on the displaced spinor. This is mathematically encoded in the
Wilson loop (related to the circulation of the magnetic fields $B_i$) along the loop $\gamma$ according to 
\be \label{wilson}
W_{\gamma}[A]={\rm P}
\exp{\int_{\gamma}\tau_i  A^i_a \frac{dx^a}{ds} ds}\ \ \in SU(2),
\ee
where $P$ denotes the path-ordered-exponential\footnote{The path ordered exponential is necessary do to the non commutativity of the
matrices $\tau^i$.}, $\tau^i$ are the
generators of $SU(2)$, and $s$ is an arbitrary parameter along along
$\gamma$.  This expression is the $SU(2)$ counterpart of the natural extended variables 
(\ref{24}) mentioned in Section \ref{ev} before the introduction of the time gauge.  In this analogy, the electric fields $E^i$ have a novel physical
interpretation: they encode (as reviewed in the previous sections) the geometry of
3-dimensional space, and define in particular  the area surfaces and volume of regions
according to \eqref{area} and \eqref{volume}.

The quantization is performed following the canonical approach,
i.e., promoting the phase space variables to self adjoint
operators in a Hilbert space $\h$ satisfying the canonical
commutation relations according to the rule $\{\ ,\ \}\rightarrow
-i/\hbar [\ ,\ ] $. 
As there is no background structure the notion of {\em particle},
as basic excitations of a {\em vacuum} representing a state of
minimal energy, does not exist. However, there is a natural {\em
vacuum} $|0\rangle_{}$ associated to the state of no geometry or vanishing electric field, i.e.  \be
{
E^i}|0\rangle_{}=0.\label{nogeo}
\ee
This state represents a very degenerate quantum
geometry where the area of any surface and the volume of any region vanishes. 
Distances are not naturally defined (some ambiguities affect its definition \cite{Thiemann:1996at, Bianchi:2008es}); however,
all definitions coincide in the statement that the distance between any pair of points is zero in the state $|0\rangle_{}$.
The operator $W_{\gamma}[A]$, the quantum version of (\ref{wilson}),  acts on the vacuum by creating a
one-dimensional flux tube of electric field along $\gamma$. These fundamental {\em
Faraday lines} represent the building blocks of a notion of {\em
quantum geometry}.

Only those excitations given by closed
Wilson lines of quantized electric field are allowed by quantum
Einstein's equations, i.e., {\em loop states}.
This is due to the  Gauss constraint (\ref{tresl}) ({\em divergence of the electric field must vanish}) 
that follows directly the equations of motion coming from (\ref{actiong}); Section \ref{haha}. 
Therefore, Faraday lines must always close and form loops.
The construction of the Hilbert space of quantum gravity is thus started by
considering the set of arbitrary multiple-loop states, which can
be used to represent (as emphasized by Wilson in the context of
standard gauge theories) the set of gauge invariant functionals of
$A^i$. Multiple-loop states can be combined to form an orthonormal basis
of the Hilbert space of gravity. The elements of this basis are
labelled by: a closed graph in space, a collection of
spins---unitary irreducible representation of $SU(2)$---assigned
to its edges, and a collection of discrete quantum numbers
assigned to intersections. As a consequence of \eqref{tresl} 
the rules of addition of angular momentum must be
satisfied at intersections: the total flux of electric field at a node is zero. They are called {\em spin-network
states}.

{\em Spin network states} are eigenstates of geometry as it
follows from the rigorous quantization of the notion of area and
volume (given by equations (\ref{area}---\ref{volume})).  In LQG the area of a surface can only
take discrete values in units of Planck scale! More precisely, given a surface $S$ and a spin-network state with edges intersecting the surface (at punctures labelled by the dummy index $p$ below) with spins $j_1,j_2\cdots$ then one has 
\be\label{area1}
a(S) |j_1,j_2\cdots\rangle=\left(8\pi \gamma \ell_p^2  \sum_{p} \sqrt{j_p (j_p+1)}\right)\  |j_1,j_2\cdots\rangle
\ee
where we have labelled the spin-network state with the relevant spins only (further details identifying the state, which are not relevant for the area eigenvalue, are not explicitly written for notational simplicity). A particularly important application of this formula is the computation of the eigenvalues of the area of a black hole horizon. A graphical representation of the situation is presented in Figure \ref{BH-quantum}: links in the figure  can be interpreted as flux lines of quantum area depending on their colouring by spins.
Similarly, the
spectrum of the volume operator $V(R)$ is 
discrete and associated to the presence of spin network
intersections inside the region $R$ (nodes in Figure \ref{BH-quantum} represent volume quanta).

\begin{figure}[h!!!!!!!]
%\begin{center}
\centerline{\hspace{0.5cm} \(\begin{array}{c}
\includegraphics[height=8cm]{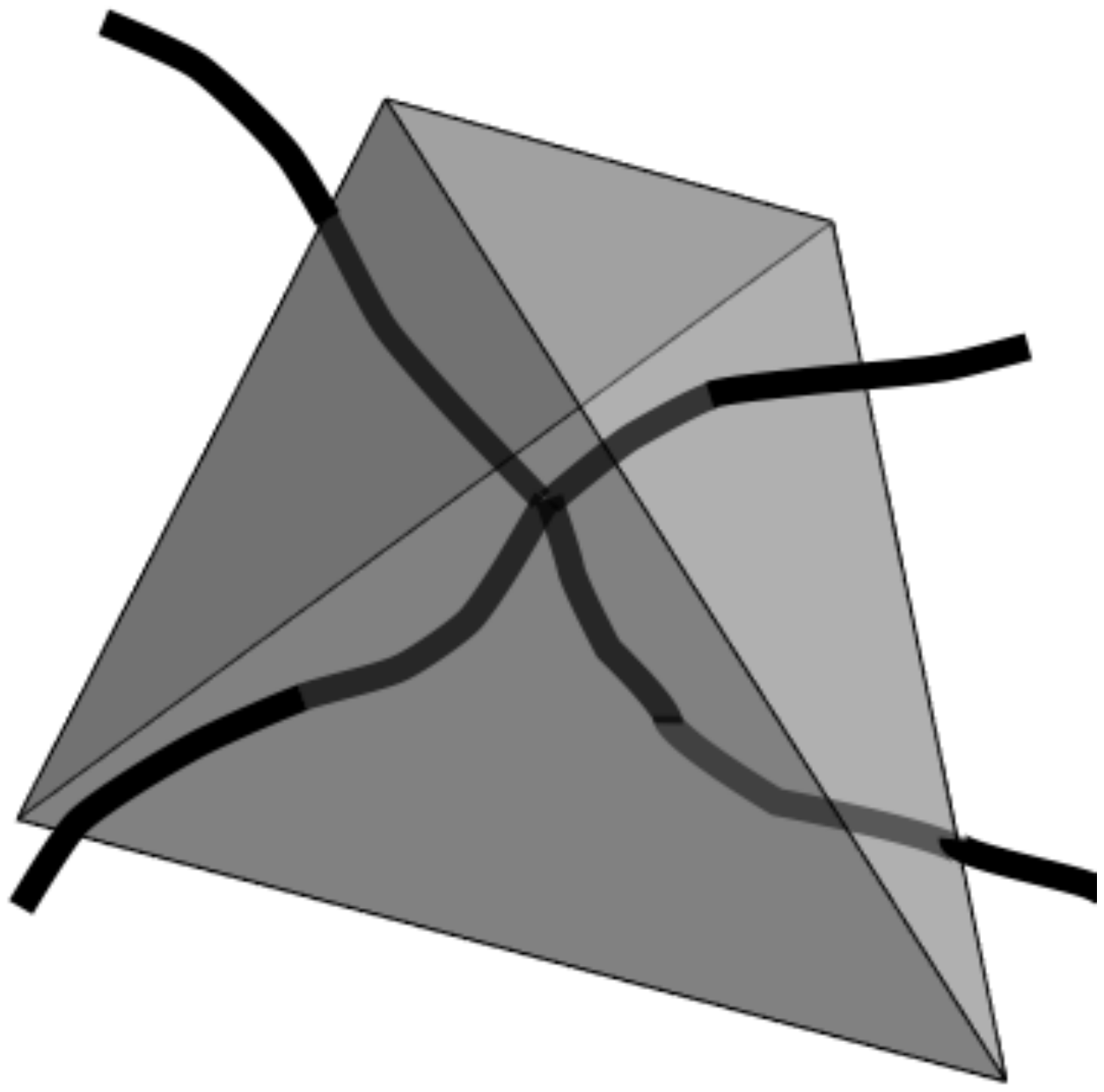} 
\end{array}\ \ \ \ \ \ \ \ \ \ \ \ \ \ \ \ \ \ \ \ \ \ \ \ \ \begin{array}{c}
\includegraphics[height=6cm]{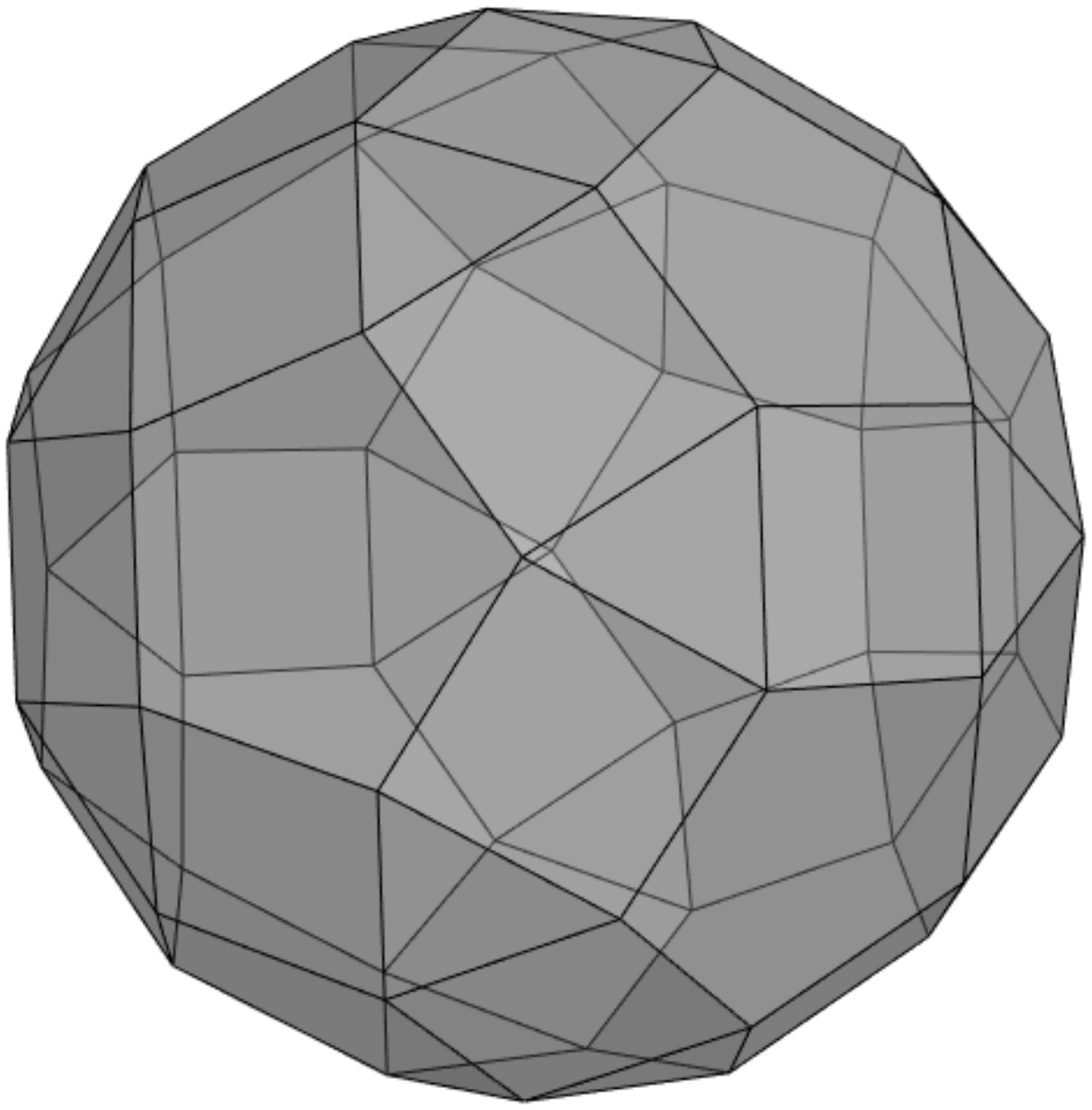} 
\end{array}\) } \caption{Minkowski theorem provides an unambiguous picture of the {\em shape} associated to a quantum state of spin-network nodes ({\em atoms of space}): it correspond to that of a convex polyhedron (not necessarily regular as in the picture) where individual faces represent the quanta of area carried by dual spin-network edges. With the exception of the four-valent case (the quantum tetrahedron on the left), the shape of the polytope depends on the value of these quanta in a global manner \cite{Bianchi:2010gc} (a generic case is represented on the right. The dual spin-network edges are not represented for simplicity).   In the Chern-Simon formulation of the black hole horizon this interpretation is also available for the horizon; however, unlike for the spin-network nodes, its physical validity is less clear as Minkowski's reconstruction works by an embedding in flat Euclidean space. Nevertheless, the picture still provides a simple intuitive visualisation of the horizon quantum states whose dual picture (the spin-network representation) is given in Figure \ref{BH-quantum}. }
\label{polytope}
%\end{center}
\end{figure}

More precisely, as mentioned before gauge invariance implies that the quantum numbers of the fluxes 
associated to the different spin-network links converging at a node must add up to zero.
This condition admits an unambiguous interpretation of nodes as quantum states of a convex polyedron.  
This interpretation is based on a theorem by Minkowski on discrete Euclidean geometry \cite{herman} (see Figure \ref{polytope}).
The properties of the quantum shape of polyedra has been studied numerically via a variational algorithm \cite{Bianchi:2010gc}. In figure \ref{BH-quantum} we represent a spin-network state including a black hole as a boundary (see below for the description of the boundary (horizon) quantum state). There are three and four valent nodes in the bulk. According to Minkowski theorem four-valent nodes represent quantum tetrahedra where the areas of the four triangles are defined by the area eigenvalues depending on the spins $j$; three-valent nodes are degenerate (zero-volume and purely quantum) excitations.

The discovery of the discrete nature of geometry at the
fundamental level has profound physical implications. In fact even
before solving the quantum dynamics of the theory one can already
answer important physical questions. The most representative
example (and early success of LQG) is the computation of black
hole entropy from first principles.

Another profound potential implication of discreteness concerns the UV
divergences that plague standard QFT's. It is well known that in
standard QFT the UV problem finds its origin in the difficulties
associated to the quantization of product of fields at the same
point (representing interactions). A first hint of the regulating
role of gravity is provided by the fact that, despite of their
non-linearity in $E_i$, area and volume are quantized without
the appearance of any UV divergences. This is  
expected to provide a universal regulating physical cut-off to the fundamental 
description of fields in LQG (see important discussion in Section \ref{LIV}).

Before the imposition of the vector constraint (\ref{unol}) two spin network differing by a tiny
modification of their graphs are orthogonal states!---that would
seem to make the theory intractable as the Hilbert space would be too large. This is where the
crucial role background independence starts becoming apparent as
the vector constraint---although is not self-evident---implies
that only the information in spin network states up to smooth
deformations is physically relevant. Physical states are given by equivalence classes
of spin networks under smooth deformations: these states are
called {\em abstract spin networks} \cite{Ashtekar:1994mh, Ashtekar:1994wa}.

{\em Abstract spin network} states represent a quantum state of
the geometry of space in a fully combinatorial manner. They can be
viewed as a collection of `atoms' of volume (given by the quanta
carried by intersections) interconnected by edges carrying quanta
of area of the interface between adjacent atoms. This is the
essence of background independence: the spin network states do not
live on any pre-established space, they define space themselves.
The details of the way we represent them on a three dimensional
`drawing board' do not carry physical information. The degrees of
freedom of gravity are in the combinatorial information encoded in
the collection of quantum numbers of the basic atoms and their
connectivity.

Finally the full non linearity of the dynamics gravity is encoded in the quantum scalar constraint (\ref{dosl}).
Quantization of this operator has been shown to be available \cite{Thiemann:1996aw, Thiemann:1997rv}; however, the process suffers from ambiguities \cite{Perez:2005fn}. 
These ambiguities are expected to be reduced if strong anomaly freeness conditions are imposed based on the 
requirement that the constraint algebra is satisfied in a suitable sense (for a modern discussion of this important problem see \cite{Laddha:book} and references therein).

%@@

\subsection{Quantum isolated horizons}\label{CSQ}

\begin{figure}[h!!!!!!!]
%\begin{center}
\centerline{\hspace{0.5cm} \(
\begin{array}{c}
\includegraphics[height=6cm]{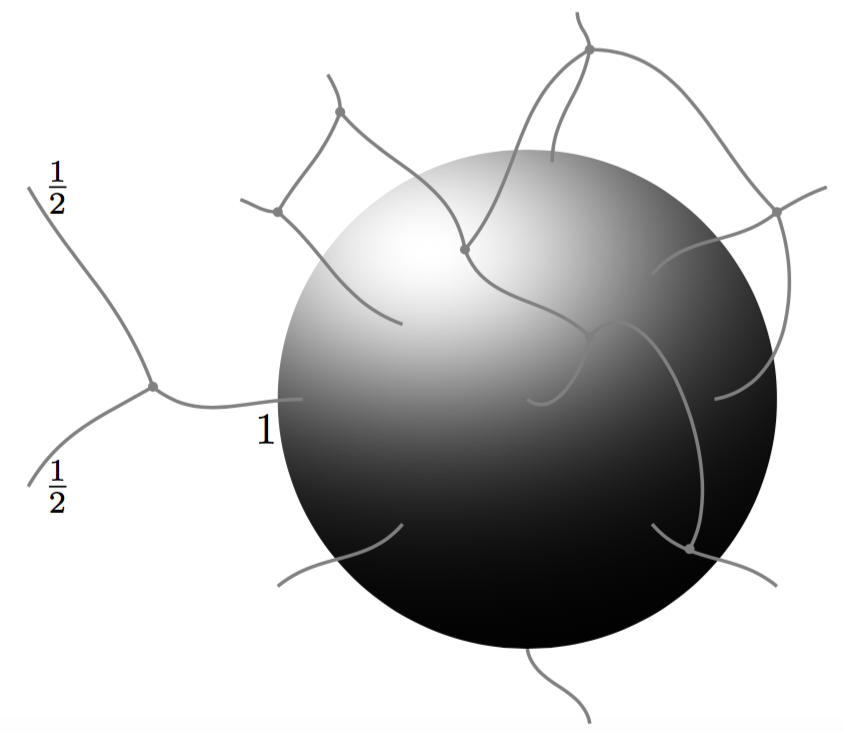} 
\end{array}\) } \caption{Quantum states of the bulk space-geometry are represented by spin network states. Spin network states are given by 
graph-like excitations with one dimensional lines representing quantized flux-lines of the $SU(2)$ electric field $E^i$ whose `intensity' $E^2$ is directly related to the geometric notion of area (see equation \eqref{area} and the quantum version \eqref{area1}). Therefore, one dimensional lines can be interpreted as carrying quantized units of transversal area. Similarly, nodes
carry quantum numbers of volume. Black holes are described by boundaries satisfying the isolated horizon boundary condition. The bulk state induces a boundary state which is given by a collection of point-like excitations (punctures) which carry quanta of boundary-area. To avoid overloading the figure with details, we have only explicitly labelled with the corresponding half-integers three particular links. At nodes, gauge invariance is encoded in the validity of the Gauss law (\ref{tresl}) which requires the net flux of electric field to vanish. This condition restricts spin labels to those satisfying the rules of addition of angular momentum in standard quantum mechanics. Spin-networks were originally proposed by Penrose as a network defined by spinning particles colliding at nodes in an ambient space. He showed that they encoded the geometry of $3$-dimensional Euclidean space combinatorially \cite{penrose-sn}. Graph states are defined modulo diffeomorphisms (graph deformations) preserving the boundary, i.e. tangent to the boundary.}
\label{BH-quantum}
%\end{center}
\end{figure}

In the presence of a boundary representing a black hole a given spin-network state intersects the boundary at punctures, Figure \ref{BH-quantum}. These punctures 
are themselves excitations in the Hilbert space of the boundary Chern-Simons theory 
in the isolated horizon model of black holes.  The Hilbert space of isolated horizons is the tensor product  $\sH_{H}\otimes\sH_{out}$ where the two factors denote the Hilbert space of the horizon (spanned by puncture-states) and the Hilbert space of the outside bulk (spanned by spin networks) respectively.  At punctures the quantum version of (\ref{ih}) must be imposed. This constraint takes the fom
\be\label{gauss-c}
\left(\frac{a_{H}}{2\pi}\  F^i(A)\otimes \mathds{1}-\mathds{1}\otimes E^i\right)|\psi\rangle_{H} |\psi\rangle_{out} =0,
\ee
where $|\psi\rangle_{H}\in \sH_{H}$ and $|\psi\rangle_{out}\in \sH_{out}$. It can be shown that $|\psi\rangle_{ out}$
breaks individually the $SU(2)$ gauge invariance when the transformation acts non trivially on the horizon \cite{Cattaneo:2016zsq}. It is precisely the addition of the Chern-Simons boundary degrees of freedom and the imposition of \eqref{gauss-c} which restores the gauge symmetry broken from the point of view of the bulk by the presence of the boundary \cite{Engle:2010kt}.  The Chern-Simons boundary degrees of freedom  can be seen in this sense as would-be-gauge excitations \cite{Benguria:1976in, Carlip:2005zn}.
The representation of a generic state is given in Figure \ref{BH-quantum}.

\subsection{The continuum limit}\label{thecontlim}

A key feature of LQG is the prediction of fundamental discreteness at the Planck scale. States of the gravitational degrees of freedom are spanned in terms of spin-network states (polymer-like excitations of quantum geometry) each of which admits the interpretation of an eigenstate of geometry which is discrete and atomistic at the fundamental level \cite{Rovelli:1994ge, Ashtekar:1996eg, Ashtekar:1997fb}. This feature is the trademark of the theory; on a rigorous basis it has been shown that the representation of the basic algebra of geometric observables as operators in a Hilbert space---containing a `vacuum' or `no-geometry' state \eqref{nogeo} which is diffeomorphism invariant and hence for which all geometric eigenvalues vanish---is unique \cite{Lewandowski:2005jk}.  
In this picture flat Minkowski space-time must be viewed as a highly exited state of the `no-geometry' state (\ref{nogeo}), where the quantum space-time building blocks are brought together to produce the (locally) flat arena where other particles interact.  Thus, there is no a priori notion of space-time unless a particular state is chosen in the Hilbert space. 
Loop quantum gravity is a concrete implementation of such non-perturbative canonical quantization of gravity \cite{Rovelli:2004tv, Thiemann:2007zz}.
Even though  important questions remain open, there are robust results exhibiting features which one might expect to be sufficiently generic to remain in a consistent complete picture. 
   
There has been an important activity in trying to construct semiclassical states in the framework. At the canonical level efforts have concentrated in the definition of coherent states of quantum geometry \cite{Thiemann:2000bw, Freidel:2010aq, Livine:2007vk, Rovelli:2010km, Bianchi:2010gc} representing a given classical configuration. The relationship between the fundamental spin-network state representation of quantum gravity and the Fock state representation of QFTs has been explored in \cite{Ashtekar:1992tm, Sahlmann:2002qj, Ashtekar:2001xp,  Conrady:2004ww}. Recently, emphasis has been given to the constructions of states that reproduce the short distance correlations of quantum field theory \cite{Bianchi:2016tmw, Bianchi:2016hmk}.  
These results provide useful insights on the nature of the low energy limit of LQG.  Clear understanding is still missing mainly because the dynamical aspect of the question---understanding the solution space of \eqref{dosl}---is still under poor control.

However, the view that arises from the above studies is that smooth  geometry should emerge  from the underlying discrete fundamental structures via the introduction of coarse observers that are insensitive to the details of the UV underlying structures. 
One expects that renormalization group techniques should be essential in such context (see \cite{Dittrich:2014ala} and references therein). The problem remains a hard one as one needs to recover a continuum limit from the underlying purely combinatorial Wheeler's {\em pre-geometric} picture \cite{Misner:1974qy}. The task is complicated further in that the usual tools---applied to more standard situations where continuity arises from fundamentally discrete basic elements (e.g. condensed matter systems or lattice regularisations of QFTs)---cannot be directly imported to the context where no background geometry is available.  

Consequently, a given classical space-time cannot directly correspond to a unique quantum state in the fundamental theory: generically, there will be infinitely many different quantum states satisfying the coarse graining criterion defined by a single classical geometry. For instance, there is no state in the LQG Hilbert space that corresponds to Minkowski space-time. Flat spacetime is expected to emerge from the contributions of a large ensemble of states all of which look flat from the coarse grained perspective. Because coarse grained observers are insensitive to Planckian details (quantum pre-geometric defects), flat space-time might be more naturally associated with a density matrix in LQG than to any particular pure state \cite{Livine:2007sy, Ariwahjoedi:2014wpa}.  The bulk entropy associated to such mix-state would be non trivial  in the sense that it would carry a non trivial entropy density (this will be important in Section \ref{loss}).

Unfortunately, due to the difficulties associated with describing the low energy limit of LQG, one cannot give a precise description of the exact nature of the pre-geometric defects that might survive in a state defining a background semiclassical geometry.  Nevertheless, there are a variety of structures at the Planck scale that do not seem to play an important role in the construction of the continuum semiclassical space-time, yet they are expected to arise in strong coupling dynamical processes as no selection rule forbids them.  For instance one has {\em close loops} and embedded {\em knots} which are the simplest solutions  to all the quantum constraint-equations including (\ref{dosl}) \cite{PhysRevLett.61.1155}. As geometric excitations they are degenerate  and carry area  but no volume quantum numbers.  More generally the semiclassical weave states in LQG can contain local degenerate contributions such as  {\em trivalent spin-network nodes} or other configurations  with vanishing volume quanta.  On the dynamical side the vertex amplitude of {\em spin foams} \cite{Engle:2007wy} is known not to impose some of the metricity constraints stronly. This allows for the possibility of pre-geometric structures to survive \cite{Speziale:2010cf} in the semiclassical limit. From the canonical  perspective the quantum Hamiltonian constraint seem to invariably produce such pre-geometric defects at the Planck scale. Finally, there is the q-bit degeneracy of the volume eigenvalues \cite{Thiemann:1996au, Ashtekar:1997fb, Bianchi:2010gc}: each non-zero eigen-space of the volume of a spin-network node is two dimensional. This local two-fold degeneracy can, by itself, be the source of a non trivial bulk entropy (a recent study of bulk entropy for a volume coarse graining see \cite{Astuti:2016dmk}).

All this indicates that the emerging picture in an approach as LQG is very different from the bulk-boundary-duality type of quantum gravity scenarios such as the one proposed by  ADS-CFT correspondence scenarios \cite{Maldacena:1997re}. In LQG the notion of smooth space-time has a capacity for infinite bulk entropy . Such non holographic behaviour at the fundamental level might seem (to some) at odds with the belief in the so-called {\em holographic principle} as a basic pillar of quantum gravity  \cite{Bousso:2002ju}.  However, further scrutiny shows that, despite its non-holographic nature, the predictions of  LQG are completely consistent with the phenomenology motivating holography. 

This is basically because, in the physical situations where holographic behaviour arises, bulk entropy only contributes as an irrelevant overall constant.  A clear example of this is the validity of covariant entropy bounds satisfied by relative (entanglement) entropy \cite{Bousso:2014sda}. Relative entropy of a state $\rho$ is defined with respect to a reference quantum state (a `vacuum' state) $\rho_0$ as
\be\label{ree}
S_{\rho_0}(\rho) = - \Tr[\rho \log \rho] + \Tr[\rho_0 \log \rho_0].
\ee
While giving non trivial information about excitations of the `vacuum' in the mean field approximation (where a space-time background makes sense) such covariant entropy bounds do not constraint the number of fundamental degrees of freedom of quantum geometry. More precisely, the bulk Planckian entropy, being a constant, just cancels out in the subtraction that regularises $S_{\rho_0}(\rho)$. The point here is that on the basis of the insights of LQG on the nature of the  fundamental quantum geometry excitations, the holographic principle is degraded from its status of fundamental principle to a low energy property of quantum fields on curved space-times. This is the perspective that follows from the LQG statistical mechanical account of black hole entropy (see \cite{Perez:2014ura} for a description in terms of entanglement and a discussion along the lines of the present paper and \cite{DiazPolo:2011np, bayo} for recent reviews). In this way the framework of LQG, without being fundamentally `holographic', can accommodate the holographic phenomenon when it is valid. The holographic behaviour is a characteristic of suitable systems (black holes, isolated horizons, null surfaces, etc.) but not a fundamental property of the theory.

Specialising to ensemble of states that all `look like' Minkowski space-time for suitably defined coarse grained observers we notice that their members must differ by hidden degrees of freedom from the viewpoint of those low energy observers. In particular they would all agree on stating that all the states (even though different at the fundamental scale) have zero ADM (or Bondi) energy. We have seen above that in the particular case of LQG a whole variety of pre-geometric structures, that are well characterized in the strong coupling regime, are potential local defects that---as they do not affect the flatness of the geometry of the semiclassical state---would carry no energy in the usual sense of the concept. These Planckian defects in the fabric of space-time are hidden to the low energy observers but represent genuine degrees of freedom to which other degrees of freedom can correlate to (this plays a key role in a recently introduced perspective for addressing the information puzzle in black hole evaporation, Section \ref{iloss}, as well as in natural mechanism for the generation of a cosmological constant, Section \ref{pheno}).  In this respect one might draw an analogy with frustrated systems in condense matter that do not satisfy the third law as they carry a non trivial residual entropy \cite{Nachtergaele1991, fannes} for states describing the continuum limit in LQG. 

\section{Black Hole entropy in LQG}\label{BHE}

We are now ready to present one of the central results about black holes in LQG: the computation of black hole entropy using statistical mechanical  methods applied to the microscopic states of the horizon as predicted by quantum geometry.  
As just discussed, a given macroscopic smooth geometry is expected to arise from the coarse graining of the underlying discrete states of LQG. This point of view is the one taken in the computation of BH entropy: for a given (spherically symmetric) isolated horizon one counts the degeneracy of fundamental states satisfying the coarse graining condition that the macroscopic area is $a_H$. 

\subsection{Direct counting}\label{dirc}

The discreteness of the area operator (\ref{area1}) predicted by LQG and the previous discussion on the nature of the continuum limit in LQG suggests an obvious definition of the statistical mechanical entropy of a BH in the theory. The microcanonical entropy
\be
S\equiv \log(\sN)
\ee
where $\sN$ is the number of geometry states of the BH horizon such that 
\begin{equation}
 {a-\epsilon} \leq{8\pi\gamma\ell_p^2} \sum_{i=1}^N\sqrt{j_i(j_i+1)}\leq {a+\epsilon},\label{cond3}
\end{equation}
where $a$ is the macroscopic area of the BH and $\epsilon$ a macroscopic coarse graining scale that does not enter the leading order contribution to the entropy as in standard statistical systems.

Initial estimates of the entropy where based on bounding the number of states \cite{Rovelli:1996dv, Krasnov:1996wc, Ashtekar:1997yu} which immediately led to indications that the entropy would be proportional to the area. The first rigorous countings (valid in the large BH limit) where proposed in \cite{Domagala:2004jt, Meissner:2004ju}. An alternative and very simple approach was given in 
\cite{Ghosh:2006ph}. These counting led to an entropy 
\be\label{liner}
S=\frac{\gamma_0}{\gamma} \frac{a}{8\pi \ell_p^2}
\ee
where $\gamma_0=0,274\cdots$\footnote{The gauge fixed $U(1)$ Chern-Simons formulation allowed for two different counting prescriptions which led to different values of $\gamma_0$; the other possibility gave $\gamma_0=0,237\cdots$. This ambiguity disappears in the $SU(2)$ (un gauge fixed) formulation. See Section \ref{CSF} for references. If for some physical mechanism \cite{Dreyer:2004jy} forces punctures to be labelled by the adjoint representation $j=1$ only, then $\gamma_0=\log(3)/(2\pi \sqrt{2})$. Such value of $\gamma_0$ has been argued to be singled out from a heuristic version of Bohr correspondence and quasinormal modes of spherical BHs \cite{Dreyer:2002vy}. This idea (which does not seem to generalized to the rotating BH in any simple way) has not become more solid since its appearance, and is mostly regarded as a strange coincidence.} is a numerical factor determined numerically by the solution of a transcendental equation
that we will discuss below (equation \eqref{condition}). This result was interpreted as a constrain on the value of the Immirzi parameter from semiclassical consistency. More precisely, consistency with the Bekenstein-Hawking value of the entropy of a black hole (which means consistency with classical general relativity coupled semiclassically with quantum fields) requires 
\be\label{gnot}
\gamma=\gamma_0.
\ee
Sophisticated and very powerful mathematical techniques based on number theory were developed \cite{Sahlmann:2007jt,Sahlmann:2007zp, Agullo:2008yv,Agullo:2010zz} for understanding in great detail the combinatorial counting problem at hand. Indeed an exact formula for the entropy can be constructed from this approach (other methods exploiting the connection with conformal field theories were studied \cite{Agullo:2009zt, Engle:2011vf}). These developments were motivated on the one hand by the interest in the computation of logarithmic corrections to the entropy which could allow comparison and contrast with other approaches. On the other hand,  numerical investigations of the state counting of small (Planck size) black holes revealed a surprising regular step structure for the entropy  \cite{Corichi:2006bs,Corichi:2006wn}, i.e. a fine structure of steps around the linear behaviour (\ref{liner}). This was initially thought it could survive the semiclassical limit (large BH limit) and so lead to strong deviations from the standard expectations for Hawking radiation \cite{Bek1,Bek2} (a strong signature of quantum gravity or an inconsistency with the semiclassical picture). The underlying mathematical reasons for this effect was understood thanks to the newly developed number theoretical techniques \cite{BV1}.  It became also clear that such fine structure had no physical relevance in the semiclassical regime as it was just a peculiarity of the area density of states for small black holes that does not survive in the thermodynamic limit \cite{G.:2011zr}.

The problem of computing the entropy of the black hole was greatly simplified by the appeal to a suitable Hamiltonian of the system. Such Hamiltonian is singled out by the notion of time flow defining the stationarity that is inherent  to the notion of equilibrium.  In what follows we explain how this idea can indeed be implemented. This complements the isolated horizon quasilocal definition of the phase space of idealized black holes (Section \ref{qg}) with the additional structure implying that they are in equilibrium with their immediate vicinity.

%\begin{enumerate}\item Conceptual discussion.  Black holes radiate due to quantum effects; one has to revise the notions discussed in \ref{two}.
%
%\item {\em The isolated horizon phase space.} Large semiclassical BHs are modeled by the notion of isolated horizons \cite{AKLR}.  
% The effective description of the surface degrees of freedom in terms of Chern-Simons theory is introduced  \cite{Ashtekar:1997yu, Engle:2009vc, Engle:2010kt}.
%\item {\em Quantum isolated horizons.}  The canonical quantization of the boundary degrees of freedom will be described as well as the imposition of the gauge symmetry constraints 
%that link the latter to the bulk d.o.f. \cite{ABK, Engle:2010kt}.
%\item {\em Holography in LQG.} Distingtion of bulk from horizon entropy:  BH entropy as horizon entropy in LQG \cite{Perez:2014ura}.  
%\item {\em The computation of Black Hole entropy.} I will review the number theoretical approach of Barbero et. al. leading to the micro canonical state counting, and I will show how these results coincide with those of the 
%canonical approach available in the quasi local framework introduced in \ref{two} (\cite{G.:2015sda} and references therein). 
%
%\item {\em Self-dual variables and Holography.} Analytic continuation of the horizon density of states taken as a function of the Immirzi parameter $\gamma$ exibits  holographic behavior when $\gamma\to \pm i$. i.e. for Ashtekar's complex self-dual formulation \cite{Frodden:2012nu, Frodden:2012dq, Ghosh:2014rra, Carlip:2014bfa}.  
%
%\end{enumerate}

\subsection{The effective area Hamiltonian from local observers}\label{nhg}

The indeterminacy, mentioned in Section \ref{lih}, of the quantities appearing in the first law for isolated horizon was due to the impossibility of finding a preferred normalization of the null generators at the horizon. Such freedom precludes the definition of a unique time evolution at the horizon and hence of a preferred notion of energy entering the first law. In the case of stationary black holes  a preferred time evolution (and energy notion) is provided by the existence of a global time translational symmetry represented by a timelike killing field whose normalization is fixed at infinity by demanding that it coincides inertial observers at rest. This global structure is absent in the isolated horizon formulation and seems at first to compromise the possibility of introducing a physical notion of energy and a standard statistical mechanical account of the BH thermal behaviour.

However, we have argued in Section \ref{intro-holo} that only the immediate vicinity of the BH horizon plays an important role in the thermodynamical description of the black hole system. It should be possible to eliminate the above indeterminacy in the first law by assuming that the near horizon geometry only is in equilibrium  without assuming stationarity of  the entire spacetime outside the black hole. 
Indeed, if the local geometry is stationary one can shift the emphasis from observers at infinity to  a suitable family of stationary nearby local observers (the local characterization of stationarity in the framework of isolated horizons see \cite{Lewandowski:2014nta}). As discussed below, these local observers provide the necessary additional structure to recover a preferred energy notion, and a thermodynamical first law. Remarkably, the new (quasi-local) first law and the associated notion of energy are extremely simple and well adapted to the structure of LQG.  

The key assumption is that the near horizon geometry is isometric to that of a  Kerr-Newman BHs\footnote{Such assumption is physically reasonable due to the implications of the no-hair theorem. At the quantum level one is demanding the semiclassical state of the BH to be picked around this solution.}.
A family of stationary observers $\mfs O$ located right outside the horizon at a small
proper distance $\ell\ll \sqrt{A}$ is defined by those following the integral
curves of the Killing vector field
\begin{eqnarray}
\chi=\xi+\Omega\,\psi=\partial_t+\Omega\,\partial_\phi,
\end{eqnarray}
where $\xi$  and $\psi$ are the Killing fields associated with the stationarity and axisymmetry of Kerr-Newman spacetime respectively, while $\Omega$ is the horizon angular velocity (the Killing field $\chi$ is timelike outside and close to the horizon). 
%\be \Omega=\frac{a}{r^2_++a^2}.\ee
%
The four-velocity of $\mfs O$ is given by
\begin{align}\label{keyy}
&u^a=\frac{\chi^a}{\|\chi\|}.
%&a^b=u^a\nabla_au^b=\frac{1}{\|\chi\|^2}\chi^a\nabla_a\chi_b+o(1).
\end{align}
It follows from this that $\mfs O$ are uniformly accelerated with an acceleration $a=\ell^{-1}+o(\ell)$ in the normal direction. 
These observers are the unique stationary ones that coincide with the {\em locally non-rotating observers} \cite{Wald:1984rg} or ZAMOs \cite{Thorne:1986iy} as $\ell\to 0$. As a result, their angular momentum is not exactly zero, but $o(\ell)$. 
Thus $\mfs O$ are at rest with respect to the horizon which makes them the preferred observers for studying thermodynamical issues from 
a local perspective.

%%
%\begin{figure}[h]
% \centerline{\hspace{0.5cm} \(
%\begin{array}{c}
%\psfrag{a}{$\sH$}
%\psfrag{b}{$\sO$}
%\includegraphics[height=4cm]{familynot.pdf}
%\end{array}
%\)}
%\caption{A family of stationary observers close to the horizons at proper distance $\ell$ is equivalent to a family of observers at fixed acceleration $a=1/\ell$ (see appendix). In this cartoon this acceleration is provided and tuned by the power of rockets.} \label{fam}
%\end{figure}

It is possible to show that the usual first law (\ref{1st}) translates into a much simpler relation among quasilocal physical quantities associated with $\mfs O$ \cite{Frodden:2011zz, Frodden:2011eb}.  As long as the spacetime geometry is well approximated by the Kerr Newman BH geometry in the local outer region between the BH horizon and the world-sheet of local observers at proper distance $\ell$, and, in the leading order approximation for $\ell/\sqrt{a}\ll 1$, the following  local 
first law holds
\begin{eqnarray}\label{tutiri}
\delta E =\frac{\overline\kappa}{8\pi}\delta a,
\end{eqnarray}
where 
$\delta E=\int_W T_{\mu\nu} u^\mu dW^\nu=\|\chi\|^{-1} \int_W T_{\mu\nu} \chi^\mu dW^\nu$ represents the flow of energy across the world-sheet $W$ defined by the local observers, and 
$\overline \kappa\equiv\kappa/(\|\chi\|)$.  This is the standard physical energy measured by the local observers; the amount of heat that a calorimeter would register if the falling matter is captured instead of let go into the black hole.  The above result follows from the conservation law $\nabla^a (T_{ab} \chi^b)=0$ that allows one to write
$\delta E$ as the flux of $T_{ab} \chi^b$ across the horizon. This, in turn, can be related to changes in its area  using Einsteins equations and perturbation theory (more precisely the optical 
Raychaudhuri equations) \cite{Frodden:2011eb}.

Two important remarks are in order:
 First, there is no need to normalize the Killing generator $\chi$ in any particular way. The calculation leading to (\ref{tutiri}) is invariant under the rescaling $\chi\to \alpha \chi$ for $\alpha$ a non vanishing constant. This means that the argument is truly local  and should be valid for more general black holes with a Killing horizon that are not necessarily asymptotically flat. This rescaling invariance of the Killing generator corresponds precisely to the similar arbitrariness of the generators of IHs as described in Section \ref{lih}. 
The fact that equation (\ref{tutiri}) does not depend on this ambiguity implies that  the local first law makes sense in the context of the IH phase space as long as one applies it to those solutions that are isometric to stationary black hole solutions in the thin layer of width $\ell$ outside the horizon. The semiclassical input is fully compatible with the notion of IHs. 

Second, the local surface gravity $\bar\kappa$ is universal $\bar\kappa=\ell^{-1}$ in its leading order behaviour for $\ell/\sqrt{a}\ll 1$. This is not surprising and simply reflects the fact that in the limit $\sqrt{a}\to \infty$, with $\ell$ held fixed, the near horizon geometry in the thin layer outside the horizon becomes isometric to the corresponding thin slab of Minkowski spacetime outside a Rindler horizon: the quantity $\bar\kappa$ is the acceleration of the stationary observers in this regime. Therefore, the local surface gravity loses all memory of the macroscopic parameters that define the stationary black hole. This implies that, up to a constant which one sets to zero, equation (\ref{tutiri}) can be integrated, thus providing an effective notion of horizon energy 
\be E=\frac{a}{8\pi G_{N}\ell},\label{ene}\ee
where $G_N$ is Newton's constant. 
Such energy notion is precisely the one 
to be used in statistical mechanical considerations by local observers.   
Similar energy formulae have been obtained in the Hamiltonian formulation of general relativity
with boundary conditions imposing the presence of a stationary bifurcate horizon \cite{Carlip:1993sa}.
The area as the macroscopic variable defining the ensemble has been
previously \cite{Krasnov:1997yt} used  in the context of BH models in loop quantum gravity. The new aspect revealed by the previous equation is its physical interpretation as energy for the local observers.

In application of this quasilocal notion of energy to quantum gravity one assumes that the quantum state of the bulk geometry in the local neigborhood of width $\ell$ outside the (isolated) horizon is well picked around a classical solution whose near horizon geometry is that of a stationary black hole.  The thermodynamical properties of quantum IHs satisfying such near horizon condition can be described using standard statistical mechanical
methods with the effective Hamiltonian that follows from equation (\ref{ene}) and the LQG area spectrum \eqref{area1}, namely
\be\label{ham}
\widehat H|j_1,j_2\cdots\rangle=\left(\gamma \frac{\ell^2_{Pl}}{2 G_N\ell}  \sum_{p} \sqrt{j_p (j_p+1)}\right)\  |j_1,j_2\cdots\rangle
\ee
where $j_p$ are positive half-integer spins of the $p$-th puncture and $\ell_p=\sqrt {G\hbar}$ is the fundamental Planck length associated with the value  gravitational coupling $G$ in the deep Planckian regime.
The analysis that follows can be performed in both the microcanonical ensemble or in the canonical ensemble; ensemble equivalence is granted
in this case because the system is simply given by a set of non interacting units with discrete energy levels. 

\subsection{Pure quantum geometry calculation}\label{newly}

In this section we compute black hole entropy first in the microcanonical ensemble
following a simplified (physicist) version \cite{Ghosh:2008jc, Ghosh:2006ph}
As  the canonical ensemble
becomes available with the notion of Hamiltonian (\ref{ham}), we will also derive the results in the canonical ensemble framework. 
The treatment in terms of the grand canonical ensemble as well as the equivalence of the three ensembles has been shown \cite{Ghosh:2011fc}.

Denote by $s_j$ the number of punctures of the horizon labelled by the spin $j$ (see Figure \ref{BH-quantum}).
The number of states
associated with a distribution of distinguishable punctures $\{s_j\}_{j=\frac{1}{2}}^\infty$ is
\be
n(\{s_j\})=\prod\limits_{j=\frac{1}{2}}^{\infty}\frac{N!}{s_j!}\,(2j+1)^{s_j},
\ee 
where $N\equiv\sum_j s_j$ is the total number of punctures. The leading term of the microcanonical entropy can be associated with
$S=\log(n(\{\bar s_j\}))$, where $\bar s_j$ are the solutions of the variational condition 
\be\label{vary}
\delta \log(n(\{\bar s_j\}))+2\pi \gamma_0\delta C(\{\bar s_j\})=0
\ee
where $\gamma_0$ (the $2\pi$ factor is introduced for later convenience) is a Lagrange multipliers for the constraint
\ba\label{c1c2}
C(\{ s_j\})&=& \sum_j \sqrt{j(j+1)} s_j-\frac{a}{8\pi\gamma\ell_p^2}=0.
\ea
In words, $\bar s_j$ is the configuration maximazing $\log(n(\{s_j\}))$ for fixed macroscopic area $a$.  
A simple calculation shows that the solution to the variational problem (\ref{vary}) is
\be\label{leading}
\frac{\bar s_j}{N}=(2j+1)\exp(- 2\pi\gamma_0 \sqrt{j(j+1)} ),  
\ee
from which it follows, by summing over $j$, that 
\be\label{condition}
1=\sum_j (2j+1)\exp(- 2\pi \gamma_0 \sqrt{j(j+1)}).
\ee
 Numerical evaluation of the previous 
condition yields $\gamma_0=0.274\cdots$.
It also follows from (\ref{leading}), and the evaluation of $S=\log(n(\{\bar s_j\}))$,  that
\be\label{92}
S=\frac{\gamma_0}{\gamma} \frac{a}{4\ell_p^2}
\ee
as anticipated in \eqref{liner}. 
In what sense the previous  result constrains the value of the Immirzi parameter  $\gamma$?
One can calculate the temperature of the system using  the thermodynamical relation 
\be\label{ttt}
\frac{1}{T}=\left.\frac{\partial S}{\partial E}\right|_N=\frac{G_N \gamma_0}{G_{} \gamma}\ {2\pi \ell},
\ee 
where $E$  is the energy (\ref{ene}) measured by quasilocal observers. Semiclassical consistency of quantum field theory  with gravity in the near horizon geometry requires the inverse temperature to be given by Unruh's value $T^{-1}=2\pi \ell$.  
This leads to the following restriction involving $\gamma_0$ the Immirzi parameter $\gamma$, $G$, and $G_N$, namely
\be\label{para} \gamma_0=\gamma \frac{G}{G_N},\ee from which it follows (when replacing it back in \eqref{92}) that
\be\label{entra}
S=\frac{a}{4\hbar G_N}.
\ee
Due to quantum effects Newton's 
constant is expected to flow from the IR regime to the deep Planckian one. On the one hand, the UV value of the gravitational coupling is defined in terms of the fundamental 
quantum of area predicted by LQG.  On the other hand, the low energy value $G_N$ appears in the Bekentein-Hawking entropy formula \cite{Jacobson:2007uj}. The semiclassical input that enters the derivation of the entropy 
through the assumption of (\ref{ene}) is the ingredient that bridges the two regimes in the present case.

\subsubsection{Freeing the value of $\gamma$ by introducing a chemical potential for punctures}

An interesting possibility is to allow for punctures to have a non trivial chemical potential.
Due to the existence of an area gap (depending on the Immirzi parameter) adding or removing a puncture from the state of the horizon is analogous to exchanging a particle carrying non zero energy (\ref{ene}) with the system. At the present stage of development of the theory (with incomplete understanding of the continuum limit) one can investigate the possibility that the number of punctures $N$ of the black hole state represent an additional conserved quantity (a genre of {\em quantum hair}) for semiclassical black holes. The consequence of such generalization is that the Immirzi parameter need non longer be fixed to a particular value: in addition to fixing the value of the area gap, the Immirzi parameter controls the value of the puncture's chemical potential \cite{Ghosh:2011fc}.

The derivation follows closely the previous one. The difference is that one needs to add a new constraint 
\be
C^{\prime}( s_j\})=\sum_j s_j-N=0,
\ee
with an additional Lagrange multiplier that we call $\sigma$.
The new extremum condition now only fixes a relationship between $\sigma$ and $\gamma_0$ which takes the form
\be
\sigma(\gamma)= \log[\sum_j (2j+1) e^{-2 \pi \gamma\frac{G}{G_N} \nn }]
\ee
once \eqref{para} is used.
The entropy becomes \be \label{entrobien}
S= \frac{a}{4\ell_p^2}+\sigma(\gamma) N.
\ee
The first term in the entropy formula is the expected Bekenstein-Hawking entropy while the second 
is a new contribution to the entropy which depends on the value of the Immirzi parameter $\gamma$.
This new contribution comes from the punctures non-trivial chemical potential which is given by
\be\label{cp}
\bar\mu=-T \left.\frac{\partial S}{\partial N}\right|_E=-\frac{\ell_p^2}{2\pi\ell}\,\sigma(\gamma)
\ee
where one is again evaluating the equation at the Unruh
temperature $T=\hbar/(2\pi \ell)$.

The above derivation can be done in the framework of the canonical and grandcanonical ensembles.
From the technical perspective it would have been simpler to do it using one of those ensembles. In particular
basic formulae allow for the calculation of the energy fluctuations 
which at the Unruh temperature are such that  
$
{(\Delta E)^2}/{\langle E\rangle^2}=\sO(1/N).
$
The specific heat at $T_{\va U}$ is
$C=N\gamma_0^2d^2\sigma/d\gamma^2$
which is  positive. This implies that, as a thermodynamic system, the IH is locally
stable. The specific heat tends to zero in the large $\gamma$ limit for fixed
$N$ and diverges as $\hbar\to 0$. The three ensembles give equivalent results \cite{Ghosh:2011fc}.

The entropy result \eqref{entrobien} might seem at first sight in conflict with, what we could call, the {\em geometric} first law \eqref{1st} (geometric because it is implied directly by Einsteins equations). 
However, when translating things back to observers at infinity, the present statistical mechanical treatment implies the following thermodynamical first law
 \be\label{globalla}
 \delta M=\frac{\kappa\hbar }{2 \pi} \delta S+\Omega \delta J+\Phi \delta Q+ \mu \delta N,
 \ee
 where $\mu=-\ell_p^2 \kappa \sigma(\gamma)/(2\pi)$ (the redshifted version of $\bar\mu$). It is now immediate to check that the exotic chemical potential term in (\ref{globalla}) cancels the term proportional to the number of punctures in the entropy formula (\ref{entrobien}). Therefore, the above balance equation is just exactly the same as (\ref{1st}). 
As in the seminal argument by Jacobson  \cite{Jacobson:1995ab}  the validity of  semiclassical consistency discussed here for general  accelerated observers in arbitrary local neighbourhoods implies the validity of Einsteins equations \cite{Smolin:2012ys}.

%\begin{table}[ht!!!!!!!!]
%\tbl{Different versions of balance equations. On the left column one has the results coming 
%from quantum geometry involving a chemical potential term. The semiclassical input of the area effective Hamiltonian in the quantum geometry statistical
% mechanics calculation leads to results that are consistent with the geometry first laws shown on the right column.}
%{\begin{tabular}{@{}cccc@{}} 
%%\toprule
%%Piston mass$^{\text a}$ & Analytical frequency & TRIA6-$S_1$ model & \% Error \\
%%& (Rad/s) & (Rad/s) \\
%\colrule
% & &  &  \\
%& Quantum Statistical  Mechanics & & Classical Einstein gravity   \\
%& &  &  \\
% Local & {\normalsize $\delta E=\frac{\bar\kappa \hbar }{2 \pi} \delta S+ \bar\mu \delta N$ }& {\normalsize$\Longleftrightarrow$} & {\normalsize $\delta E=\frac{\bar\kappa}{8\pi} \delta A $} \\
%& &  &  \\
%&{\normalsize $\Updownarrow$}& &{\normalsize$\Updownarrow$}  \\
%& &  &  \\
%Global &{\normalsize$\delta M=\frac{\kappa \hbar }{2 \pi} \delta S+\Omega \delta J+\Phi \delta Q+ \mu \delta N$} & {\normalsize$\Longleftrightarrow$} & {\normalsize$\delta M=\frac{\kappa \hbar }{2 \pi} \delta A+\Omega \delta J+\Phi \delta Q$} \\ 
%& &  &  \\
%\botrule
%\end{tabular}
%}
%\begin{tabnote}
%$^{\text a}$Moving along horizontally in this table is a trivial identity; moving vertically requires the background 
%geometry to be a stationary black hole solution.
%\end{tabnote}
%\label{one}
%\end{table}

\subsection{Changing statistics to include matter }\label{holi}

In the previous sections only pure geometric excitations have been taken into account.
However, from the local observers perspective matter fields are highly exited close to the horizon.
More precisely, the quantum state of all non-geometric excitations is seen as a highly excited state at inverse temperature $\beta=2\pi \ell/\ell_p^2$.  This is a necessary condition on the UV structure of the quantum state so that it just looks like the vacuum state  for freely falling observers (at scales smaller than the size of the BH). This is related to the regularity condition of the quantum state granting that the expectation value of the energy momentum tensor is well defined \cite{Wald:1995yp}: a statement about the two-point correlation function called the Hadamard condition that is intimately related to the UV behaviour of entanglement across the horizon.  This has been used to argue \cite{tHooft:1984kcu} that  quasilocal stationary observers close to the horizon would find that the number of matter degrees of freedom contributing to the entropy grows exponentially with the horizon area according to 
\be\label{holos}
D \propto \exp(\lambda a/(\hbar G_N)),
\ee
where $\lambda$ is an unspecified dimensionless constant that cannot be determined (within the context of quantum field theory) due to two related issues:
On the one hand UV divergences of standard QFT introduce regularization ambiguities affecting the value of $\lambda$; on the other hand, the value of $\lambda$ depends on the number of species of fields considered. The degeneracy of states corresponds to the number of matter degrees of freedom that are entangled across the horizon \cite{Solodukhin:2011gn}. 

All this implies that matter degrees of freedom might play an important role in the entropy computation, as
for each and every state of the quantum geometry considered in the previous section there is a large degeneracy in the matter sector that has been neglected in the counting. Can one take this aspect into account in LQG?
At first the question seems a difficult one because of the lack of a complete understanding of the matter sector in the theory. For instance, because $\lambda$ in \eqref{holos} depends on the number of species one would seem to need a complete unified understanding of the matter sector to be able to begin answering this question.
However, further analysis shows \cite{Ghosh:2013iwa} that the discrete nature quantum geometry combined with the assumption of the regularity of the quantum state of matter fields across the horizon (embodied in the form of the degeneracy of states (\ref{holos}) for an undetermined $\lambda$) plus the additional assumption of indistinguishability of puncture exitations  is sufficient to recover Bekenstein-Hawking entropy.

In the treatments mentioned so far punctures were considered distinguishable \footnote{In the pure gravity $U(1)$ Chern-Simons formulation the necessity of distinguishability of punctures follows from a technical point in the quantization \cite{Ashtekar:2000eq}.}. Let us see here what indistinguishability would change. Instead of the microcanonical ensemble, we use now the grand canonical ensemble as this will  considerably shorten the derivations 
(keep in mind that all ensembles are equivalent). Thus we start from the canonical partition function which for a system 
of non interactive punctures is $Q(\beta,N)=q(\beta)^N/N!$ where the $N!$ in the denominator is the Gibbs factor
that effectively enforces indistinguishability, and the one-puncture partition function $q(\beta)$ is given by  
\be\label{qqq}
q(\beta)=\sum_{j=\frac{1}{2}}^{\infty} d_j \exp(-\frac{\hbar \beta \gamma_0}{\ell} \nn ),
\ee
where $d_j$ is the degeneracy of the spin $j$ state (for instance $d_j=(2j+1)$ as in 
the $SU(2)$ Chern-Simons treatment) and $\gamma_0$ is given by \eqref{para}. The grand canonical partition function is
\be\label{zzz}
\sZ(\beta,z)=\sum_{N=1}^{\infty} \frac{z^Nq(\beta)^N}{N!}=\exp(zq(\beta)).
\ee
From the equations of state $E=-\partial_{\beta} \log(\sZ)$, and $N=z\partial_z\log(\sZ) $ one gets
\ba\label{eqsta}
\frac{a}{8\pi G_N \ell}&=&-z\partial_{\beta} q(\beta)\n \\
N &=& z q(\beta)=\log(\sZ).
\ea
In thermal equilibrium at the Unruh temperature one has $\beta=2\pi\ell\hbar^{-1}$ and the $\ell$ dependence disappears from the previous equations.
However, for $d_j$ that grow at most polynomially in $j$, the BH area predicted by the equation is just Planckian and the number of punctures $N$ of order one \cite{Ghosh:2013iwa}.
Therefore, indistinguishability with degeneracies $d_j$ of the kind we find in the pure geometry models of Section \ref{newly} is ruled out because it cannot acommodate BHs that are large in Planck units. 

If instead we assume that matter degrees of freedom contribute to the degeneracy factor then regularity of the quantum state of matter near the horizon takes the form (\ref{holos}) which in the quantum geometry language translates into $D[\{s_j\}]=\prod_j d_j$ with $d_j=\exp(\lambda 8\pi \gamma_0\nn)$. For simplicity lets take $\nn\approx j+1/2$ \cite{FernandoBarbero:2009ai}. We also introduce 
two dimensionless variables $\delta_{\beta}$ and $\delta_{h}$ and write
$\beta=\beta_{\va U}(1+\delta_\beta)$---where $\beta_{U}=2\pi \ell/\hbar$---and $\lambda=(1-\delta_h)/4$.
A direct calculation of the geometric series that follows from (\ref{qqq}) yields
\be
q(\beta)
%=e^{-\pi\gamma_0\delta}\sum_{j} e^{-2\pi\gamma_0\delta j}
=\frac{\exp(-\pi\gamma_0\delta(\beta))}{\exp(\pi\gamma_0\delta(\beta))-1},
\ee
where $\delta(\beta)=\delta_h+\delta_{\beta}$. The equations of state \eqref{eqsta} now predict large semiclassical 
BHs as follows: for large $a/(\hbar G_N)$ and by setting $\beta=2\pi \ell\hbar^{-1}$ in (\ref{eqsta}) one can determine $\delta_h\equiv \delta(2\pi\ell\hbar^{-1})$ as a function of $a$ and $z$. The result is
$\delta_h=2\sqrt{{G_N \hbar z}/({\pi\gamma_0 a})}\ll 1$. 
%For semiclassical BHs $\delta_{\beta}\ll 1$  since the temperature measured by quasilocal observers must be close to the Unruh temperature. This, together with the previous equation for $\delta$, implies $\delta_h\ll1$.
In other words semiclassical consistency implies that the additional degrees of freedom producing the degeneracy (\ref{holos}) must saturate the {\em holographic bound}  \cite{Ghosh:2013iwa}, i.e. we get 
\be\label{llalla} \lambda=\frac{1}{4}\ee up to quantum corrections. The entropy is given by the formula $S=\beta E-\log(z) N+\log(\sZ)$ which upon evaluation yields
\be
S=\frac{a}{4G_N\hbar} - \frac{1}{2}(\log(z)-1) \left({\frac{z a}{\pi \gamma \ell_p^2}}\right)^{\frac{1}{2}}
\ee
This gives the Bekenstein-Hawking entropy to leading order plus 
quantum corrections. If one sets the chemical potential of the punctures to zero (as for photons or gravitons)
then these corrections remain. One can get rid of the corrections by setting the chemical potential 
$\mu=T_U$. Such possibility is intriguing, yet the physical meaning of such a choice is not clear at this stage. 
The thermal state of the system is dominated by large spins as the mean spin $\langle j\rangle = a/(N\ell_p^2)$ grows like $\sqrt{a/\ell_p^2}$.
The conclusions of this subsection hold for arbitrary puncture statistics. This is to be expected because the system behaves as if it were at a very high effective temperature  (the Unruh temperature is the precise analog of 
    the Hagedorn temperature \cite{Hagedorn:1965st} of particle physics). A similar result can be obtained by using Bosonic or Fermionic statistics for the punctures \cite{Ghosh:2013iwa}. The leading term remains the same, only corrections change.
    In the case of Bosons the square root correction can be understood as coming from the Hardy-Ramanujan formula giving the asymptotic form of the number of partitions of an integer $a$ in LQG Planck units \cite{Asin:2014gta}.

\subsection{Bosonic statistics and the correspondence with the continuum limit} \label{contlim}

The partition function for Bosonic statistics and for $z=1$ is specially interesting because it produces an expression of the partition function that coincides with the formal continuum path integral partition function \cite{Ghosh:2013iwa}. Explicitly, from \eqref{qqq} and \eqref{zzz} it follows that
\be\label{z=1}
\sZ(\beta)=\prod_{j=\frac{1}{2}}^\infty \sum_{s_j} \exp(2\pi\ell-\beta)\frac{a_j}{8\pi\ell G_N},
\ee
where $a_j=8\pi\gamma\ell_p^2\nn$ are the area eigenvalues, and we have assumed for simplicity $\lambda=1/4$ in (\ref{holos}), namely $d_j=\exp(a_j/(4G_N\hbar))$. There is a well known relationship between the statistical mechanical partition function and the Euclidean path integral on a flat background.
One has that
\be
Z_{sc}(\beta)=\int D\phi \exp\{-S[\phi]\}
\ee
where field configurations are taken to be periodic in Euclidean time with period $\beta$.
Such expression can be formally extended  to the gravitational context at least in the treatment of stationary black holes.
One starts from the formal analog of the previous expression and immediately uses the stationary phase approximation to make sense of it on the background of a stationary black hole.
%\footnote{Even when the partition function can be computed via the Euclidean path integral for free theories on flat space times, and can be made sense of for interactive theories on flat space times via perturbation theory and renormalisation, the inclusion of gravitational degrees of freedom leads to serious mathematical difficulties. One reason is that the Euclidean action is not bounded from below\cite{Gibbons:1978ac}, another reason is that the Wick rotation from Lorentzian to Euclidean makes only sense for static backgrounds where real sections of complexified space-time exist.}. 
Namely 
\ba\label{EPI}
Z_{sc}(\beta)&=&\int Dg D\phi \exp\{-S[g, \phi]\}\n \\
&\approx& \exp\{-S[g_{cl}, 0]\} \int D\eta \exp\left[-\int dx dy \eta(x) \left(\frac{\delta^2 \sL}{\delta \eta(x)\delta \eta(y)}\right) \eta(y)\right]
\ea
where the first term depends entirely on the classical BH solution $g_{cl}$ 
while the second term represents the path integral over fluctuation fields, both of the metric as well as the matter, that we here schematically denote by $\eta$.  
For local field theories $\delta_{\eta(x)}\delta_{\eta(y)}{\sL}=\delta(x,y) \square_{gc}$ where $\square_{gc}$ is a the Laplace like operator 
(possible gauge symmetries, in particular diffeomorphisms must be gauge fixed to make sense of such formula). 

In the analytic continuation one sends the Killing parameter $t\to -i\tau$ and the space time tube-like region outside the horizon up to the local stationary observers at distance $\ell$ inherits a positive definite Euclidean metric (for rotating black holes this is true only to first order in $\ell$ \cite{Frodden:thesis}). The $S^2\times \R$ representing the black hole horizon shrinks down to a single $S^2$ in the euclidean and the time translation orbits become compact rotations around the Euclidean horizon with $0\le \tau\le \beta$.  The tube-like region becomes $D\times S^2$ where $D$ is a disk in the plane orthogonal to and centred at the Euclidean horizon with proper radius $0\le R\le \ell$ (in the Euclidean case the BH horizon shrinks to a point, represented here by the center of $D$). Recall the Einstein-Hilbert action
\be
S[g_{cl}, 0]=\frac{1}{8\pi G_{N}} \ \int\limits_{D\times S^2} \!\!\!\!\!\sqrt{g} \, R +{\rm boundary\ \ terms}
\ee
On shell the bulk term in the previous integral would seem at first to vanish. However, when $\beta\not=2\pi \ell$, the geometry has a conical singularity at the centre of the disk and $R$ contributes with a Dirac delta distribution multiplied by the factor $(2\pi \ell-\beta)$. Using the Gibbons-Hawking prescription  boundary terms \cite{Gibbons:1976ue,York:1986it} one can see that they cancel to leading order in $\ell$.
A direct calculation gives the semi-classical free energy
\be\label{holographic}
-S[g_{cl}, 0]=\log(Z_{cl})=(2\pi \ell-\beta) \frac{a}{8\pi G_{N} \ell}.
\ee
Replacing \eqref{holographic} in \eqref{EPI} and comparing with the form the partition function (\ref{z=1}) we conclude 
that  the inclusion of the holographic degeneracy (\ref{holos}) plus the assumption of 
Bosonic statistics for punctures makes the results of section \ref{holi} compatible with the continuous 
formal treatment of the Euclidean path integral. Equation (\ref{z=1}) is thus  compatible with the continuum limit.

%{\bf The replica trick vs. the euclidean path integral! There is a link between the geometric input in the replica trick and gravity (conical singularity). This needs to be clarified.}

\subsection{Logarithmic corrections}

The equation of state $E=-\partial_{\beta} \log(Z_{cl})$ reproduces the quasilocal energy (\ref{ene}).
The entropy is $S=\beta E+\log(Z)=A/(4\ell_p^2)$ when evaluated at the inverse Unruh temperature $\beta_U=2\pi\ell$.
Notice that in the quasi-local framework used here, entropy grows linearly with energy (instead of quadratically as in the usual Hawking treatment). This means that the usual ill behaviour of the canonical ensemble of the standard 
global formulation \cite{Hawking:1976de} is cured by the quasilocal treatment. Quantum corrections to the entropy come from the fluctuation factor 
which can formally be expressed  in terms of the determinant of a second order local (elliptic) differential operator $\square_{g_{cl}}$ 
\ba
&& F%\n \int D\eta \exp\left[-\int dx dy \eta(x) \left(\frac{\partial^2 \sL}{\partial \eta(x)\partial \eta(y)}\right) \eta(y)\right]=\\
=\int D\eta \exp\left[-\int dx \eta(x) \square_{g_{cl}} \eta(x)\right]=\left[{\det(\square_{g_{cl}})}\right]^{-\frac{1}{2}}.
\ea
The determinant can be computed from the identity (the heat kernel expansion)
\be
\log\left[{\det(\square_{g_{cl}})}\right]
%={\rm Tr}[\log\left[{\det(\square_{g_{cl}})}\right]]
=\int_{{\epsilon}^2}^{\infty} \frac{ds}{s} {\rm Tr}\left[\exp(-s \,\square_{g_{cl}})\right],
\ee
where $\epsilon$ is a UV cut-off needed to regularize the integral. We will assume here that it is proportional to $\ell_p$.
In the last equality we have used the heat kernel expansion in $d$ dimensions \be{\rm Tr}\left[\exp(-s\, \square_{g_{cl}})\right]=(4\pi s)^{-\frac{d}{2}} \sum_{n=0}^{\infty} a_n s^{\frac{n}{2}}, \ee where the coefficients $a_n$ are given by integrals in $D\times S^2$ of local quantities.  

%\subsubsection{Logarithmic corrections}

At first sight the terms $a_n$ with $n\le 2$ produce potential important corrections to BH entropy. All of these suffer from regularisation ambiguities with the exception of the term $a_2$ which leads to logarithmic corrections.
Moreover, contributions coming from $a_0$ and $a_1$ can be shown to contribute to the renormalization of various couplings in the underlying Lagrangian \cite{Sen:2012dw}; for instance $a_0$ contributes to the cosmological constant renormalization. 
True loop corrections are then encoded in the logarithmic term $a_2$ and for that reason it has received great attention in the literature (see \cite{Sen:2012dw} and references therein). Another reason is that its form is regularisation independent.   
According to \cite{G.:2011zr} there are no logarithmic corrections in the $SU(2)$ pure geometric 
model once the appropriate smoothing is used (canonical ensemble). From this we conclude that the only possible source of logarithmic corrections in the $SU(2)$ case must come from the non-geometric degrees of freedom that produce the so called matter degeneracy that plays a central role in Section \ref{holi}. A possible way to compute these corrections is to compute the heat kernel coefficient $a_2$ for a given matter model. This is the approach taken in reference \cite{Sen:2012dw}. 
One can argue \cite{Ghosh:2013iwa} that logarithmic corrections in the one-loop effective action are directly  reflected as logarithmic corrections in the LQG  BH entropy. 

\subsection{Holographic degeneracy from LQG}\label{stringy}

The key assumption that led to the results of  Section \ref{holi}  was that matter degeneracy satisfies \eqref{holos} which was motivated by the regularity Hadamard condition on the vacuum state in the vicinity of the horizon.  
Can one actually predict such degeneracy directly from the fundamental nature of quantum geometry?
Even when a complete model of matter at the Planck scale would seem necessary to answer this question there are indications that the fundamental structure of LQG might indeed allow for such degeneracy when coupling with matter fields.
Notice that according to \ref{contlim} this question might be directly related to the question of the continuum limit \ref{thecontlim}. 

The exact holographic behaviour of the degeneracy of the area spectrum has been obtained from the analytic continuation of the  dimension of the boundary Chern-Simons theory by sending the spins $j_i\to i s-1/2$  with $s\in \R^+$\cite{Frodden:2012dq, Han:2014xna, Achour:2014eqa, Geiller:2013pya}.  The new continuous labels correspond to $SU(1,1)$ unitary representations that solve the $SL(2,\C)$ self(antiself)-duality constraints 
$L^i\pm K^i=0$ (see \cite{Perez:2012wv}), which in addition comply with the necessary reality condition $E\cdot E\ge 0$ for the fields $E^a_i$ \cite{Frodden:2012dq}.
All this suggests that the holographic behaviour postulated in (\ref{holos}) with $\lambda=1/4$ would naturally follow from the definition of LQG in terms of self(antiself)-dual variables, i.e.
$\gamma=\pm i$.  The same holographic behaviour of the number of degrees of freedom available at the horizon surface
is found from a conformal field theoretical perspective for $\gamma=\pm i$ \cite{Ghosh:2014rra}. A relationship between the termal nature of BH horizons and self dual variables seems also valid according to similar analytic continuation arguments \cite{Pranzetti:2013lma}.    
The analytic continuation technique has also been applied in the context of lower dimensional BHs \cite{Frodden:2012nu}. 
However, these results are at the moment only indications on an interesting behaviour. A clear understanding of the quantum theory in terms of complex Ashtekar variables is desirable on these grounds but unfortunately still missing.

Recent investigation of the action of diffeomorphisms on boundaries \cite{Ghosh:2014rra, Freidel:2015gpa, Freidel:2016bxd} revealed the
existence of potentially new degrees of freedom associated to broken residual diffeomorphism around the defects defined by the spin network punctures (as in Figure \ref{BH-quantum}). The associated generators are shown to satisfy a Virasoro algebra with central charge $c=3$. Such CFT degrees of freedom could
naturally account for the Bekenstein-Hawking area law and provide a microscopic explanation of \eqref{llalla} . The central feature that makes this
possible in principle is the fact that the central charge of
the CFT describing boundary degrees of freedom is proportional  to  the  number  of  punctures  that  itself  grows
with  the  BH  area.   This  is  a  feature  that  resembles  in
spirit previous descriptions \cite{Carlip:1998wz,Carlip:1999cy, Carlip:2005zn, Carlip:2002be}.  However,  an important
advantage of the present treatment is the precise identification of the underlying microscopic degrees of freedom.
Preliminary results (based on the use of the Cardy formula) indicate that the correct value of BH entropy 
could be obtained without the need of tuning the Immirzi parameter to any special value.

It is worth mentioning here the approaches where a holographic degeneracy of the BH density of states arises naturally form symmetry considerations in transverse `$r$-$t$ plane'  of the near horizon geometry \cite{Carlip:1998wz, Carlip:1999cy, Carlip:2002be, Dreyer:2013noa, Dreyer:2001py, Carlip:2014bfa}. A  clear connection or synthesis between these seemingly dual ideas remains open (see \cite{Pranzetti:2013lma} for a hint of a possible link).

Finally, in the related {\em group field theory approach} (GFT) to quantum gravity \cite{Oriti:2013aqa, Oriti:2014uga} the continuum limit is approached via the notion of condensate states (Bosonic statistics plays here a central role \cite{Gielen:2013naa}). The problem of calculating black hole entropy has been explored in \cite{Oriti:2015rwa}.

\subsection{Entanglement entropy perturbations and black hole entropy}

Starting from a pure state $|0\rangle\langle 0|$ (``vacuum state") in QFT one can define a reduced 
density matrix $\rho={\rm Tr}_{in}(|0\rangle\langle 0|)$ by taking the trace over the degrees of freedom
{\em inside} the BH horizon.  
The entanglement entropy is defined as $S_{ent}[\rho]=-{\rm Tr}(\rho\log(\rho))$.
In four dimensions \cite{Solodukhin:2011gn}  the leading order term of entanglement entropy in standard QFT goes like \be\label{entro} S_{ent}= \lambda \frac{a}{\epsilon^2}+corrections\ee 
where $\epsilon$ is an UV cut-off, and  $\lambda$ is left undetermined in the standard QFT calculation
due to UV divergences and associated ambiguities (recall discussion in Section \ref{holi}).  An important one is that 
$\lambda$ is proportional to the number of fields considered; this is known as the {\em species problem}. 
These ambiguities seem to disappear if one studies perturbations of (\ref{entro}) 
when gravitational effects are taken into account \cite{Bianchi:2012br, Bianchi:2013rya}.
The analysis is done in the context of perturbations of the vacuum state in Minkowski spacetime
as seen by accelerated Rindler observers. Entanglement entropy is defined by tracing out degrees of freedom
outside the Rindler wedge. Such system reflects some of the physics of stationary black holes
in the infinite area limit.  A key property \cite{Wald:1995yp} is that, 
\be\label{eeqq}
\rho=\frac{\exp (-2\pi \int_{\Sigma} \hat T_{\mu\nu}\chi^\mu d\Sigma^\nu)}{{\rm Tr}[\exp (-2\pi \int_{\Sigma} \hat T_{\mu\nu}\chi^\mu d\Sigma^\nu)]},
\ee
%where 
%\be\label{kk} \hat K=,
%\ee
where $\Sigma$ is any Cauchy surface of the Rindler wedge.
If one considers a perturbation of the vaccum state $\delta \rho$
then the first interesting fact is that the (relative entropy) $\delta S_{ent}=S_{ent}[\rho+\delta\rho]-S_{ent}[\rho]$ is UV finite and hence free of regularization ambiguities \cite{Casini:2008cr}. 
The second  property that follows formally (see below) from (\ref{eeqq}) is that 
\ba \label{ff}\delta S_{ent}=2\pi {\rm Tr}(\int_{\Sigma}  \delta\langle T_{\mu\nu}\rangle\chi^\mu d\Sigma^\nu).\ea
Now from semiclassical Eintein's equations $\nabla^{\mu} \delta \langle T_{\mu\nu}\rangle=0$, this (together with the global properties of the Rindler wedge) implies that one can replace the Cauchy surface $\Sigma$ by the Rindler horizon $H$ in the  previous equation. As in the calculation leading to (\ref{tutiri}) one can use the Raychaudhuri equation (i.e. semiclassical Eintein's equations)  to relate the flux of $\delta \langle T_{\mu\nu}\rangle$ across the Rindler horizon to changes in its area. The result is that $\delta S_{ent}={\frac{ \delta A}{4G_N\hbar}}$ independently of the number of species.
The argument can be generalized to static black holes \cite{Perez:2014ura} where a preferred vacuum state exists (the Hartle-Hawking state). In this case the perturbation can send energy out to infinity as well, and the resulting balance equation is \be\label{dent}
\delta S_{ent}={\frac{ \delta a}{4G_N\hbar}}+\delta S_{\infty},
\ee
where $\delta S_{\infty}=\delta E/T_{H}$, and $\delta E$ is the energy flow at $\sI^+\cup i^+$.
Changes of entanglement entropy match changes of Hawking entropy plus an entropy flow to infinity.
These results shed light on the way the species problem could be resolved in quantum gravity. The key point being that $a$ is dynamical in gravity and thus grows with the number of gravitating fields. However,  as the concept of relative entropy used here is insensitive to the UV degrees of freedom,
one key question  is whether the present idea can be extrapolated to the Planck scale (for some results in this direction see \cite{Bianchi:2012vp, Chirco:2014saa, Chirco:2014naa}). Another important limitation of the previous analysis is that property \eqref{ff} is only valid in a very restrictive sense (see for example \cite{Blanco:2013joa}). Indeed, as shown in \cite{Varadarajan:2016kei} generic variations involving for instance coherent states will violate \eqref{ff}. Thus this remains a formal remark pointing in an interesting direction that deserves further attention.

\subsection{Entanglement entropy vs. statistical mechanical entropy}\label{sf}

One can argue that the perspective that BH entropy should be accounted for in terms of entanglement entropy  \cite{Bombelli:1986rw}  (for a review see \cite{Solodukhin:2011gn})  and the
statistical mechanical derivation presented sofar are indeed  equivalent in a suitable sense \cite{Perez:2014ura}.  The basic reason for such equivalence resides in the microscopic structure predicted by LQG \cite{Bianchi:2012ui, Bianchi:2012ev, Chirco:2014saa}.  
In our context, the appearance of the UV divergence in (\ref{entro}) tells us that the leading contribution to $S_{ent}$ must come
from the UV structure of LQG close to the boundary separating the two regions. 
Consider a basis  of the subspace of the horizon Hilbert space characterised by condition (\ref{cond3}), and assume the discrete index $a$ labels the elements of its basis. 
Consider the  state 
\be \label{here}
|\Psi\rangle=\sum\limits_{a} \alpha_{ a } \,  |\psi^{a}_{int}\rangle\, |\psi^{a}_{ext}\rangle, \ee 
where $|\psi^{a}_{int}\rangle$ and $|\psi^{a}_{ext}\rangle$  denote physical states compatible with the IH boundary data $a$, and describing  the interior and the exterior state 
of matter and geometry of the BH respectively. The assumption that such states exist is a basic input of Section \ref{CSQ}. 
%We concentrate on correlations of physical states describing the semiclassical near boundary geometry and matter fields in a region of with $\ell\ll\epsilon$ (where $\epsilon$ represents the scale up to which 
%standard QFT can be trusted). 
In the form of the  equation above we are assuming that correlations  between the outside and the inside at Planckian scales  are  mediated by the
spin-network links puncturing the separating boundary.
% More precisely we assume that    
%\be\label{key} \alpha_{iao}= a_{ia} b_{ao}\ee 
%for observables $a, i$ in the region $\ell\ll\epsilon$. 
This encodes the idea that vacuum correlations are ultra-local at the Planck scale. The proper understanding of the solutions of (\ref{dosl}) might reveal a richer entanglement across the horizon (the exploration of this important question is underway \cite{Bianchi:2016tmw}). 
This assumption is implicit in the recent treatments \cite{Bianchi:2012ui} based on the analysis of a single quantum of area correlation and it is related to the
(Planckian) Hadamard condition as defined in \cite{Chirco:2014saa}. 
We also assume states to be normalized as follows: $\langle \psi_{ext}^a|\psi_{ext}^a\rangle=1$, $\langle \psi_{int}^a|\psi_{int}^a\rangle=1$, and  $\langle \Psi|\Psi\rangle=1$.
The reduced density matrix obtain from the pure state by tracing over the interior observables is
%\ba
%\n \rho_{ext}&\equiv&\sum_{i,a} \langle a,i |\Psi\rangle\langle\Psi| a,i\rangle\\
%%\n &=&\sum_{i,a,i', a', o', i'',a'',o''} \alpha_{i',o'}\bar\alpha_{i'',o''} 
%%\langle \psi_{int}^{a,i}|\psi^{a',i'}_{int}\rangle\, |\psi^{a',o'}_{ext}\rangle\langle\psi^{a'',o''}_{ext}|\, \langle\psi^{a'',i''}_{int}| \psi_{int}^{a,i}\rangle\\
% &=&\sum_{a,i} |\alpha_a|^2\beta_{ia}\bar\beta_{ia}\, |\psi^{a}_{ext}\rangle\langle\psi^{a}_{ext}| 
%%\\
%%&=&\sum_{a, o',o''} A_a \, b_{ao'}\bar b_{ao''}\, |\psi^{a,o'}_{ext}\rangle\langle\psi^{a,o''}_{ext}|+corrections,
%\ea
%where $\beta_{ia}\equiv \langle a,i |\psi^a_{int}\rangle$. From this one gets 
\ba
 \rho_{ext}
&=&\sum\limits_a p_a |\psi^{a}_{ext}\rangle\langle \psi^{a}_{ext}|, 
\ea
with $p_a=|\alpha_a|^2$. It follows from this that the entropy $S_{ext}\equiv -{\rm Tr}[\rho_{ext} \log(\rho_{ext})]$ is  bounded by micro-canonical entropy of the ensemble as discussed in Section \ref{newly}. 
If instead one starts from a mixed state encoding an homogeneous statistical mixture of quantum states compatible with  (\ref{cond3}), then the reduced density matrix leads to an entropy that matches the microcanonical one \cite{Perez:2014ura}. Such equipartition of probability is a standard assumption in the statistical mechanical description of standard systems in equilibrium. It is interesting to contemplate the possibility of arguing for its validity from a more fundamental level using the ideas of typicality \cite{popescu}.

\subsection{Hawking radiation}\label{HR}

The derivation of Hawking radiation from first principles in LQG remains an open problem, this is partly due to the difficulty associated with the definition of semiclassical states approximating space-time backgrounds. Only heuristic models based on simple analogies exist at the moment  \cite{Heidmann:2016yfz}. 
Without a detailed account of the emission process it is still possible to obtain information from a spectroscopical approach (first applied to BHs in \cite{Bekenstein:1995ju, Krasnov:1997yt}) that uses as an input the details of the area spectrum in addition to some semiclassical assumptions \cite{Barrau:2011md, Barrau:2015ana}. 
The status of the question has improved with the definition and quantisation of spherical symmetric models
\cite{Gambini:2013nea, Gambini:2013hna, Gambini:2013exa}. The approach resembles the hybrid quantisation techniques used in loop quantum cosmology \cite{Agullo:2016tjh, Ashtekar:2011ni}. More precisely, the quantum spherical background space-time is defined using LQG techniques, whereas perturbations, accounting for Hawking radiation, are described by a quantum test field (defined by means of a Fock Hilbert space). 

A fundamental microscopic account of the evaporation in detail would require dynamical considerations where the solutions of (\ref{dosl}) describing a semiclassical BH state will have to enter. For an attempt to include dynamics in the present framework see \cite{Pranzetti:2012pd}; and \cite{Pranzetti:2012dd} for a related model of evaporation.

\section{Insights into the hard problem: Black Hole quantum dynamics}\label{loss}

\begin{figure}[t]
%\begin{center}
\centerline{\hspace{0.5cm} \(
\begin{array}{c}
\includegraphics[height=12cm]{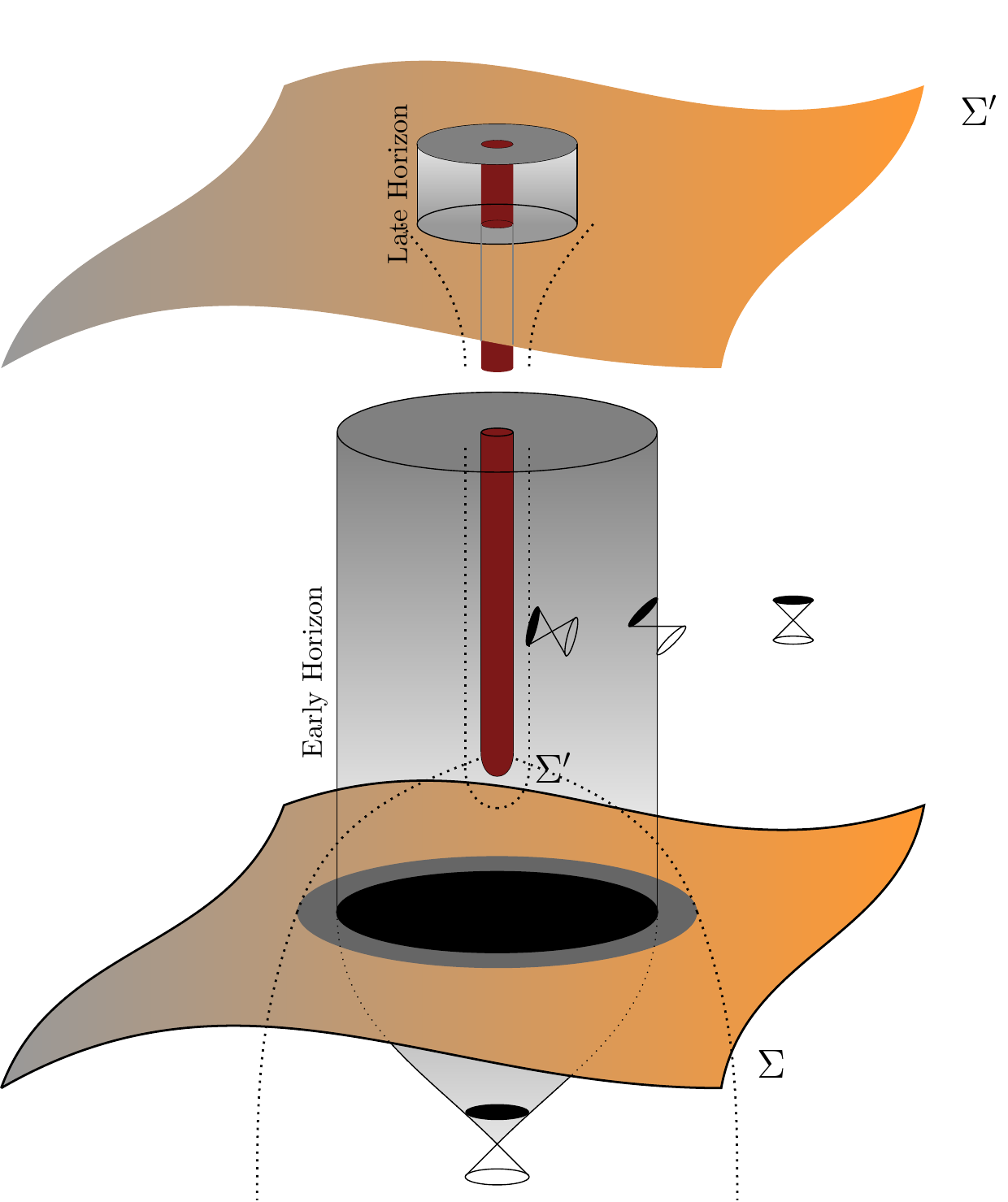} 
\end{array}\) } \caption{A Cauchy surface $\Sigma^\prime$ sufficiently to the future of the gravitational collapse: A key feature of the BH interior is that the singularity is a space like surface, e.g. in the spherically symmetric case this means that the $r=$constant surfaces are spacelike. The light cones are tilted towards the singularity inside the BH. This means that a Cauchy surface $\Sigma^{\prime}$ representing an  {\em instant} of the BH spacetime sufficiently to the future contains a large volumetric extension inside the BH region. This volume would be unbounded if BHs did not evaporate (we could push $\Sigma^\prime$ arbitrarily to the future to make its extension inside the BH grow without limits). In the figure the portion of $\Sigma^\prime$ inside the BH is represented by the dotted lines going down around the high curvature region. One can show that the volume inside measured at the latest possible instant where curvatures remain sub-Planckian (notice that moving $\Sigma^{\prime}$ to the future implies getting closer to  the singularity inside the horizon) is greater than a lowe bound that goes like $(M/M_p)^{\alpha} \ell_p^3$ with $\alpha>3$ and hence diverges in the $\hbar\to 0$ limit. }
\label{figu-info-bis}
%\end{center}
\end{figure}

\subsection{The information loss problem}\label{iloss}
 
Classically black holes are causal sinks. According to classical general relativity anything crossing a BH event horizon---Figure \ref{figu}, notice light cone structure---is constrained by the causal structure of spacetime to end up at the singularity, which in the classical theory is regarded as an endpoint of spacetime. When quantum effects are considered the situation changes as black holes evaporate through Hawking radiation. On the one hand, the classical notion of singularity is expected to be described by Planckian scale new physics and so the question of what is the dynamical fate of what falls into this region is expected to have a well defined description. On the other hand, this high curvature region, initially hidden by the BH event horizon, could become eventually visible to outside observers  at the end of evaporation or remain forever causally disconnected from the outside region. As these questions concern the physics in the Planckian regime, these questions can only be settled in the framework of a full theory of quantum gravity. Semiclassical considerations meet their end.

 According to the Hawking effect a black hole in isolation slowly lowers its initial mass $M$  by the emission of radiation which is very well approximated by thermal radiation (at least while the curvature at the horizon is far from Planckian, or equivalently, while the mass of the BH is much larger than $m_p$).  During this semiclassical-era of evaporation---which lasts an extremely long time $\tau_{evap}\approx M^3$ in Planck units \footnote{In more graphical terms $$\tau_{evap}\approx 10^{54} \frac{M^3}{M^3_{\odot}} \tau_{univ.}$$
where $M_{\odot}$ is the solar mass and $\tau_{univ}$ is the age of the universe.}---the black hole also works as an information sink. According to classical gravity and quantum field theory on curved spacetimes, whatever falls into the BH becomes causally disconnected from the outside at least for times $\tau \le \tau_{evap}$. Evolution of the quantum state of matter fields from one instant defined by a Cauchy surface $\Sigma$ (see Figure \ref{figu-info-bis}) to another defined by a Cauchy surface $\Sigma^\prime$ in its future (embedded in the lower than Planckian curvature region) is unitary. For the spacetime regions which can be well approximated by classical gravity there is part of the Cauchy surfaces $\Sigma^\prime$ that is trapped inside the BH and whose future is the classical singularity. As $\Sigma^\prime$ is pushed towards the future, the portion inside the BH grows (for instance in terms of its volume) as it approaches more and more the singularity (see dotted portion of $\Sigma^\prime$ in Figure \ref{figu-info-bis}). 

It is easier to emphasize one aspect of the information lost paradox by concentrating only on the Hawking radiation produced by the BH during its history, and thus neglecting for simplicity of the analysis all the other things that have fell into the BH during its long life (in particular those that led to its formation in the first place).
Hawking particles are created by the gravitational tidal interaction of the BH geometry with the vacuum state $|0\rangle$ of matter fields. This state can be expressed as a  pure state density matrix $|0\rangle\langle 0|$. It can be precisely shown that when a particle is created by this interaction and send out to the outside, another correlated excitation falls into the singularity \cite{Wald:1975kc, Hawking:1974sw, Hotta:2015yla}. It is because these correlated excitations that fell into the BH cannot affect any local experiment outside that an outside observer can trace them out and in this way get a mixed state
\be
\rho_{T_{\va \rm BH}}={\rm Tr}_{\rm \va BH}[|0\rangle\langle 0|].\ee
The previous mixed state is thermal state with Hawking temperature $T_{\rm \va BH}$ to an extremely good accuracy while the BH is large in Planck units (the trace is taken at a given instant defined by a Cauchy surface $\Sigma$). Yet the overall state of matter, when we do not ignore the fallen correlated excitations, is a pure state! The question of the fate of information in BH physics can then be stated in terms of the question of whether the quantum state of the system after complete evaporation of the BH is again a pure state (the initially lost correlations emerge somehow from the ashes of the end result of evaporation) or the state remains mixed and the information carried by the excitations that fell into the BH are forever lost. If the second scenario is realized then the unitarity of the description of the gravitational collapse and subsequent evaporation would be compromised.
This in itself would seem to have very little practical importance for local physics considering the time scales involved for BHs with macroscopic masses. But it could lead to an in principle detectable phenomenology in contexts where small BHs could be created in large numbers classically (such as for primordial BHs in cosmology) or by quantum fluctuations (yet in this case deviations from the semiclassical expectation could become important). Finally, the consideration of such purely theoretical questions can lead to new perspectives possibly useful for understanding the theory we seek.

\begin{figure}[t]
%\begin{center}
\centerline{\hspace{0.5cm} \(
\begin{array}{c}
\includegraphics[height=12cm]{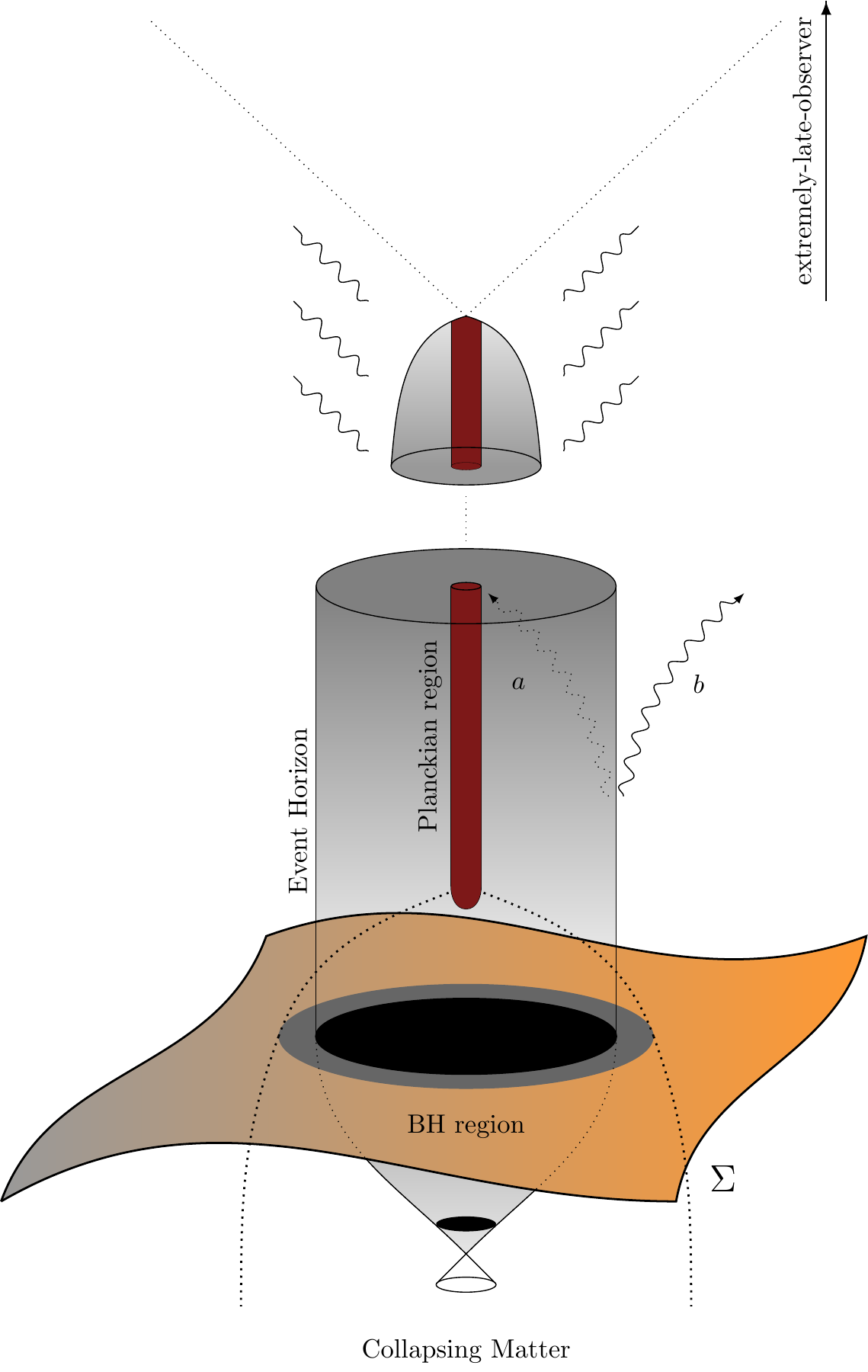} 
\end{array}\ \ \ \begin{array}{c}
\includegraphics[height=12cm]{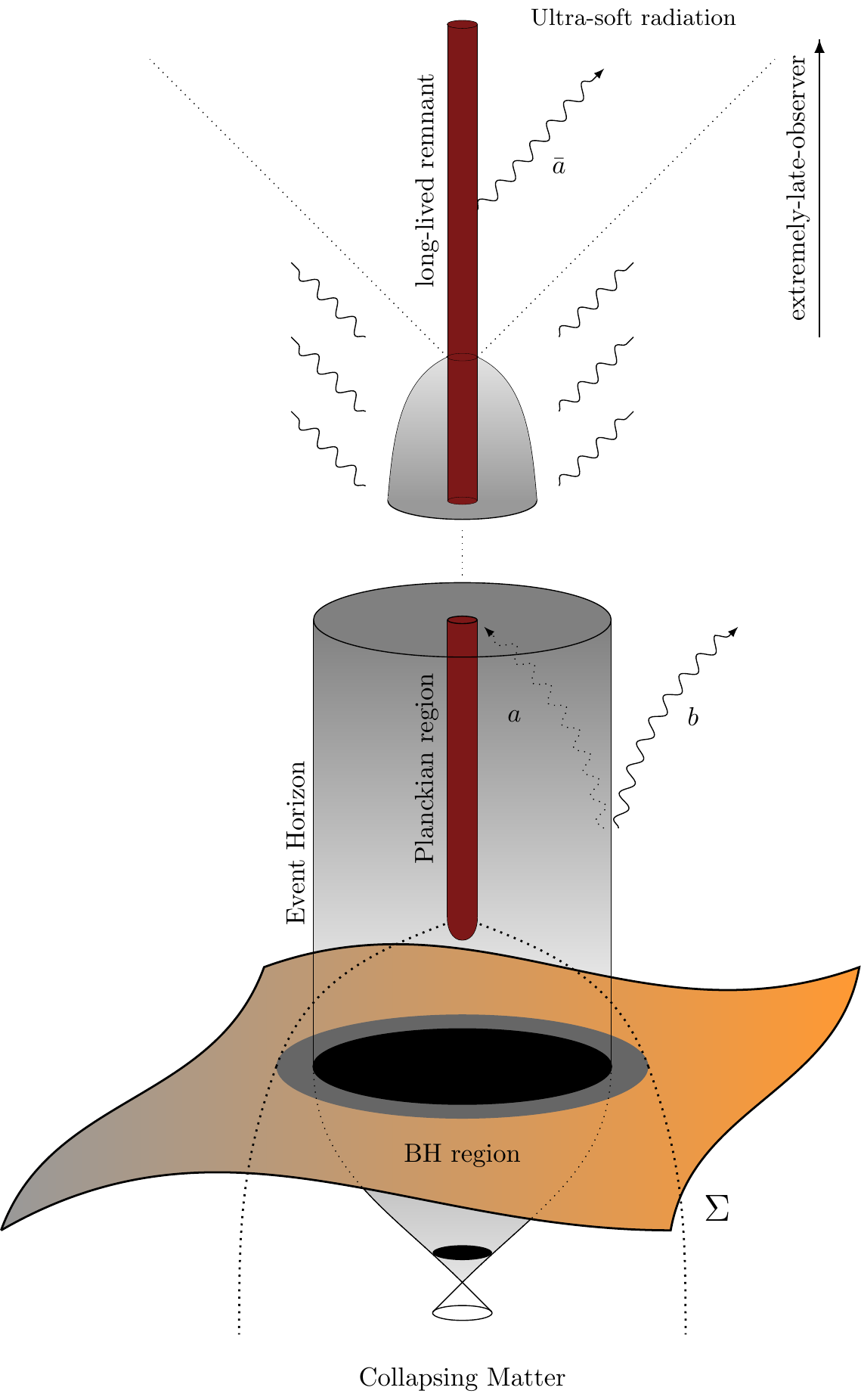} 
\end{array}\) } \caption{The left panel shows the situation as initially pictured by Hawking: a Hawking particle $b$ and its partner $a$, created from the vacuum by the interaction with the gravitational field when the black hole can still be considered semiclassical are maximally correlated. The in-falling particle $a$ enters the strong quantum fluctuation region (the {\em would-be-singularity} of classical gravity) and the information it carries is lost behind the horizon forever for outside observers. The right panel illustrates the remnant scenario: the information carried by particle $a$ remains in a quantum gravity region that becomes a remnant  after the horizon shrinks via Hawking evaporation to Planckian size and disappears. The information either remains forever inside a stable remnant, or is recovered via the emission of ultra-soft radiation $\bar a$ in a astronomically huge time of the order of $M^4$ in Planck units. }
\label{figu-info}
%\end{center}
\end{figure}

How much information has fallen into the BH at the end of evaporation? The answer really depends on the entire history of the BH. As an extreme example one could think of feeding a BH with matter continuously to compensate for the energy loss via Hawking radiation. In this way a BH can have a lifetime as large as wanted and thus can swallow an unlimited amount of information (independently of its apparent size as seen from the outside). Thus the answer would be {\em infinite} in this case. One could think of the opposite scenario where the BH is created quickly by gravitational collapse and left unperturbed in isolation until it completely evaporates. In this case we would expect that it has absorbed at least all the information that would be necessary to purify the hawking radiation that has been emitted during its evaporation process. Assuming that this process is close to stationary for the most part of the history of the BH and using the generalized second law of thermodynamics one expects the information lost to be of the order of the initial Bekenstein-Hawking entropy 
$S_{\rm BH} = a_0/(4\ell_p^2)$.    
   
The volume inside the BH right before hitting the singularity is huge when calculated at an ultimate instant defined by a constant curvature 
slice before the radius of curvature becomes Planckian inside the BH. At that instant the area of the BH as seen from the outside is Planckian. Nevertheless its internal volume defined by this spacelike slice can be as large as the the volume of a ball with a radius
$R\approx 10^6 \times (M/M_{\odot}) R_{\rm univ}$ where $R_{\rm univ}$ denotes the radius of the observable universe (for discussion of the volume inside a BH see \cite{ Christodoulou:2014yia, Christodoulou:2016tuu}). Such trapped volume is not bounded in any way by the area of the BH, for instance it can be made as big as desired by feeding the black hole with matter to compensate for its evaporation as in the situation evoked before where an unlimited amount of information would be absorbed by the BH. This is not 
surprising as in a curved geometry the volume of a region is not bounded in any way by the apparent size of the region from the outside (the black hole area for instance); in the present case this corresponds to Wheelers {\em bag of gold} scenario \cite{Whee}. 

Hawking 1976 \cite{Hawking:1976ra} formulation of the information paradox can be stated in the questions: is the information falling into the BH region forever lost? 
or can it be recovered at the end of BH evaporation? It is clear that the answer to these questions is tight to the fate of the causal structure of spacetime across the BH singularity, and is therefore a quantum gravity question.  These questions will only be clarified when a solid understanding of the Planckian dynamics becomes 
available. The central interest of Hawking's information paradox is the theoretical challenge it represents; it tells us that one cannot ignore the physics of the singularity.

The following scenarios represent some of the main ideas that have been put forward during the last 4 decades:
\begin{enumerate}
 \item {\em Black holes are information sinks:} A simple possibility is that even when the singularity is replaced by its consistent Planckian 
 description one finds that the excitations that are correlated with the outside can never interact again with it and remain in some quantum gravity sense forever causally disconnected from the outside.  There are two possibilities evoked in the literature: The first possibility is that lost information could end entangled in a pre-geometric quantum substrate (where large quantum fluctuations \cite{Ashtekar:1996yk} prevent any description in terms of geometry); this would be described as a singularity from the point of view of spacetime physics in which case the place where informations ends could be seen as a boundary of spacetime description \cite{Wald2001}. The second is that to the future of the singularity (a region of large quantum fluctuations at the Planck scale) a new spacetime description becomes available but that the 
 newly born spacetime regions remain causally disconnected from the BH outside: a baby universe \cite{Frolov:1989pf, Frolov:1988vj}.

\item {\em Information is stored in a long-lasting remnant:} 
  
 A concrete proposal consists of assuming that a {\em remnant} of a mass of the order of Planck mass at the end of the Hawking evaporation can carry the missing information \cite{Aharonov:1987tp, Giddings:1992hh}.
As the final phase of evaporation lies outside of the regime of applicability of the semiclassical
analysis such hypothesis is in principle possible.  Notice that this might be indistinguishable from the outside from the baby universe possibility if no information is allowed to come out of this remnant.  The Planckian size remnant will  look as a point-like particle to outside observers.

In order to purify the state of fields in the future,  the remnant must have a huge number of internal degrees of freedom which correlate with those of the radiation emitted during evaporation in addition to those related to the formation history of the BH. If one traces out these degrees of freedom one has a mixed state that represents well the physics of local observers in the future right before the end of evaporation. The entropy of such mixed state is expected to be at least as big as the one of the initial black hole $S_{BH}(M)$ before Hawking evaporation starts being important. The value of $S_{BH}(M)$ is a lower bound of the number of such internal states, which, as pointed out above, it can be virtually infinite depending of the past history of the BH.
If this particle like object admits a description in terms of an effective field theory (which in itself is not so clear \cite{Hossenfelder:2009xq}) this huge internal degeneracy would lead to an (unobserved) very large pair production rate in standard particle physics situations. One can contemplate the possibility that these remnants could decay via emission of soft photons carrying the missing information back to the outside. However,  as the energy available for this is of the order of $M_p$,  remnants would have to be very long lived (with lifetimes of the order of $M^4/M_p^4 t_{p}$ \cite{Bianchi:2014bma}) in order to evacuate all the internal information in electromagnetic, gravitational or any other field-like radiation.
Hence, they would basically behave as stable particles and one would run into the previous difficulties with large pair creation rates. The possibility that such remnants  can lead to finite rate production despite of the large degeneracy of their spectrum has been suggested \cite{Banks:1992is, Banks:1992mi}. More discussion of remnants and references see \cite{Hossenfelder:2009xq}.  Some aspects of the previous  two scenarios is illustrated in Figure \ref{figu-info}.

\item {\em Information is recovered in Hawking radiation:}

Another proposed scenario for purification of the final state of black hole evaporation consist of postulating that information comes out with the Hawking radiation. This view has been advocated by  't Hooft in \cite{'tHooft:1990fr} and raised to a postulate by Susskind et al. \cite{Susskind:1993if} where one declares that ``there exists a unitary $S$-matrix which describes the evolution from infalling matter to outgoing Hawking-like radiation''. See also Page \cite{Page:1993wv}. Such view cannot accommodate with the spacetime causal structure representing the BH within the framework of quantum field theory on curved spacetimes (see light cones in Figure \ref{figu-info-bis}). New physics at low energy scales is invoked to justify that information of the in-falling modes is somehow `registered' at the BH horizon and sent back out to infinity.
More precisely, if standard QFT on a curved space-time is assumed to be a valid approximation when the curvature around the black hole horizon is low (for large BHs) then no information on the in-falling modes can leak out the horizon without violating causality. Yet, as argued by Page \cite{Page:1993up}, in order for unitarity to hold such leaking of information must be important when the BH is still large and semiclassical (at Page time corresponding to the time when the BH has evaporated about half of its initial area $a$). Some peculiar quantum gravity effect must take place at the BH horizon. Further tensions arise when trying to describe the physics from the point of view of freely falling observers who (according to the equivalence principle)  should not feel anything special when crossing the BH horizon. In particular, they must find all the information that fell through the horizon right inside. Thus, there is a troublesome doubling of information  in this scenario:  a so-called principle of  {\em complementarity} is  evoked in trying to address these issues \cite{Susskind:1993if}. Hence, the above postulate implies that  quantum gravity effects would be important where they are not expected to be opening the door for paradoxical situations with theoretically convoluted proposed resolutions.

\begin{figure}[t]
%\begin{center}
\centerline{\hspace{0.5cm} \(
\begin{array}{c}
\includegraphics[height=8cm]{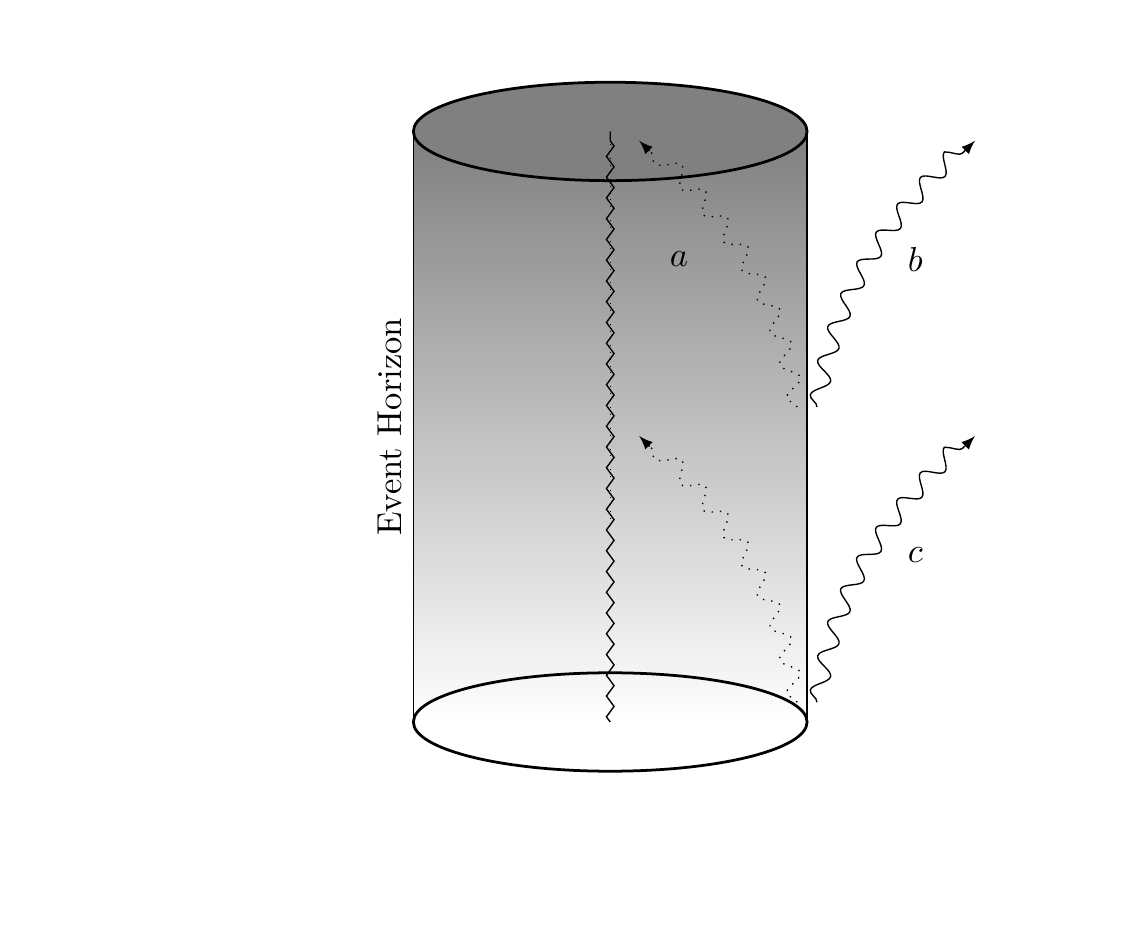} 
\end{array}\) } \caption{ According to quantum field theory the pair of particles $a$ and $b$ are created from the vacuum in a maximally entangled state \cite{Unruh:1976db, Mathur:2009hf}. Monogamy of entanglement \cite{Coffman:1999jd} precludes the possibility of having a non trivial entanglementent between the early Hawking radiation $c$ and the late Hawking radiation $b$. Therefore, according to standard quantum field theory (which should be valid for macroscopic black holes) information cannot come out with the Hawking radiation during the semiclassical phase. This invalidates the heuristics of the Page curve \cite{Page:1993up, Page:1993wv}. Moreover, if one postulates deviations from maximal entanglement between $a$ and $b$ at later times, in order to allow for non trivial correlations between $c$ and $b$, then a divergence of the energy momentum tensor develops at the location of the black hole horizon \cite{Braunstein:2009my, Almheiri:2012rt}.}
\label{AMPS}
%\end{center}
\end{figure}

The existence of such puzzling {\em large quantum gravity effects} in the present scenario was made manifest by the analysis of \cite{Braunstein:2009my, Almheiri:2012rt} where it is explicitly 
shown that (assuming the validity of semiclassical QFT in the vicinity of a large BH) one cannot have information go out of a large BH and across its horizon without a catastrophic violation of the equivalence principle at the BH horizon!

A cartoon description of the phenomenon can be given with the help of Fig. \ref{AMPS} as follows. According to the formalism of QFT on curved space-times the UV structure of the two-point correlation functions is universal for well behaved (Hadamard) states. Physically, this implies that the state of fields looks like  `vacuum' to freely falling observers crossing the horizon with detectors sensitive to wavelengths much shorter that the BH size. In the context of the Hawking effect this implies  that a pair of particles $a$ and $b$ created at the horizon by the interaction of the field with the background geometry must be maximally correlated \cite{Mathur:2009hf}.  The statement that the final state of the Hawking radiation is pure (and thus that information has sneaked out of the horizon during the evaporation process) necessitates the existence of non trivial correlations between the early radiation (particle $c$ in Fig. \ref{AMPS})  and late radiation (particle $b$) \footnote{The analogy with the cooling of a hot body in standard situations is often drawn. An initially hot body can be described by a pure state. While it cools down it emits radiation that looks close to thermal at any particular time. Unitarity implies (among other things) that the early radiation be correlated with the late one to maintain the purity of the state. Such view is misleading when applied to BHs because it disregards the causal structure.}. But correlations between $c$ and $b$ are forbidden by the fact that $a$ and $b$ are maximally correlated. 
This is due to the so-called monogamy of entanglement in quantum mechanics \cite{Coffman:1999jd}.
Relaxing correlations between $a$ and $b$  implies a deviation from the notion of `vacuum' at the point where $a$ and $b$ are created. This perturbation leads to a divergence of the energy-momentum tensor  at the horizon in the past due to its unlimited blue-shift along the horizon towards the past: a `firewall' detectable by freely falling observers. If one is not ready to accept such flagrant violation of the equivalence principle one must admit the inviability of the complementarity scenario.

\item {\em Information is degraded via decoherence with Planckian degrees of freedom:}

This is a natural proposal where the fundamental discreteness of quantum geometry at the Planck scales plays a central role in understanding the puzzle of information. The information loss is viewed as a simple phenomenon of decoherence with the quantum gravity substratum reflected in an increase of the von Neumann entropy describing the state of Hawking radiation. This perspective puts in equal footing  the apparent lost of information in the BH context with the degrading of information taking place in the more familiar situations described by standard thermodynamics and captured by the second law.

The second law of thermodynamics is not a fundamental principle in physics but rather a statement
about the (illusory) apparent asymmetry of time evolution when sufficiently complicated systems are put in special initial conditions and later described statistically in terms of coarse physical variables that are unable to discern all the details of the fundamental system. The idea is easily illustrated in classical mechanics. On the one hand, Liouville's theorem implies that the support of the phase space distribution of the system spans a volume that is time independent; on the other hand, the shape of the support is not restricted by the theorem. An initially simple distribution supported in a ball in $\Gamma$ will (in suitably complicated systems) evolve into a more and more intricate shape whose apparent phase space volume, when measured with a devise of resolution lower than that of the details of the actual distribution, will grow with time.    
In a practical sense, the second law implies that information is degraded (yet not lost) in time when encoded in 
coarse variables. The words in a newspaper are gone when the newspaper is burned but the information they carry continue to be encoded in the correlation among microscopic molecular degrees of freedom that become unavailable in practice. 

At the quantum level information can be degraded in addition due to decoherence through the entanglement with degrees of freedom that are not accessible to the observer \cite{thibaut}. In fact this view leads to a beautiful statement of the foundations of statistical mechanics and sets the fundamental basis for thermodynamics \cite{popescu, thibaut}. It is the view of the author that this perspective offers the possibility of a simple solution of the information loss paradox in the context of a quantum gravity theory where spacetime geometry is granular or discrete at the fundamental level \cite{Perez:2014xca}. 

We have seen that BHs behave like thermodynamical systems. The validity of the laws of BH mechanics and their strict relationship with thermodynamics points to an underlying fundamental description where spacetime is made of discrete granular structures. Without adhering to any particular approach to quantum gravity, the solid theoretical evidence coming from general relativity and quantum field theory on curved spacetimes strongly suggest that BH Horizons are made of Planckian size building blocks: they carry and entropy given by $A/4$ in Planck units and satisfy the generalized second law (total entropy of matter plus BH entropy can only increase). In the framework of LQG we have seen in Section \ref{BHE} that it is precisely the huge multiplicity microscospic quantum states of the BH geometry that can account for its   thermal properties. Such microscopic degeneracy of states is also expected in the description of the continuum limit in LQG as argued in Section \ref{thecontlim}.  If these expectations are correct then the information puzzle must be understood in terms that are basically equivalent to those valid in familiar situations. Information is not lost in BH evaporation but degraded in correlations with these underlying `atoms' of geometry at Planck scale. In this scenario BH evaporation is represented by Figure \ref{figu-info-2} and will be presented in more detail in Section \ref{info-lqg}.

\end{enumerate}

\subsection{Information loss resolutions suggested by LQG}\label{info-lqg}

The means for the resolution of the information puzzle, advocated here, can be formulated in the context of the scenario proposed by Ashtekar and Bojowald (AB) in \cite{Ashtekar:2005cj}. 
The central idea in the latter paper is that the key to the puzzle of information resides in understanding the fate of the classical {\em would-be-singularity} in quantum gravity. This view has enjoyed from a steady consensus in the non perturbative quantum gravity literature \cite{Hossenfelder:2009xq}.

The scenario was initially motivated by the observed validity of the unitary evolution across the initial big-bang singularity in symmetry reduced models in the context of {\em loop quantum cosmology} \cite{Bojowald:2001xe} (see \cite{Ashtekar:2011ni} and references therein for a modern account). Similar singularity avoidance results due to the underlying discreteness of LQG  have been reported  recently in the context of spherically symmetric black hole models \cite{Gambini:2014qga, Gambini:2014lya} (see also \cite{Saini:2016vgo}).   The consistency of the AB paradigm is supported by the analysis of \cite{Ashtekar:2008jd} in two dimensional CGHS black holes \cite{PhysRevD.45.R1005} where still some assumptions about the validity of quantum dynamics across the singularity are made.  Numerical investigations of the CGHS model in the mean field approximation \cite{Ashtekar:2010hx, Ashtekar:2010qz} strongly suggest the global causal picture proposed in the AB paradigm as well.

\begin{figure}[t]
%\begin{center}
\centerline{\hspace{0.5cm} \(
\begin{array}{c}
\includegraphics[height=12cm]{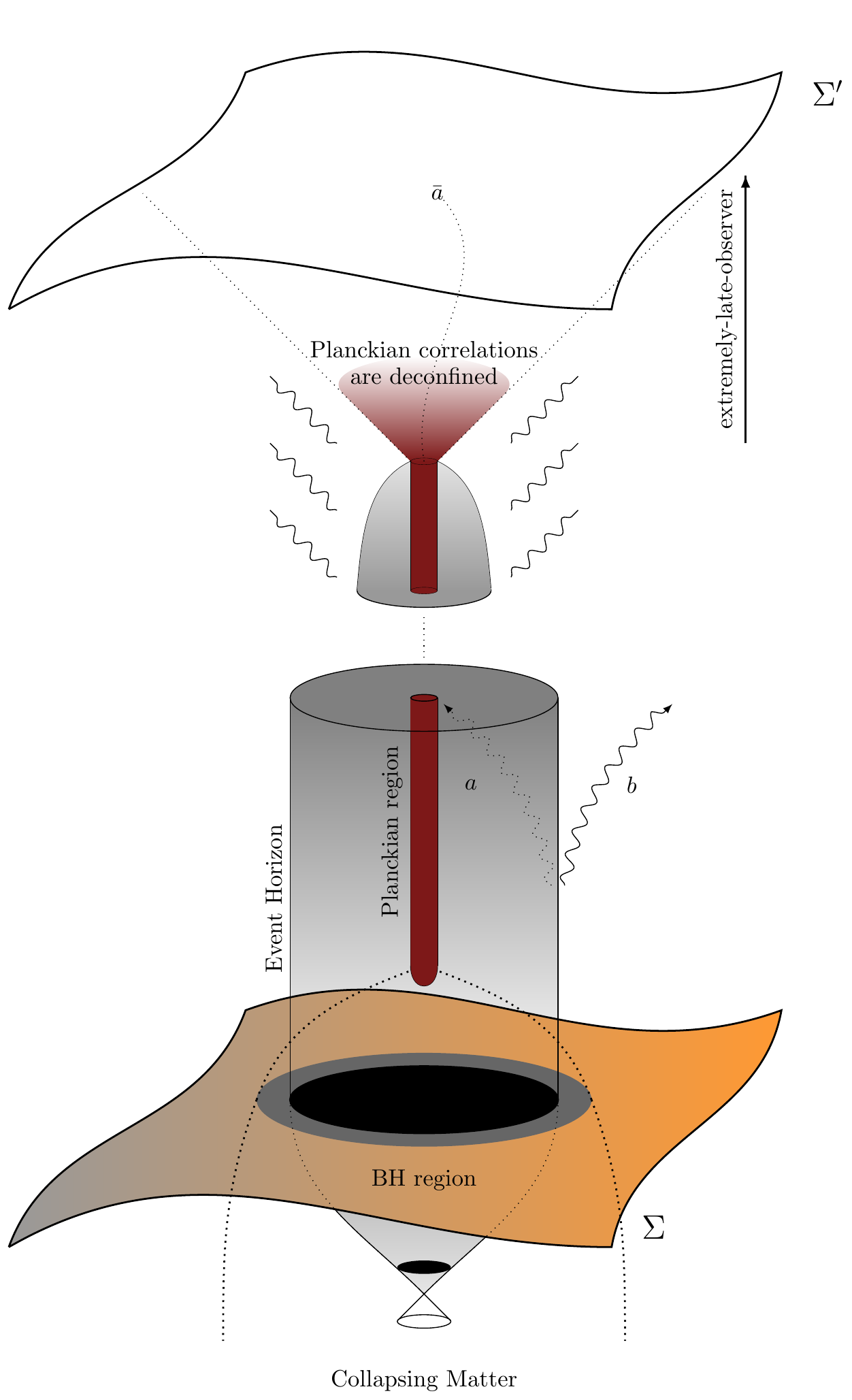} 
\end{array}\) } \caption{Possible scenario for unitarity in loop quantum gravity. Correlations between a Hawking particle $b$ and its partner $a$, created from the vacuum by the interaction with the gravitational field when the black hole can still be considered semiclassical, are not lost. The in-falling particle enters the strong quantum fluctuation region (the {\em would-be-singularity} of classical gravity) and interacts with the microscopic quantum granular structure of the spacetime geometry; the quantum geometry of Section \ref{qg}. The correlations between $a$ and $b$ are not lost they are transferred to Planckian degrees of freedom denoted by $\bar a$  in the strong quantum region. They become in principle accessible after the BH has completely evaporated. The Hawking radiation is purified by correlations with these Planckian  micro-states which cannot be described in terms of the usual matter excitations. Information is not lost but simply degraded;  just as when burning a news paper the information in the text becomes inaccessible in practice as it has been transferred  to correlations between the molecules of gas produced by the combustion. }
\label{figu-info-2}
%\end{center}
\end{figure}

The spacetime of the AB framework is represented in Figure \ref{figu-info-2}. Hypothetical observers falling into the black hole unavoidably meet the {\em would be singularity}. Quantum gravity evolution in the Planckian region takes us across the singularity to the future where the BH has evaporated.  In the AB proposal the space-time may rapidly become semiclassical so that our test observer emerges into a flat space-time future above the {\em would-be-singularity} where spacetime is close to a flat spacetime (because all of the mass sourcing gravity has been radiated away to infinity via Hawking radiation).  The region where it emerges is in causal contact with the outside. From the view point of an external observer the black hole slowly evaporates until it becomes Planckian. At this final stage the semiclassical approximation fails, curvature at the black hole horizon becomes Planckian and  external observers become sensitive to the strong quantum gravitational effects which are responsible for the resolutions of the classical  {\em would-be-singularity}. In practical terms this means that external observers become in causal contact with the strong quantum gravity region and the singularity becomes naked for them.

In this framework there is a natural resolution of the question of the fate of information. The full quantum dynamics is unitary when evolving from $\Sigma$ to $\Sigma^{\prime}$ (see Figure \ref{figu-info-2}). The correlations in a Hawking pair $a-b$ created in the vicinity of the BH horizon (Figure \ref{figu-info-2}) are maintained by the evolution. The field excitation $a$ falls into the Planckian region where it interacts with the fundamental discrete spacetime foam structure and gets imprinted into the Planckian fabric in what we call $\bar a$. Correlations that make the state pure are not lost, at the end of evaporation the quanta $b$ in the Hawking radiation are entangled with Planckian degrees of freedom $\bar a$ which cannot be encoded in a smooth description of the late physics. These degrees of freedom are associated with a large degeneracy of the flat Minkowski spacetime expects to arise from the continuum limit of LQG via coarse graining: $\bar a$ is a defect in the fundamental structure not detectable for the low energy probes for whom the spacetime seems smooth.  
The granular structure predicted by LQG can realise the idea of decoherence without dissipation evoked in \cite{Unruh:1995gn, Unruh:2012vd}. The account of the fate of information in the context of BHs evaporation would, in this scenario, be very similar to what one believes it happens when burning the newspaper.  After burning, the articles in the newspaper remains written in the correlations of the gas molecules diffusing in the atmosphere. After evaporation the information initially available for low energy probes in the initial data that lead to the gravitational collapse is encoded in the correlations with Planckian degrees of freedom which are harder to access. Information gets degraded but not lost: the `fire' of the singularity is a place where the initially low energy smooth physics excitations are forced, by the gravitational collapse, to interact with the Planckian fabric where a new variety of degrees of freedom are exited.     

The viewpoint developed in considering the question of information in quantum gravity leads to some phenomenological proposals that we briefly describe in what follows.

\section{Discreteness and Lorentz invariance}\label{LIV}

A central assumption behind all the results and perspectives discussed in this article is the compatibility of the prediction of loop quantum gravity of a fundamental discreteness of quantum geometry at the Planck scale with the continuum description of general relativity. As emphasized before the problem of the continuum limit of LQG remains to a large extend open  partly  due to the technical difficulties in reconstructing the continuum from the purely  combinatorial structures of quantum geometry, but also due to the difficulties associated with the description of dynamics in the framework (spacetime is a dynamical question involving the solutions of \eqref{dosl}). 

A problem that immediately comes to mind is the apparent tension between discreteness and the Lorentz invariance (LI) of the continuum low energy description. Is the notion of a minimum length compatible with Lorentz invariance? The apparent tension was initially taken as an opportunity for quantum gravity phenomenology as such a conflict would immediately lead to observable effects (see \cite{Mattingly:2005re} and references therein). Given the lack of clear understanding of the continuum limit,  it was initially assumed that the discreteness of quantum gravity would select a preferred rest frame  breaking LI  `only' at the deep Planckian regime. However, it was later shown \cite{Perez:2003un, Collins:2004bp,  Collins:2006bw, Polchinski:2011za} that such naive violation of LI would not be compatible with standard QFT at familiar energy scales: violations of LI at Planck scale would generically get amplified via radiative corrections and thus `percolate' from the Planck scale to low energy scales producing effects that would be of the same order of magnitude as the phenomenology predicted by the standard model of particle physics. This is in sharp conflict with observations.

These results indicate that discreteness in quantum gravity does not admit a naive interpretation as some granular structure similar to molecules or atoms in a lattice. Compatibility with LI requires a more subtle relation expected to be clarified via the precise understanding of the continuum limit and the solutions of the scalar constraint \eqref{dosl} (the quantum nature of such discreteness is probably one aspect of its elusive nature \cite{Rovelli:2002vp}). The key point seems dynamical physical discreteness should be associated to gauge invariant quantities commuting with the scalar constraint. Let us illustrate this with the simpler case of the area operator which is not a gauge invariant observable (a Dirac observable) unless further structure is provided.  In the case of the BH models considered here the area of the event horizon is a gauge invariant notion (thanks to the restrictions imposed by the isolated horizon boundary condition) and its discreteness is justified. Notice also that only normal Lorentz transformations preserve the boundary condition and for such the area is an invariant notion. 

The apparent tension of the discreteness as predicted by our calculations in Section \ref{qg} in view of the expected LI at low energies 
can be attenuated with general dynamical considerations as well. Unfortunately, unlike the argument for BHs we just gave, in the general situation the discussion will remain at a more heuristic level until more control on the dynamical question is gained. However, we can be precise if we use the concrete scenario provided by models where time-reparametrization invariance (the gauge symmetry associated to \eqref{dosl}) is eliminated by the use dust or other suitable (massive) matter degrees of freedom as a physical gauge fixing \cite{Brown:1994py, Giesel:2007wn, Domagala:2010bm, Giesel:2012rb}.  Discreteness of geometry at the Planck scale realizes in relational observables \cite{Rovelli:1990ph, Rovelli:2001bz} like area and volume of regions in the rest frame of matter degrees of freedom. Compatibility, with Lorentz invariance comes from the fact that the discreteness of geometric observables is associated with such preferred `observers' selected by the gauge fixing  degrees of freedom.
In these models the Planck length enters in a way that is similar to the mass $m$ of a field in relativistic field theory. The presence of a scale does not break Lorentz symmetry because the meaning of $m$ is that of the rest-mass of the associated particle (it means a definite scale in a special reference frame). Similarly, the discreteness of geometry in the deparametrized context has a meaning in a reference frame determined by the physical degrees of freedom. 
These models are simplistic in that the matter `rulers' that provide the gauge fixing that eliminates (\eqref{dosl}) are not properly quantized but it illustrates clearly the way in which the apparent tension between discreteness and Lorentz-FitzGerald contraction could be resolved.

Waiting for a more detailed understanding we also mention that, in the context of applications to black holes, discreteness of the geometry of null surfaces (themselves a LI object) is the key feature behind the results we have discussed. Another important point concerning BHs is that the results presented here would  all be preserved (only with a possible modification of the the value of $\gamma_0$ in (\ref{gnot})) as long as a non trivial area gap remains (the area spectrum can be continuous as long a there is a minimum non-vanishing area eigenstate \cite{Ghosh:2013iwa, Frodden:2012dq}). In the context of spin foams \cite{Rovelli:2010ed} (which provides the framework for understanding the continuum limit dynamically) there are indications that the area gap is a LI feature of quantum geometry \footnote{There is for instance the presence of the area gap in the covariantly derived area operator \cite{Alexandrov:2002br}, and the persistence of the gap in the LI definition of the area operator in self dual variables \cite{Frodden:2012dq}.
See also \cite{Dittrich:2007th} for a general discussion on dynamics versus discreteness.}. Although in such case the physical interpretation in the previous terms seems elusive.

\subsection{Phenomenology}\label{pheno}

The discussion of black hole issues in quantum gravity suggests interesting avenues for phenomenology based on the possible observational implications of Planckian discreteness. Some years ago there was an initial surge of interest in quantum gravitational effects associated with violations of Lorentz invariance mediated by effects associated with a preferred frame where discreteness would realize \cite{AmelinoCamelia:1997gz, AmelinoCamelia:1998ax, Ahluwalia:1999aj}.  However, it  has by now  become  quite clear  that this idea  faces  severe  problems. From   the  direct  observational side one   can conclude  that,   if  that kind of    effects   exist  at all,  they  must  be far more suppressed than initially  expected \cite{Jacobson:2002ye, Ackermann:2009aa}.  From the theoretical side these effects are forbidden by the no-go argument of \cite{Collins:2004bp}.   However,  this  result has created a great puzzle:  in what way  space-time Planckian  scale  discreteness (as predicted by LQG) could actually be realized in consistency with the observed Lorentz invariance? Reference \cite{Collins:2004bp} rules out the  direct  and  naive atomistic view of a spacetime made of pieces stuck together in some sort of space-lattice, but it does not offer  a  clear  answer   to  the  question. The answer must come from the dynamical understanding of the theory (the solution of (\ref{dosl})) and the construction of physical observables. 

A related idea that avoids this no-go argument is that spacetime discreteness can lead to violations of conservation of the energy momentum of matter fields when idealized as propagating in the continuum (no violation of lorentz invariance in the previous naive sense is necessary; see for instance \cite{Dowker:2003hb, Philpott:2008vd}). The idea is that discreteness would naturally lead to ``energy diffusion'' form the low energy field theoretical degrees of freedom to the micro-Planckian structure of spacetime. 
Such diffusion is generically unavoidable if the decoherence scenario evoked in Section \ref{info-lqg} is realized. 
Therefore, this phenomenological idea is partly motivated by our considerations of the information puzzle in BH evaporation.

Violations of energy momentum conservation are incompatible with Einstein's equations; however, in the context of cosmology, unimodular gravity can be shown to be a good effective description of violations that respect the cosmological principle \cite{Josset:2016vrq}. In that case, the effect of the energy leakage is the appearance of a term in Einstein's equations satisfying the dark energy equation of state with contributions that are of the order of magnitude of the observed cosmological constant.  Dimensional analysis, together with the natural hypothesis that Planckian discreteness would primarily manifest  in interactions with massive matter (see Section \ref{LIV}) in a way that is best captured by the scalar curvature $R$ (vanishing via Eintein's equation for conformally invariant massless matter),  lead to the emergence of a cosmological constant in agreement with observations without fine tuning \cite{nosotros}.  These results are new and poorly understood from the perspective of a fundamental theory of quantum gravity. Nevertheless, they are encouraging and present a fresh view on the dark energy problem that seems promising.  

Finally, another phenomenological aspect that follows from the discussion of BHs in LQG  is the suggestion that quantum effects in gravitational collapse might be stronger than those predicted by the semiclassical framework that leads to Hawking evaporation. These hypothetical strong quantum gravity effects would be important in regions of low curvatures near the event horizon and could actually dominate at some stage of the black hole collapse. The models are motivated by heuristic considerations based on bouncing cosmologies in LQG \cite{Rovelli:2014cta} and later refined in \cite{Haggard:2014rza, Rovelli:2014cta}. The initially proposed spacetimes suffer of certain instabilities \cite{DeLorenzo:2015gtx, Bianchi:2014bma}. In these scenarios black holes would explode in time scales of order $M^2$ (in Planck units) \cite{Christodoulou:2016vny}  and, as argued, they might lead to precise observational signatures \cite{Barrau:2014yka, Barrau:2015uca, Barrau:2014hda}.

\section{Future directions and discussion }\label{outlook}

At present there is no complete understanding of that unified framework of quantum mechanics and gravity that we  call quantum gravity. Several theoretical approaches exist with their advantages and disadvantages depending on the judgement of what physical phenomena are the most relevant guiding principles.   Loop quantum gravity is not an exception to such assertion. Important implications of the formalism remain unclear such as the (dynamical) question of the continuum limit or that of the nature of matter at the fundamental scale.
In such context, black hole physics is a challenge and an opportunity  where phenomenology, firmly rooted in predictions of general relativity and quantum field theory on curved spacetimes, guides our steps for the construction of a consistent theory. In this sense Black holes are cosmic  microscopes of the fundamental structure of space and time. They hide the key for solving the puzzle of quantum gravity.

In this article we have reviewed the main achievements of the formalism of loop quantum gravity when applied to black holes. The central feature  behind all these results is the discreteness of geometry at the Planck scale that follows directly (as explained in Section \ref{qg}) from the canonical uncertainty relations of gravity in the first order variables. We argued that there is a finite dimensional ensemble of possible gravitational actions in these variables, Section \ref{caca}, and that the Immirzi parameter arises from the associated coupling constants.  In the quantum theory the Poisson non-commutativity of geometric variables implies the discreteness of area and volume whose eigenvalues  
are modulated by the Immirzi parameter $\gamma$ (see for instance equation \eqref{area1}). The parameter $\gamma$ is thought of as labelling inequivalent quantizations.

We have seen that the approach succeeds in explaining the proportionality of black hole entropy with its horizon area without the need of invoking  holographic ideas at the fundamental level. Consistency with the low energy semiclassical limit requires a very definite value of the proportionality constant between area and entropy. Two competing perspectives coexist at present. On the one hand, there is the view (motivated by the formalism of quantum isolated horizons) that only geometry degeneracy must contribute to the entropy, Section \ref{dirc}. In this case semiclassical consistency is achieved by fixing the value of the Immirzi parameter as in equation \eqref{gnot}. On the other hand,  if matter degrees of freedom are taken into account and punctures are assumed to be indistinguishable, we have seen, Section \ref{holi}, that it is possible to achieve semiclassical consistency for arbitrary values of the Immirzi parameter ($\gamma$ only appears in subheading quantum corrections to the entropy). Moreover, if in addition Boson statistics is postulated for punctures then correspondence with the continuum limit holds; Section \ref{contlim}.   

At present there is no consensus on which of the previous alternative views is more appropriate.  
The second perspective is more challenging as it demands deeper understanding of the of the nature of matter degrees of freedom at the Planck scale. This is a difficult yet potentially promising direction where the properties of black holes can teach us about some aspects of matter coupling of LQG at high energies.  In Section \ref{stringy} we mention some ideas which can be considered first steps in this direction. 

We have seen in this article that black holes are modelled in terms of boundaries and the imposition of boundary conditions at the classical level. This approach is natural in the semiclassical context where black holes are large in Planck units and thus radiate so little that  can be idealised as stationary.  Quantum aspects are explored via canonical quantization of the phase space to general relativity restricted by these boundary conditions. 
In the dynamical regime black holes are more elusive notions. Indeed it is likely that the very notion of black hole (as a trapped region) makes no sense in the full quantum gravity regime (recall discussion in Section \ref{loss}). We have also discussed how  some of the most puzzling issues such as the emergence of the Lorentz invariant continuum, or the fate of information in gravitational collapse requires the full dynamical description of the evaporation process and, what classically would be regarded as, the BH singularity. At present one can argue for possible scenarios on the basis of general features such as discreteness of geometry at the Planck scale. However, the precise treatment of  these hard questions necessitate full control of the quantum theory and its dynamics at the Planck scale. There is active research on basically two fronts trying to address the dynamical question: the spin foam approach towards the path integral representation of LQG \cite{Perez:2012wv}, and the canonical Dirac program of regularisation and quantisation of the quantum Einsteins equations \cite{Laddha:2014xsa, Henderson:2012ie, Laddha:2011mk}. In the near future, perhaps the reader will contribute with new insights  into these pressing questions.

\section*{Acknowledgments}
I would like to thank Tommaso De Lorenzo, Thibaut Josset, Daniele Pranzetti, Carlo Rovelli, Daniel Sudarsky and Madhavan Varadarajan for discussions. 
This work was supported in part by 
OCEVU Labex (ANR-11-LABX-0060) and the A*MIDEX project (ANR-11-IDEX-0001-02) funded by the ``Investissements d'Avenir" French government program managed by the ANR.

%\bibliography{referencias}
%\bibliographystyle{h-elsevier}

\end{document}